\documentclass[reprint, amsmath, amssymb, aps, pre]{revtex4-1}
\usepackage{graphicx, dcolumn, bm, amsmath, amssymb, xcolor, enumitem, float}
\usepackage[spaces]{grffile}
\allowdisplaybreaks

\let\tempvec\vec
\renewcommand{\vec}[1]{\tempvec{{}#1}}
\let\tempbar\bar
\renewcommand{\bar}[1]{\tempbar{{}#1}}

\newcommand{\pmat}[2]{\def\arraystretch{#2}\begin{pmatrix}#1\end{pmatrix}}
\newcommand{\bmat}[2]{\def\arraystretch{#2}\begin{bmatrix}#1\end{bmatrix}}
\newcommand{\re}{\textrm{Re}}

\begin{document}
	\title{On the effect of boundaries on noninteracting weakly active particles in different geometries}
	\author{Michael Wang}
		\email{mw3189@nyu.edu}
	\affiliation{Department of Physics and Center for Soft Matter Research, New York University, 726 Broadway, New York, New York 10003, USA}
	
	\begin{abstract}
		We study analytically how noninteracting weakly active particles, for which passive Brownian diffusion cannot be neglected and activity can be treated perturbatively, distribute and behave near boundaries in various geometries.  In particular, we develop a perturbative approach for the model of active particles driven by an exponentially correlated random force (active Ornstein-Uhlenbeck particles).  This approach involves a relatively simple expansion of the distribution in powers of the P\'{e}clet number and in terms of Hermite polynomials.  We use this approach to cleanly formulate boundary conditions, which allows us to study weakly active particles in several geometries: confinement by a single wall or between two walls in 1D, confinement in a circular or wedge-shaped region in 2D, motion near a corrugated boundary, and finally absorption onto a sphere.  We consider how quantities such as the density, pressure, and flow of the active particles change as we gradually increase the activity away from a purely passive system.  These results for the limit of weak activity help us gain insight into how active particles behave in the presence of various types of boundaries.
	\end{abstract}
	
	\maketitle
	\section{Introduction}
		Active particles consume fuel locally to propel and generate persistent motions \cite{Ramaswamy,Bechinger et al}.  Examples of such self-propelled particles range from humans \cite{Silverberg et al} down to microorganisms \cite{Berg et al,Polin et al} and artificial swimmers \cite{Palacci et al,Paxton et al,Walsh et al}.  The propulsion and, most importantly, the persistence or correlation time of the direction of propulsion are responsible for out-of-equilibrium phenomena such as phase separation without attractive interactions \cite{Buttinoni et al} and preferential motion through funnel-shaped walls \cite{Galajda et al} or around gear-like objects \cite{Leonardo et al, Sokolov et al}.
		
		When an active particle collides with a solid boundary, it often surfs along the boundary until eventually turning around and propelling away \cite{Galajda et al,Volpe et al,Li Tang}.  This behavior is difficult to analyze mathematically because it often leads to singular behavior at a boundary.  One approach is to treat a system of active particles (without passive diffusion) in the presence of a wall as two coupled populations of particles: those stuck at the wall and those in the bulk \cite{Lee, Wagner et al,Ezhilan et al}.  This results in additional terms in the equations for density that capture the fluxes of particles from the bulk to the wall and similarly, from the wall back into the bulk.  This formulation is related to a class of models known as two-way diffusion equations in which one must specify how particles at the wall reenter the bulk \cite{Fisch Kruskal,Beals,Wagner Beals}.  An alternative approach is to represent boundaries as soft confining potentials, which has been useful for studying the pressure and distribution of active particles \cite{Solon et al pressure,Caprini Marconi,Marconi et al}.  However, compared to passive Brownian particles, for which we have the Boltzmann distribution, it is considerably more difficult to determine the distribution of active particles in arbitrary potentials.
		
		To gain insight into how activity affects the behavior of active particles near different types of boundaries, we consider the limit of weakly active particles, when passive Brownian diffusion cannot be neglected and activity can be treated pertubatively.  Passive Brownian diffusion due to thermal fluctuations is always present in any physical system.  As we will see from a mathematical perspective, including passive diffusion makes it easier for us to deal with boundaries by allowing us to define familiar Neumann, Dirichlet, or Robin type boundary conditions.  Physically, these may represent impenetrable or absorbing boundaries.  The limit of weak activity is also particularly useful because it allows us to apply perturbation theory to known results in the limit of passive particles or zero activity.
		
		The paper is structured as follows.  In Section \ref{sec:AOUP}, we introduce and summarize the active particle model of a Brownian particle driven by an exponentially correlated random force.  In Section \ref{sec:method outline}, we show how the Fokker-Planck equation describing the distribution of these active particles can be solved perturbatively by expanding the distribution in powers of the P\'{e}clet number and in terms of Hermite polynomials.  In Sections \ref{sec:cartesian}, \ref{sec:polar}, and \ref{sec:corrugated wall}, we use this approach to study several problems of noninteracting weakly active articles near impenetrable boundaries.  This includes simple confinement in 1D, confinement to a circular or wedge-shaped region, and confinement by a corrugated boundary.  Finally, in Section \ref{sec:spherical}, we consider an absorbing boundary problem of weakly active particles around a spherical absorber.  For each example geometry, we start the section with a brief description of the equations we solve and the boundary conditions we apply to obtain the distribution of active particles.
				
	\section{\label{sec:AOUP}Active Particles Driven By An Exponentially Correlated Random Force}	
		We start by describing a Brownian particle driven by an exponentially correlated random force.  The equation of motion for its position is given by the overdamped Langevin equation
		\begin{equation}
			\gamma\dot{\boldsymbol{r}}=\boldsymbol{\eta}+\sqrt{2D_p\gamma^2}\boldsymbol{\xi}_r\label{eq:langevin r},
		\end{equation}
		where $\gamma$ is the friction coefficient, $D_p$ is the passive diffusivity, and $\boldsymbol{\xi}_r$ is a zero mean Gaussian white noise with $\langle\xi_{r,\alpha}(t)\xi_{r,\beta}(t')\rangle=\delta_{\alpha\beta}\delta(t-t')$.  Note that the passive diffusivity can be related to temperature through the Einstein relation $D_p\gamma=k_BT$.  The variable $\boldsymbol{\eta}$ is the active force that propels the particle.  We assume this force has first and second moments
		\begin{subequations}
			\begin{align}
				\langle\eta_{\alpha}(t)\rangle&=0\label{eq:1st moment eta},\\
				\langle\eta_{\alpha}(t)\eta_{\beta}(t')\rangle&=\delta_{\alpha\beta}\frac{\gamma^2v^2}{d}\exp\left(-\frac{1}{\tau}|t-t'|\right)\label{eq:2nd moment eta},
			\end{align}
		\end{subequations}
		where $\tau$ is the persistence time of the propulsion, $v$ is the swim speed, and $d$ is the spacial dimension.  In other words, this propulsion force on average has no preferred direction and is exponentially correlated in time.  The exponential correlation in time means that the active particle will have memory of its propulsion direction for roughly a time $\tau$ before orienting in a new direction.  Note that the magnitude of the correlations are chosen so that the characteristic propulsion force is $\sqrt{\langle\boldsymbol{\eta}^2\rangle}=\gamma v$, which is simply the force needed to move through a viscous environment at a speed $v$.
		
		One common way to generate an exponentially correlated random force is through an Ornstein-Uhlenbeck process given by
		\begin{equation}
			\tau\dot{\boldsymbol{\eta}}=-\boldsymbol{\eta}+\sqrt{\frac{2v^2\tau\gamma^2}{d}}\boldsymbol{\xi}_{\eta},
			\label{eq:langevin eta}
		\end{equation}
		where $\boldsymbol{\xi}_{\eta}$ is a zero mean Gaussian white noise independent of $\boldsymbol{\xi}_r$ with $\langle\xi_{\eta,\alpha}(t)\xi_{\eta,\beta}(t')\rangle=\delta_{\alpha\beta}\delta(t-t')$.  Eqs.\ (\ref{eq:langevin r}) and (\ref{eq:langevin eta}) thus describe the dynamics of an active particle driven by an exponentially correlated random force.  The mean-squared displacement of such a particle is given by
		\begin{equation}
			\langle\boldsymbol{r}^2(t)\rangle=2dD_pt+2v^2\tau^2\left[\frac{t}{\tau}-1+e^{-\frac{t}{\tau}}\right]
		\end{equation}
		There are two timescales: $\tau$ and $dD_p/v^2$.  The latter is the crossover between passive diffusion and ballistic motion.  On timescales longer than the persistence time $t\gg\tau$, the propulsion force becomes uncorrelated and the active particle effectively diffuses with $\langle\boldsymbol{r}^2\rangle\simeq2dD_{\textrm{eff}}t$, where the effective diffusivity $D_{\textrm{eff}}=D_p+v^2\tau/d$ is the sum of the passive and active diffusivities $D_p$ and $D_a=v^2\tau/d$.  For $t\ll dD_p/v^2$, the particle undergoes passive Brownian diffusion with $\langle\boldsymbol{r}^2\rangle\simeq2dD_pt$.  Finally for $dD_p/v^2\ll t\ll\tau$, the particle undergoes ballistic motion with $\langle\boldsymbol{r}^2\rangle\simeq v^2t^2$.  Note that this ballistic regime disappears when $\tau\ll dD_p/v^2$ or equivalently $v\tau\ll \sqrt{dD_p\tau}$, that is, when transport due to propulsion is much smaller than transport due to passive diffusion.  We refer to such particles as ``weakly active'', which will be our main focus here.
		
		The Langevin equations (Eqs.\ (\ref{eq:langevin r}) and (\ref{eq:langevin eta})) can be recast into a Fokker-Planck equation.  The distribution of a noninteracting system of these active particles satisfies the Fokker-Planck equation
		\begin{align}
			\begin{split}
				\frac{\partial\rho}{\partial t}&=-\frac{1}{\gamma}\boldsymbol{\eta}\cdot\boldsymbol{\nabla}_{r}\rho+D_p\nabla_r^2\rho+\frac{1}{\tau}\boldsymbol{\nabla}_{\eta}\cdot(\boldsymbol{\eta}\rho)+\frac{\gamma^2v^2}{d\tau}\nabla_{\eta}^2\rho\\
				&=-\boldsymbol{\nabla}_r\cdot\boldsymbol{J}_r-\boldsymbol{\nabla}_{\eta}\cdot\boldsymbol{J}_{\eta},
				\label{eq:fokker-planck}
			\end{split}
		\end{align}
		where $\rho=\rho(\boldsymbol{r},\boldsymbol{\eta},t)$ is the distribution of the active particles and
		\begin{subequations}
			\begin{align}
				\boldsymbol{J}_r&=\frac{1}{\gamma}\boldsymbol{\eta}\rho-D_p\boldsymbol{\nabla}_r\rho,\label{eq:current r}\\
				\boldsymbol{J}_{\eta}&=-\frac{1}{\tau}\boldsymbol{\eta}\rho-\frac{\gamma^2v^2}{d\tau}\boldsymbol{\nabla}_{\eta}\rho\label{eq:current eta}.
			\end{align}
		\end{subequations}
		are the currents for positions and propulsions.  Note that passive diffusion introduces the spacial gradient $\boldsymbol{\nabla}_r\rho$ in $\boldsymbol{J}_r$, which will be extremely useful for formulating boundary conditions.
		
		Our goal is to find a way to solve Eq.\ (\ref{eq:fokker-planck}) for the distribution of active particles.  Note that in the bulk far from any boundary, the spacial density of noninteracting active particles should be uniform and the propulsion force $\boldsymbol{\eta}$ for this model will be Gaussian distributed in steady-state.  The exact steady-state distribution in the bulk satisfying Eq.\ (\ref{eq:fokker-planck}) is
		\begin{equation}
			\rho(\boldsymbol{r},\boldsymbol{\eta})=\frac{\rho_{\textrm{bulk}}}{(2\pi\gamma^2v^2/d)^{d/2}}\exp\left(-\frac{\boldsymbol{\eta}^2}{2\gamma^2v^2/d}\right).
			\label{eq:exact bulk distribution}
		\end{equation}
		We wish to determine how this distribution changes near a boundary given certain conditions on the current or density at that boundary.
		
	\section{\label{sec:method outline}Perturbation Theory and Eigenfunction Expansion}
		To simplify the problem, let us define the dimensionless position and propulsion force to be
		\begin{subequations}
			\begin{align}
				\tilde{\boldsymbol{r}}&=\frac{\boldsymbol{r}}{\sqrt{2D_p\tau}}=\frac{\boldsymbol{r}}{\lambda},\\
				\tilde{\boldsymbol{\eta}}&=\frac{\boldsymbol{\eta}}{\sqrt{2\gamma^2v^2/d}}=\frac{\boldsymbol{\eta}}{\sigma},
			\end{align}
		\end{subequations}
		where the length scale $\lambda=\sqrt{2D_p\tau}$ is how far the particle passively diffuses in a persistence time and the force scale $\sigma=\sqrt{2\gamma^2v^2/d}$ is roughly the force needed to move through a viscous environment at speed $v$.  This length scale $\lambda$ is important and will show up again and again in subsequent sections.  In our context of boundaries, we can interpret it as the distance over which active particles will be persistent and still interact with a boundary through diffusion and propulsion.  In other words, within this distance, we will observe the influence of a boundary on, for example, the distribution of active particles.  Beyond this distance, however, the active particles may reorient many times without colliding with a boundary and thus behave as if they are in bulk.  
		
		The resulting dimensionless Fokker-Planck equation in steady-state is
		\begin{equation}
			\nabla_r^2\tilde{\rho}+\nabla_{\eta}^2\tilde{\rho}+2\boldsymbol{\nabla}_{\eta}\cdot(\tilde{\boldsymbol{\eta}}\tilde{\rho})=2\epsilon\tilde{\boldsymbol{\eta}}\cdot\boldsymbol{\nabla}_r\tilde{\rho},
			\label{eq:dimensionless fokker-planck}
		\end{equation}
		where the dimensionless distribution is $\tilde{\rho}(\tilde{\boldsymbol{r}},\tilde{\boldsymbol{\eta}})=\rho(\lambda\tilde{\boldsymbol{r}},\sigma\tilde{\boldsymbol{\eta}})\lambda^d\sigma^d$.  Throughout the work, tildes will indicate a dimensionless quantity.  The only parameter that remains is
		\begin{equation}
			\epsilon=\sqrt{\frac{v^2\tau}{dD_p}}=\sqrt{\textrm{Pe}},
			\label{eq:epsilon}
		\end{equation}
		where Pe is the P\'{e}clet number which is the ratio of advective transport (swimming) to diffusive transport.  This parameter controls the level of activity of the particles; for example, $\epsilon=0$ corresponds to a passive particle while $\epsilon\gg1$, a strongly active particle.  As mentioned earlier, our main focus here will be on weakly active particles or $\epsilon\ll1$.
		
		To make further progress, we perform two expansions on the distribution: an expansion in powers of $\epsilon$ and an eigenfunction expansion in Hermite polynomials (Appendix \ref{subapp:hermite polynomials}).  To keep things simple here, we will only show the series solution for 1D, though it should be emphasized that the result can easily be extended to arbitrary dimensions (Appendix \ref{app:expansion arbitrary dimension}).  We start by writing the distribution in powers of $\epsilon$ as
		\begin{equation}
			\tilde{\rho}(\tilde{x},\tilde{\eta})=\sum_{n=0}^{\infty}\epsilon^n\tilde{\rho}^{(n)}(\tilde{x},\tilde{\eta}),
		\end{equation}
		which gives us for each order
		\begin{equation}
			\frac{\partial^2\tilde{\rho}^{(n)}}{\partial\tilde{x}^2}+\frac{\partial^2\tilde{\rho}^{(n)}}{\partial\tilde{\eta}^2}+2\frac{\partial}{\partial\tilde{\eta}}\Big(\tilde{\eta}\tilde{\rho}^{(n)}\Big)=2\tilde{\eta}\frac{\partial\tilde{\rho}^{(n-1)}}{\partial\tilde{x}}.
			\label{eq:dimensionless fokker-planck order 1D}
		\end{equation}
		The zeroth order solution $\tilde{\rho}^{(0)}$ is related to the density of passively diffusing particles, which is usually easy to find.  Thus, we can determine the effect of activity by iteratively computing higher-order terms starting from the solution for passive Brownian particles.
		
		There should be no active particles with an arbitrarily large propulsion force $\boldsymbol{\eta}$.  In other words, the distribution in $\boldsymbol{\eta}$ should decay sufficiently quickly as $|\boldsymbol{\eta}|\rightarrow\infty$.  We can therefore simplify the second and third terms on the left-hand side of Eq.\ (\ref{eq:dimensionless fokker-planck order 1D}) by expanding the distribution in terms of Hermite polynomials.  By writing each order of the distribution as
		\begin{equation}
			\tilde{\rho}^{(n)}(\tilde{x},\tilde{\eta})=\sum_{m=0}^{\infty}C_m^{(n)}(\tilde{x})e^{-\tilde{\eta}^2}H_m(\tilde{\eta}),
		\end{equation}
		we reduce the entire problem to solving for the coefficients $C_m^{(n)}(\tilde{x})$, which satisfy in 1D the simple ordinary differential equation
		\begin{equation}
			\frac{d^2C_m^{(n)}}{d\tilde{x}^2}-2mC_m^{(n)}=\frac{d}{d\tilde{x}}\left[C_{m-1}^{(n-1)}+2(m+1)C_{m+1}^{(n-1)}\right].
			\label{eq:1D coef equation}
		\end{equation}
		We will focus on two kinds of boundaries: impenetrable and absorbing.  At an impenetrable boundary, particles cannot pass through it and so the current normal to the boundary must be zero.  At an absorbing boundary, particles are removed from the system and so the density at the boundary is maintained to be zero.  By writing the current $\tilde{J}_x=2\epsilon\tilde{\eta}\tilde{\rho}-\frac{\partial\tilde{\rho}}{\partial\tilde{x}}$ or density $\tilde{\rho}$ in terms of the coefficients $C_m^{(n)}$, we can use the orthogonality of Hermite polynomials to determine the boundary conditions for the coefficients.  We now illustrate this approach with several examples of noninteracting weakly active particles in different geometries.
	
	\begin{figure*}
		\centering
		\includegraphics[scale=0.50]{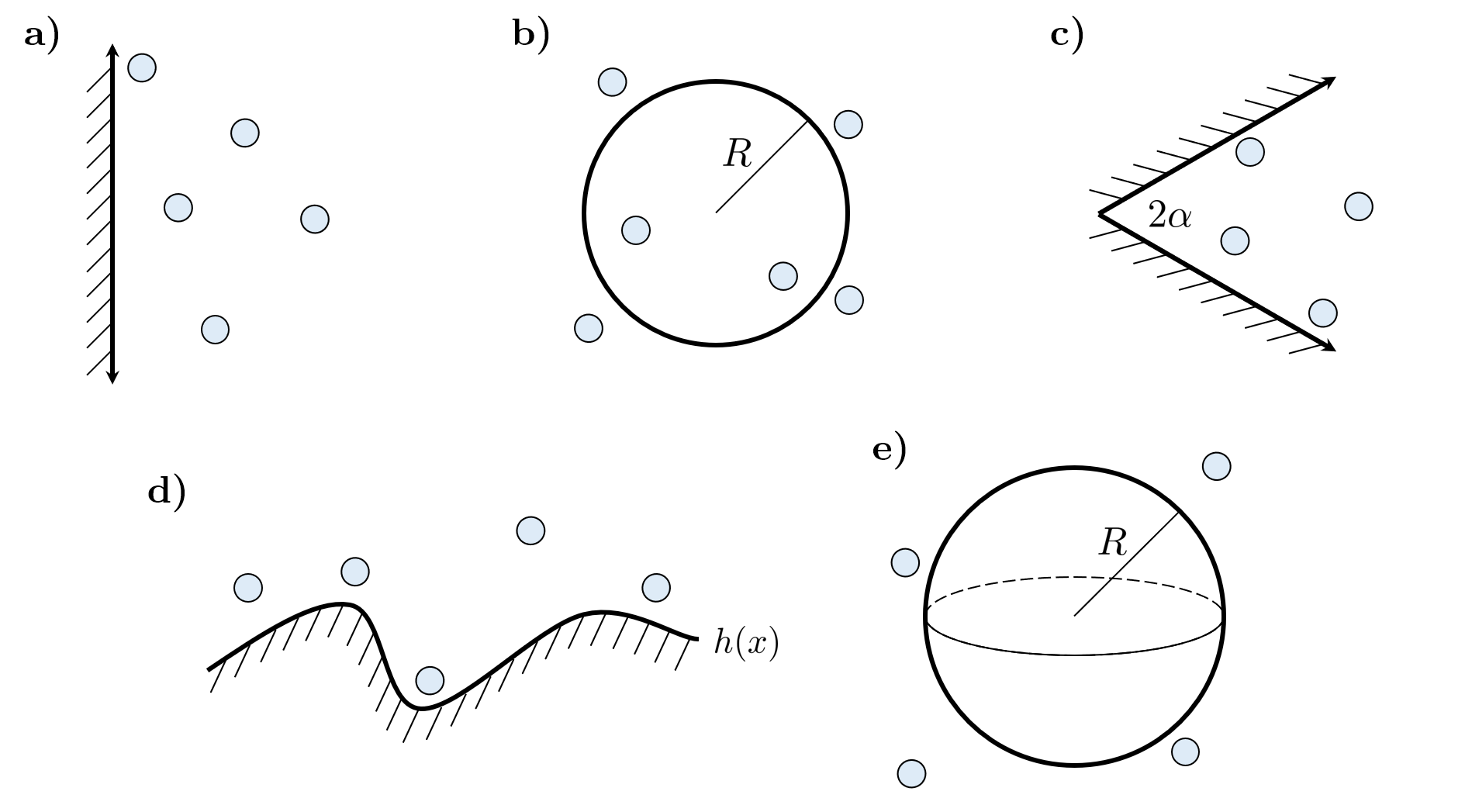}
		\caption{Several of the geometries we consider: \textbf{a)} active particles near a single hard wall, \textbf{b)} active particles inside and outside a circular boundary, \textbf{c)} active particles confined to a wedge-shaped region, \textbf{d)} active particles near a corrugated boundary, and finally \textbf{e)} active particles near a spherical absorber.}
		\label{fig:various geometries}
	\end{figure*}
	\section{\label{sec:cartesian}Problems in Cartesian Coordinates}
		\subsection{\label{subsec:particles on line}Active particles on a line}
			Consider the simplest example of noninteracting weakly active particles freely propelling left or right on a line until they collide with an impenetrable wall.  We are interested in how the presence of such a wall modifies the bulk distribution of active particles (Eq.\ \ref{eq:exact bulk distribution}).  As outlined in Section \ref{sec:method outline}, the dimensionless distribution in 1D can be written as
			\begin{equation}
				\tilde{\rho}(\tilde{x},\tilde{\eta})=\sum_{n=0}^{\infty}\epsilon^n\sum_{m=0}^{\infty}C_m^{(n)}(\tilde{x})e^{-\tilde{\eta}^2}H_m(\tilde{\eta}),
			\end{equation}
			where the coefficients satisfy Eq.\ (\ref{eq:1D coef equation}).  We consider the cases of active particles confined by one solid wall and between two walls.
		
			\subsubsection{\label{subsec:semi-infinite}1D semi-infinite domain: one wall}
				We start with the case of an impenetrable wall at $x=0$ that confines the active particles to the region $x>0$.  The zero current condition at this wall $J_x(0,\eta)=0$ gives us the condition on the coefficients
				\begin{equation}
					\frac{dC_m^{(n)}(0)}{d\tilde{x}}=C_{m-1}^{(n-1)}(0)+2(m+1)C_{m+1}^{(n-1)}(0).
				\end{equation}
				Details of the solution can be found in Appendix \ref{subapp:semi-infinite}.  We are interested in how the presence of an impenetrable wall affects the distribution $\rho(x,\eta)$ and the currents $J_x(x,\eta),J_{\eta}(x,\eta)$ of the active particles.  In the bulk, the distribution is given by Eq.\ (\ref{eq:exact bulk distribution}) with $d=1$.  As we approach the wall, the distribution will no longer be independent of $x$.  Up to $\epsilon^2$, the distribution is
				\begin{widetext}
					\begin{equation}
						\rho(x,\eta)\simeq\frac{\rho_{\textrm{bulk}}}{\sigma\sqrt{\pi}}e^{-\frac{\eta^2}{\sigma^2}}\left\{1-\epsilon\sqrt{2}e^{-\frac{\sqrt{2}x}{\lambda}}\frac{\eta}{\sigma}+\epsilon^2\left[e^{-\frac{\sqrt{2}x}{\lambda}}+\left(\sqrt{2}e^{-\frac{2x}{\lambda}}-e^{-\frac{\sqrt{2}x}{\lambda}}\right)\left(\frac{2\eta^2}{\sigma^2}-1\right)\right]\right\},
						\label{eq:1d full solution}
					\end{equation}
				\end{widetext}
				where $\lambda=\sqrt{2D_p\tau}$ and $\sigma=\sqrt{2\gamma^2v^2}$.  Integrating out the active force $\eta$, we have for the spacial density
				\begin{align}
					\begin{split}
						\frac{\rho(x)}{\rho_{\textrm{bulk}}}\simeq&\,1+\epsilon^2e^{-\frac{\sqrt{2}x}{\lambda}}\\
						&+\epsilon^4\left[2\sqrt{2}e^{-\frac{2x}{\lambda}}+2\sqrt{2}\left(\frac{x}{4\lambda}-1\right)e^{-\frac{\sqrt{2}x}{\lambda}}\right].
					\end{split}
					\label{eq:1d density up to 4th}
				\end{align}
				\begin{figure}
					\centering
					\includegraphics[scale=0.4]{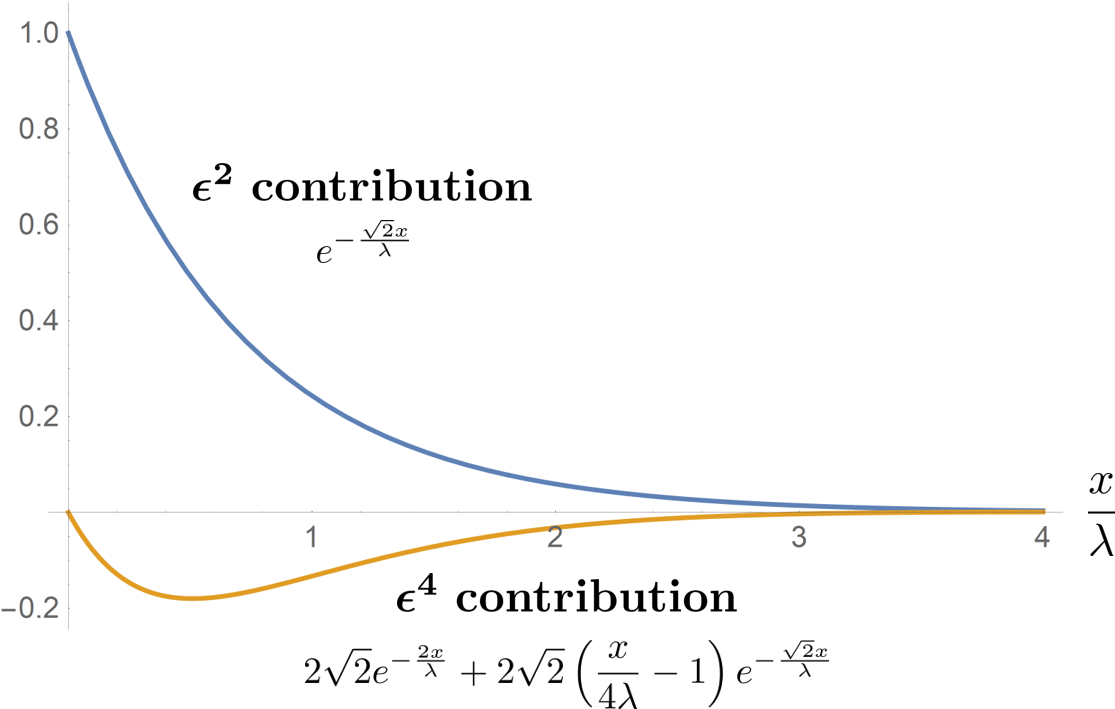}
					\caption{$\epsilon^2$ (blue) and $\epsilon^4$ (orange) contributions to the density of weakly active particles near a single impenetrable wall (Eq.\ \ref{eq:1d density up to 4th}).  The $\epsilon^2$ contribution gives an elevated density while the $\epsilon^4$ contribution slightly depletes the density near the wall.}
					\label{fig:single wall epsilon 2 4}
				\end{figure}%
				The $\epsilon^2$ and $\epsilon^4$ contributions to this density are shown in Figure \ref{fig:single wall epsilon 2 4}.  The density is elevated over the length scale $\lambda$, which, as we discussed earlier, is the distance over which the presence of the wall will be felt by the active particles.  Another way of thinking of this is that persistence of the active particles causes them to spend more time near the wall, thus elevating the density.  The excess number of particles near the wall is given by
				\begin{align}
					\begin{split}
						N_{\textrm{excess}}&=\int_{0}^{\infty}\left[\rho(x)-\rho_{\textrm{bulk}}\right]dx\\
						&\simeq\rho_{\textrm{bulk}}\sqrt{D_p\tau}\left[\epsilon^2-\frac{4\sqrt{2}-5}{2}\,\epsilon^4\right].
					\end{split}
					\label{eq:excess N}
				\end{align}
				It is interesting to note that the $\epsilon^4$ correction does not further increase the density near the wall and actually depletes it, as there is a decrease in the excess number of particles near the wall.  This suggests that at higher P\'{e}clet numbers, the accumulated density may become steeper.
				
\begin{figure}
					\centering
					\includegraphics[scale = 0.55]{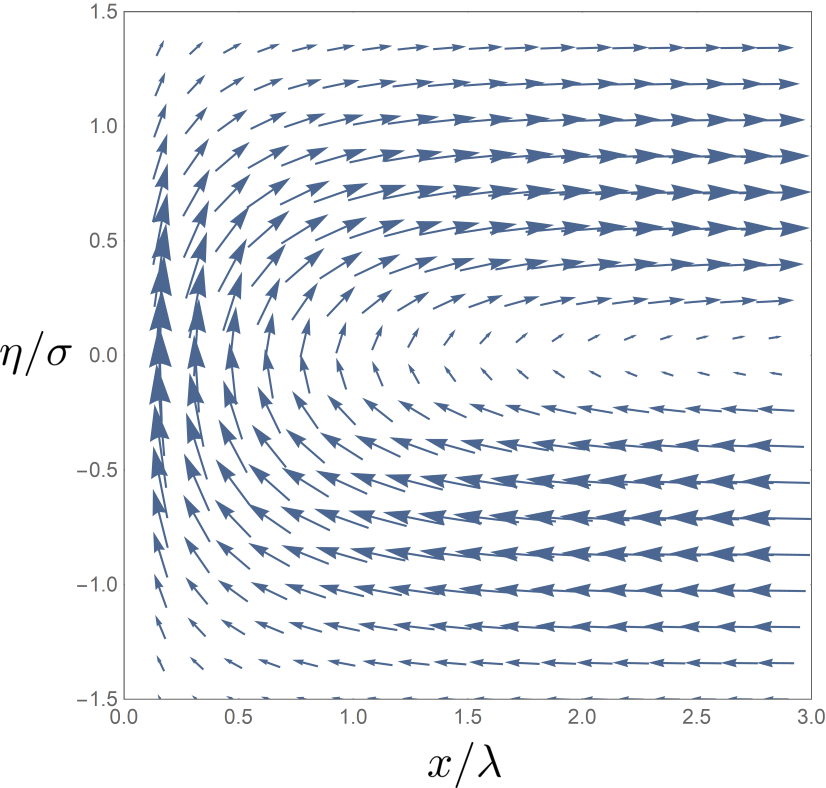}
					\caption{Currents $J_x(x,\eta),J_{\eta}(x,\eta)$ (Eqs.\ (\ref{eq:1D Jx current}) and (\ref{eq:1D Jeta current})) resulting from activity near a single impenetrable wall.  The circulation shows the simple behavior of active particles propelling towards the wall, spending time turning around, and then propelling away.}
					\label{fig:1D currents}
				\end{figure}

				The currents in the position $x$ and active force $\eta$ are given by
				\begin{subequations}
					\begin{align}
						J_x(x,\eta)&\simeq\epsilon\frac{\lambda}{\tau}\left(\frac{\rho_{\textrm{bulk}}}{\sigma}\right)\frac{1}{\sqrt{\pi}}\left(1-e^{-\frac{\sqrt{2}x}{\lambda}}\right)\frac{\eta}{\sigma} e^{-\frac{\eta^2}{\sigma^2}},\label{eq:1D Jx current}\\
						J_{\eta}(x,\eta)&\simeq\epsilon\frac{\sigma}{\tau}\left(\frac{\rho_{\textrm{bulk}}}{\sigma}\right)\frac{1}{\sqrt{2\pi}}e^{-\frac{\sqrt{2}x}{\lambda}}e^{-\frac{\eta^2}{\sigma^2}}.\label{eq:1D Jeta current}
					\end{align}
				\end{subequations}
				These characterize how the positions and propulsions of our active particles change on average as they move near a wall.  There is circulation in the $x\eta$-plane (Figure \ref{fig:1D currents}), a signature of out-of-equilibrium systems \cite{Battle et al}.
				In this case, the behavior is quite simple: active particles on average swim towards the wall, spend some time turning, and then swim away.  It is interesting to note that while we observe currents at order $\epsilon$, we do not observe any deviations from the passive density $\rho(x)=\rho_{\textrm{bulk}}$ until order $\epsilon^2$.  The currents result in an asymmetric distribution in $\eta$ at the wall.  Up to order $\epsilon$, the distribution at the wall is
				\begin{align}
					\rho(0,\eta)&\simeq\frac{\rho_{\textrm{bulk}}}{\sigma\sqrt{\pi}}e^{-\frac{\eta^2}{\sigma^2}}\left(1-\epsilon\frac{\sqrt{2}\eta}{\sigma}\right).
					\label{eq:wall distribution}
				\end{align}
				The correction shifts the mean to $\langle\eta\rangle_{x=0}\simeq-\epsilon\gamma v=-\gamma v\sqrt{\frac{v^2\tau}{D_p}}$.  Physically, this occurs because particles with $\eta<0$ swim towards the wall and have an increased density when they are slowed while particles with $\eta>0$ swim away and have a decreased density.  The result for $\langle\eta\rangle$ can also be obtained from a simple balancing of currents.  Recall that the density (Eq.\ (\ref{eq:1d density up to 4th})) is elevated by $\epsilon^2\rho_{\textrm{bulk}}$ over a length scale $\sqrt{D_p\tau}$, which gives a diffusive flux $J_{\textrm{diff}}\sim D_p\frac{\epsilon^2\rho_{\textrm{bulk}}}{\sqrt{D_p\tau}}$ away from the wall.  Setting this equal to the swim flux $J_{\textrm{swim}}\sim\frac{1}{\gamma}\langle\eta\rangle\rho_{\textrm{bulk}}$, we get an average propulsion of $\langle\eta\rangle\sim \gamma D_p\frac{\epsilon^2}{\sqrt{D_p\tau}}=\gamma v\sqrt{\frac{v^2\tau}{D_p}}$ towards the wall.  At higher P\'{e}clet numbers, we expect that the distribution at the wall will shift more towards $\eta<0$ and that the density $\rho(0,\eta>0)$ will become significantly depleted.  This occurs because the moment an active particle turns around, it immediately propels away from the wall and no longer contributes to $\rho(0,\eta>0)$.
			
			\subsubsection{\label{subsec:finite}1D finite domain: two walls}				
				We now consider the case of active particles confined between two walls located at $x=\pm L$.  Unlike for a single wall, there is no bulk where the distribution of active particles is unaffected by the wall.  When there are two walls, each wall can have an effect on the distribution at the other.  The zero current boundary conditions $J_x(\pm L,\eta)=0$ give the following relation for the coefficients
				\begin{equation}
					\frac{dC_m^{(n)}(\pm\tilde{L})}{d\tilde{x}}=C_{m-1}^{(n-1)}(\pm\tilde{L})+2(m+1)C_{m+1}^{(n-1)}(\pm\tilde{L}).
				\end{equation}
				The steps for finding the solution (Appendix \ref{subapp:finite domain}) are similar to those of the single wall.  The main difference is that we now have a finite number of particles trapped between the two walls instead of an infinite bulk with constant density.  Taking $\rho(x,\eta)$ and integrating out $\eta$, we have for the density
				\begin{equation}
					\frac{\rho(x)}{N/2L}=1+\epsilon^2\left(\frac{\cosh\frac{\sqrt{2}x}{\lambda}}{\cosh\frac{\sqrt{2}L}{\lambda}}-\frac{\tanh\frac{\sqrt{2}L}{\lambda}}{\frac{\sqrt{2}L}{\lambda}}\right),
					\label{eq:density two walls}
				\end{equation}
				where $N$ is the number of particles between the walls.  Note that as a result of the accumulation at walls, there is a depletion of particles around the center of the confinement.  This is captured by the second term in the parenthesis, which vanishes in the limit of large separation of the walls.
				
				For large separations, each wall should not influence the other and we should obtain the result for a single wall in the previous section.  Indeed, taking $L\gg\lambda$, we find
				\begin{equation}
					\rho(x)\simeq\frac{N}{2L}\left[1+\epsilon^2\left(e^{\frac{\sqrt{2}(x-L)}{\lambda}}+e^{-\frac{\sqrt{2}(x+L)}{\lambda}}\right)\right],
					\label{eq:density 2 walls large L}
				\end{equation}
				which is simply the sum of the accumulations due to each wall if it were by itself.  We can also examine the opposite limit of two walls that are very close to each other ($L\ll\lambda$), for which we find the the parabolic profile
				\begin{equation}
					\rho(x)\simeq\frac{N}{2L}\left[1+\epsilon^2\left(\frac{x^2}{\lambda^2}-\frac{L^2}{3\lambda^2}\right)\right].
				\end{equation}
				
				We now look at the distribution of particles at each wall to see how one wall may influence the other.  The distribution at the walls to order $\epsilon$ is
				\begin{equation}
					\rho(\pm L,\eta)=\frac{N}{2L}\cdot\frac{e^{-\frac{\eta^2}{\sigma^2}}}{\sigma\sqrt{\pi}}\left(1\pm\epsilon\frac{\sqrt{2}\eta}{\sigma}\tanh\frac{\sqrt{2}L}{\lambda}\right),
				\end{equation}
				which gives for the average propulsion at each wall
				\begin{equation}
					\langle\eta\rangle(\pm L)\simeq\pm\epsilon\gamma v\tanh\frac{\sqrt{2}L}{\lambda}.
					\label{eq:avg eta two walls}
				\end{equation}
				Interestingly, we can interpret this average propulsion as a weighted average $\langle\eta\rangle(\pm L)\simeq\mp\epsilon\gamma vP_c\pm\epsilon\gamma vP_f$, where
				\begin{equation}
					P_c=\frac{1}{1+e^{-\frac{2\sqrt{2}L}{\lambda}}},\hspace{0.2in}P_f=\frac{e^{-\frac{2\sqrt{2}L}{\lambda}}}{1+e^{-\frac{2\sqrt{2}L}{\lambda}}},
				\end{equation}
				are the weights for the closest and farthest walls, respectively.  Note that these weights are proportional to how much the closest and farthest walls contribute to the accumulation.  When the two walls are far apart ($L\gg\lambda$), we get back the single wall result $\langle\eta\rangle(\pm L)\simeq\mp\epsilon\gamma v$ since the farthest wall contributes nothing ($P_f\ll P_c\approx1$).
		
			\subsubsection{Pressure on solid boundaries}
				One question of interest is how one relates the density of active particles to the pressure they exert on walls.  For an equilibrium system of noninteracting particles, the density is $\rho(x)=\rho_0$ everywhere and the pressure is simply the ideal gas pressure $P=\rho_0k_BT$.  To compute the pressures in our present case of noninteracting active particles, we start by replacing the impenetrable walls with soft confining potentials.  This approach of using soft potentials has been useful for computing the mechanical properties of active particles near boundaries \cite{Solon et al pressure,Caprini Marconi,Marconi et al,Zakine et al,Grosberg and Joanny,Duzgan Sellinger}.  The idea is that a confining potential, just like a wall, can prevent particles from moving a certain direction.  The pressure can then be computed by simply summing up the forces the potential exerts on the particles.  One can then take the limit as the potential becomes steep to obtain the pressure for an impenetrable wall.  For our purposes, we consider the ramp potentials (Appendix \ref{subapp:pressure})
				\begin{equation}
					U(x)=\begin{cases}
						-fx, & x < 0\\
						0, & x > 0
					\end{cases}
					\label{eq:U 1 ramp}
				\end{equation}
				for one wall at $x=0$ and
				\begin{equation}
					U(x)=\begin{cases}
						-fx-fL, & x<-L\\
						0, & -L<x<L\\
						fx-fL, & x>L
					\end{cases}
					\label{eq:U 2 ramp}
				\end{equation}
				for two walls at $x=\pm L$.  The mechanical pressure for both these cases is simply given by the integral
				\begin{equation}
					P=-\int U'(x)\rho(x)dx.
				\end{equation}
				over one of the wall regions where $U(x)\ne0$.  The distributions and the pressures obtained for these ramp potentials are left for Appendix \ref{subapp:pressure}.  We are interested in the limit of impenetrable walls or $f\rightarrow\infty$.  In this limit, we find the pressures
				\begin{equation}
					P=\rho_{\textrm{bulk}}D_p\gamma(1+\epsilon^2),
					\label{eq:P 1 wall}
				\end{equation}
				for one wall and
				\begin{equation}
					P\simeq\frac{ND_p\gamma}{2L}\left[1+\epsilon^2\left(1-\frac{\tanh\frac{\sqrt{2}L}{\lambda}}{\frac{\sqrt{2}L}{\lambda}}\right)\right],
					\label{eq:P 2 wall}
				\end{equation}
				for two.  We start by noting that the pressure for one wall (Eq.\ (\ref{eq:P 1 wall})) is actually exact even though it was obtained from perturbation theory.  A proof of this can be found in Appendix \ref{subapp:pressure}.  There are two interpretations of this pressure.  The first interpretation is $P=\rho_{\textrm{bulk}}[D_p(1+\epsilon^2)]\gamma=\rho_{\textrm{bulk}}D_{\textrm{eff}}\gamma$, which can be thought of as the ideal gas pressure of particles with an effective diffusivity $D_{\textrm{eff}}=D_p+v^2\tau$.  The second interpretation is $P=[\rho_{\textrm{bulk}}(1+\epsilon^2)]D_p\gamma\simeq\rho_{\textrm{wall}}D_p\gamma$, where $\rho_{\textrm{wall}}=\rho(0)$ is the density at the boundary.  This is just the the passive pressure due to the elevated density of particles (Eq.\ (\ref{eq:1d density up to 4th})) close to the wall.  For the case of confinement between two walls, the situation is different as there is no longer a bulk.  However, notice that we still have the relation $P\simeq\rho_{\textrm{wall}}D_p\gamma$, where $\rho_{\textrm{wall}}=\rho(\pm L)$.  Using Einstein's relation $D_p\gamma=k_BT$, we may write $P\simeq\rho_{\textrm{wall}}k_BT$ for noninteracting active particles.  While we have only shown that this relation for pressure holds up to order $\epsilon^2$, it is not inconceivable that it should hold in general.  In Appendix \ref{subapp:exactly solvable}, we show that for the exactly solvable model of noninteracting 1D run-and-tumble particles with passive diffusion, this relation for pressure holds without any approximations.  It is worth noting that a similar relation has been shown to hold for noninteracting active Brownian particles \cite{Duzgan Sellinger}, a different model of active particles.
		
		\subsection{\label{subsec:corner}Confinement in 2D: right-angled corner}
			In previous sections, we considered weakly active particles in 1D.  Those results can easily be extended to flat walls in higher dimensions such as an infinite line (2D) or infinite plate (3D).  In fact, the density profiles, particularly the exponential decays of density away from a flat wall, are exactly the same as those obtained in 1D.
			
\begin{figure}
				\centering
				\includegraphics[scale=0.55]{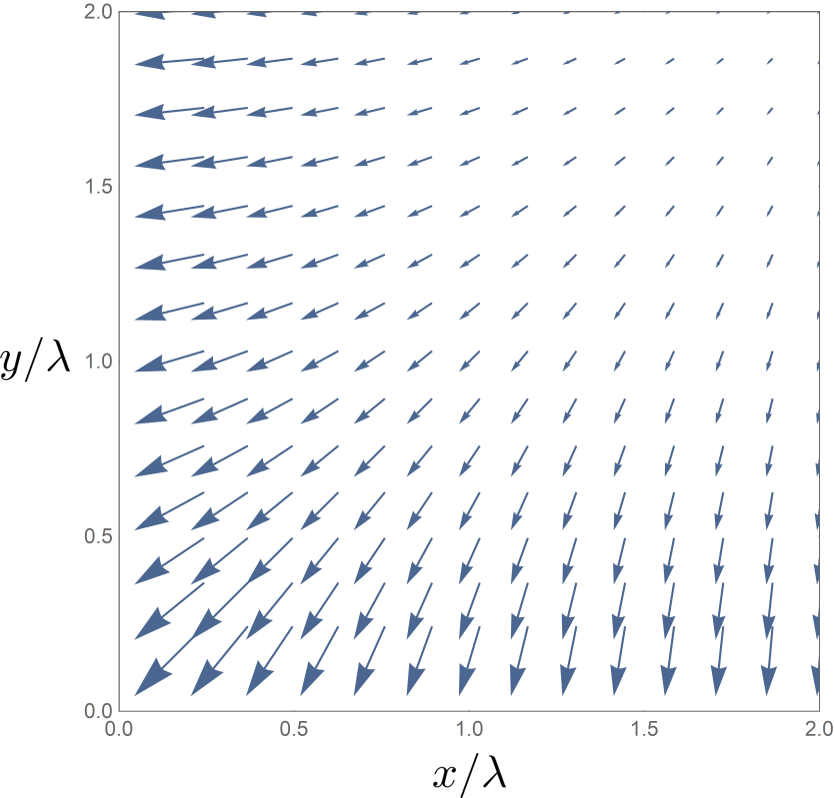}
				\caption{Average propulsion $\langle\boldsymbol{\eta}\rangle$ (Eq.\ (\ref{eq:average eta corner})) near a right-angled corner that confines active particles to $x>0,y>0$.  This orientation towards the corner leads to additional accumulation given by the last term of the density Eq.\ (\ref{eq:corner density up to 4th}).}
				\label{fig:corner average propulsion}
			\end{figure}

			We are interested in going beyond these simple geometries and studying cases where the walls may be curved or are not parallel and meet at certain angles.  In this particular section, we will focus on the simpler case of two flat walls that meet at a right angle and confine active particles to a region $x>0,y>0$.  We write the dimensionless distribution as
			\begin{equation}
				\rho(\tilde{\boldsymbol{r}},\tilde{\boldsymbol{\eta}})=\sum_{n=0}^{\infty}\epsilon^n\sum_{\boldsymbol{m}}C_{\boldsymbol{m}}^{(n)}(\tilde{\boldsymbol{r}})e^{-\tilde{\boldsymbol{\eta}}^2}H_{m_x}(\tilde{\eta}_x)H_{m_y}(\tilde{\eta}_y),
			\end{equation}
			Notice that going to higher dimensions simply requires additional Hermite polynomials.  The coefficients satisfy
			\begin{align}
				\begin{split}
					\frac{\partial^2C_{\boldsymbol{m}}^{(n)}}{\partial\tilde{x}^2}+\frac{\partial^2C_{\boldsymbol{m}}^{(n)}}{\partial\tilde{y}^2}-2(m_x+m_y)C_{\boldsymbol{m}}^{(n)}=\frac{\partial w_x}{\partial\tilde{x}}+\frac{\partial w_y}{\partial\tilde{y}},
				\end{split}
			\end{align}
			where
			\begin{subequations}
				\begin{align}
					w_x&=C_{m_x-1,m_y}^{(n-1)}+2(m_x+1)C_{m_x+1,m_y}^{(n-1)},\\
					w_y&=C_{m_x,m_y-1}^{(n-1)}+2(m_y+1)C_{m_x,m_y+1}^{(n-1)}.
				\end{align}
			\end{subequations}
			The zero current boundary conditions $J_x(0,y,\boldsymbol{\eta})=J_y(x,0,\boldsymbol{\eta})=0$ give the following conditions on the coefficients
			\begin{subequations}
				\begin{align}
					\frac{\partial C_{\boldsymbol{m}}^{(n)}(0,\tilde{y})}{\partial\tilde{x}}=w_x(0,\tilde{y})
					\\
					\frac{\partial C_{\boldsymbol{m}}^{(n)}(\tilde{x},0)}{\partial\tilde{y}}=w_y(\tilde{x},0)
				\end{align}
			\end{subequations}
			Details of the solution can be found in Appendix \ref{subapp:2D corner}.  Integrating out $\boldsymbol{\eta}$ from $\rho(\boldsymbol{r},\boldsymbol{\eta})$, we find the density near the corner is
			\begin{align}
				\begin{split}
					\frac{\rho(\boldsymbol{r})}{\rho_{\textrm{bulk}}}\simeq&\ 1+\epsilon^2\left(e^{-\frac{\sqrt{2}x}{\lambda}}+e^{-\frac{\sqrt{2}y}{\lambda}}\right)\\
					&+\epsilon^4\left[2\sqrt{2}e^{-\frac{2x}{\lambda}}+2\sqrt{2}\left(\frac{x}{4\lambda}-1\right)e^{-\frac{\sqrt{2}x}{\lambda}}\right]\\
					&+\epsilon^4\left[2\sqrt{2}e^{-\frac{2y}{\lambda}}+2\sqrt{2}\left(\frac{y}{4\lambda}-1\right)e^{-\frac{\sqrt{2}y}{\lambda}}\right]\\
					&+\epsilon^4e^{-\frac{\sqrt{2}(x+y)}{\lambda}}\\
					=&\ \frac{\rho_{\textrm{1D}}(x)+\rho_{\textrm{1D}}(y)}{\rho_{\textrm{bulk}}}+\epsilon^4e^{-\frac{\sqrt{2}(x+y)}{\lambda}},
				\end{split}
				\label{eq:corner density up to 4th}
			\end{align}
			where $\rho_{\textrm{1D}}$ is the 1D density (Eq.\ (\ref{eq:1d density up to 4th})) if each wall were by itself.  The last term is new and enhances the accumulation near the corner.  This arises due to correlations between $\eta_x,\eta_y$ near the corner.  Indeed, if we study the full distribution, we find at order $\epsilon^2$ the new term
			\begin{align}
				\begin{split}
					\rho(\boldsymbol{r},\boldsymbol{\eta})\simeq&\ \textrm{single wall contributions}\\
					&+2\epsilon^2\frac{\rho_{\textrm{bulk}}}{\pi\sigma^2}e^{-\frac{\boldsymbol{\eta}^2}{\sigma^2}}\frac{\eta_x\eta_y}{\sigma^2}e^{-\frac{\sqrt{2}(x+y)}{\lambda}},
				\end{split}
			\end{align}
			which gives a nonzero correlation $\langle\eta_x\eta_y\rangle$ near the corner and is responsible for the extra accumulation at order $\epsilon^4$.  If we consider the average propulsion $\langle\boldsymbol{\eta}\rangle$ of the active particles near the corner, which is given by
			\begin{equation}
				\langle\boldsymbol{\eta}\rangle(x,y)=-\frac{\epsilon\gamma v}{\sqrt{2}}\left(e^{-\frac{\sqrt{2}x}{\lambda}},e^{-\frac{\sqrt{2}y}{\lambda}}\right),
				\label{eq:average eta corner}
			\end{equation}
			we also see that there is an increased tendency for active particles to orient and propel towards the corner (Figure \ref{fig:corner average propulsion}).
			
	\section{\label{sec:polar}Problems in Polar Coordinates}
		We have thus far considered simple examples of weakly active particles near flat walls or a right-angled corner.  We now want to study how the curvature of a surface or the sharpness of a corner affects the distribution of these particles.  We focus our attention on two cases: a circular boundary and a corner with an angle other than $\pi/2$.  
		
		These examples are best studied using polar coordinates.  In polar coordinates, we write the dimensionless distribution as
		\begin{align}
			\tilde{\rho}(\tilde{r},\theta,\tilde{\boldsymbol{\eta}})=\sum_{n=0}^{\infty}\epsilon^n\sum_{\boldsymbol{m}}C_{\boldsymbol{m}}^{(n)}(\tilde{r},\theta)e^{-\tilde{\boldsymbol{\eta}}^2}H_{m_x}(\tilde{\eta}_x)H_{m_y}(\tilde{\eta}_y).
		\end{align}
		The coefficients satisfy
		\begin{align}
			\begin{split}
				&\frac{1}{\tilde{r}}\frac{\partial}{\partial\tilde{r}}\left(\tilde{r}\frac{\partial C_{\boldsymbol{m}}^{(n)}}{\partial\tilde{r}}\right)+\frac{1}{\tilde{r}^2}\frac{\partial^2C_{\boldsymbol{m}}^{(n)}}{\partial\theta^2}-2(m_x+m_y)C_{\boldsymbol{m}}^{(n)}\\
				&=\frac{1}{\tilde{r}}\frac{\partial}{\partial\tilde{r}}\left(\tilde{r}w_r\right)+\frac{1}{\tilde{r}}\frac{\partial w_{\theta}}{\partial\theta},
			\end{split}
			\label{eq:coef eq polar}
		\end{align}
		where the components of $\boldsymbol{w}$ in polar coordinates are
		\begin{subequations}
			\begin{align}
				\begin{split}
					w_r=&\ \left[C_{m_x-1,m_y}^{(n-1)}+2(m_x+1)C_{m_x+1,m_y}^{(n-1)}\right]\cos\theta\\
					&+\left[C_{m_x,m_y-1}^{(n-1)}+2(m_y+1)C_{m_x,m_y+1}^{(n-1)}\right]\sin\theta,
				\end{split}
				\\
				\begin{split}
					w_{\theta}=&\ -\left[C_{m_x-1,m_y}^{(n-1)}+2(m_x+1)C_{m_x+1,m_y}^{(n-1)}\right]\sin\theta\\
					&+\left[C_{m_x,m_y-1}^{(n-1)}+2(m_y+1)C_{m_x,m_y+1}^{(n-1)}\right]\cos\theta.
				\end{split}
			\end{align}
		\end{subequations}
		
		\subsection{\label{subsec:circular}Active particles around a circular boundary}
			Consider an impenetrable circular boundary of radius $R$ with active particles both inside and outside.  We will separately obtain the densities in both regions.  The radial current on both sides of the boundary must be zero, or $J_r(R,\theta,\boldsymbol{\eta})=0$.  We thus have the boundary condition for the coefficients
			\begin{equation}
				\frac{\partial C_{\boldsymbol{m}}^{(n)}(\tilde{R},\theta)}{\partial\tilde{r}}=w_x(\tilde{R},\theta)\cos\theta+w_y(\tilde{R},\theta)\sin\theta.
			\end{equation}
			Details of the solutions both inside and outside the region can be found in Appendix \ref{subapp:circular}.  The densities outside and inside the boundary to order $\epsilon^2$ are
			\begin{subequations}
				\begin{align}
					\rho_{\textrm{out}}(r,\theta)\simeq&\ \rho_{\textrm{bulk}}\left[1+2\epsilon^2\frac{K_0\left(\frac{\sqrt{2}r}{\lambda}\right)}{K_0\left(\frac{\sqrt{2}R}{\lambda}\right)+K_2\left(\frac{\sqrt{2}R}{\lambda}\right)}\right],\label{eq:circular density outside}\\
					\begin{split}
						\rho_{\textrm{in}}(r,\theta)\simeq&\ \frac{N}{\pi R^2}\left[1+2\epsilon^2\vphantom{\frac{I_0\left(\frac{\sqrt{2}r}{\lambda}\right)}{I_0\left(\frac{\sqrt{2}R}{\lambda}\right)}}\right.\\
						&\left.\times\frac{I_0\left(\frac{\sqrt{2}r}{\lambda}\right)-I_0\left(\frac{\sqrt{2}R}{\lambda}\right)+I_2\left(\frac{\sqrt{2}R}{\lambda}\right)}{I_0\left(\frac{\sqrt{2}R}{\lambda}\right)+I_2\left(\frac{\sqrt{2}R}{\lambda}\right)}\right],
					\end{split}
				\end{align}
			\end{subequations}
			where $I_{\mu},K_{\mu}$ are the modified Bessel functions of the first and second kinds and $N$ is the number of particles inside the circular region.  
			
			Let us consider some limiting behaviors as $R\rightarrow\infty$ or $R\rightarrow0$.  Defining $\delta r=r-R$ as the radial distance from the circular boundary and taking $|\delta r|\ll R$, we obtain for the density near a large circular boundary ($R\rightarrow\infty$)
			\begin{subequations}
				\begin{align}
					\rho_{\textrm{out}}(R+\delta r,\theta)&\simeq\rho_{\textrm{bulk}}\left(1+\epsilon^2e^{-\frac{\sqrt{2}\delta r}{\lambda}}\right),\\
					\rho_{\textrm{in}}(R+\delta r,\theta)&\simeq\frac{N}{\pi R^2}\left(1+\epsilon^2e^{\frac{\sqrt{2}\delta r}{\lambda}}\right).
				\end{align}
			\end{subequations}
			Note that these are just the density profiles near a flat wall (Eqs.\ (\ref{eq:1d density up to 4th}) and (\ref{eq:density 2 walls large L})) since the curvature of the wall becomes negligible as $R\rightarrow\infty$.  In the opposite limit of a small circular boundary ($R\rightarrow0$), we have
			\begin{subequations}
				\begin{align}
					\rho_{\textrm{out}}(r\ll\lambda,\theta)&\simeq\rho_{\textrm{bulk}}\left[1+\frac{2\epsilon^2R^2}{\lambda^2}\left(\ln\frac{\sqrt{2}\lambda}{r}-\gamma_{em}\right)\right],\\
					\rho_{\textrm{in}}(r,\theta)&\simeq\frac{N}{\pi R^2}\left[1+\epsilon^2\left(\frac{r^2}{\lambda^2}-\frac{R^2}{2\lambda^2}\right)\right],
				\end{align}
			\end{subequations}
			where $\gamma_{em}\approx0.577$ is the Euler-Mascheroni constant.  The weak dependence of $\rho_{\textrm{out}}$ on $r$ is a result of the particles outside interacting with a small circular boundary, which should not affect the density much.  Finally, just as in the one dimensional case, the density inside the small region $\rho_{\textrm{in}}$ takes on a parabolic profile.
			
\begin{figure}
				\centering
				\includegraphics[scale=0.45]{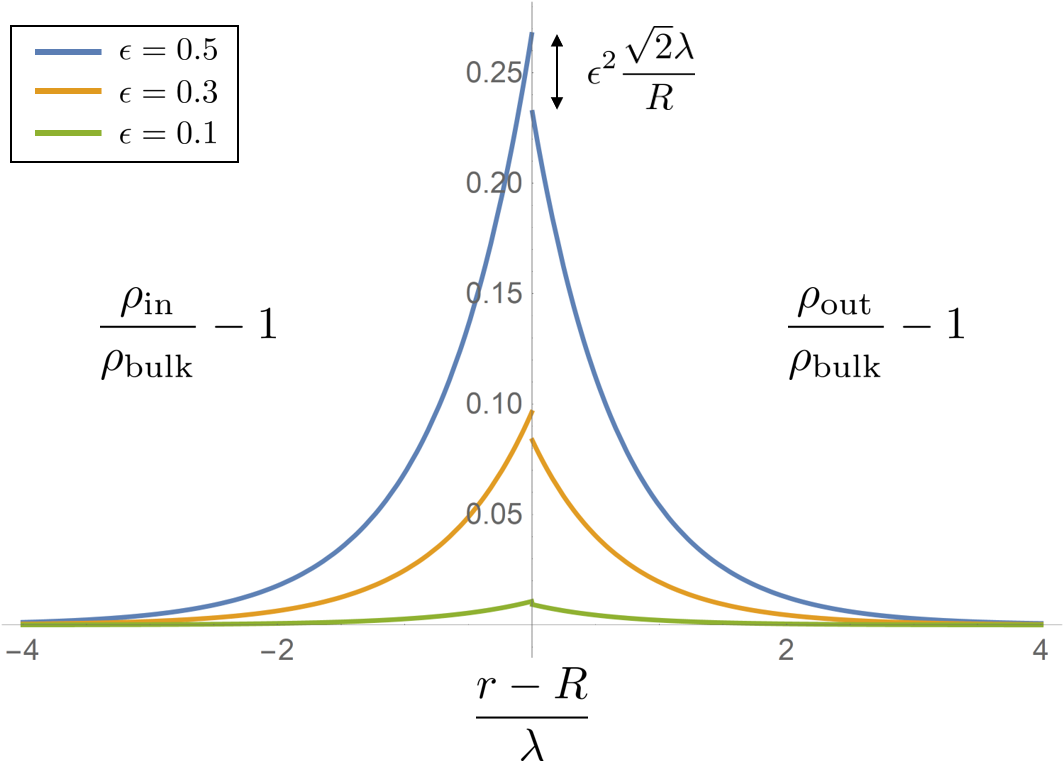}
				\caption{$\epsilon^2$ contributions to the density inside (Eq.\ (\ref{eq:circular density inside})) and outside (Eq.\ (\ref{eq:circular density outside})) an impenetrable circular boundary where the density inside the boundary at $r=0$ is maintained to be $\rho_{\textrm{bulk}}$.  Here, $R=10\lambda$.  Active particles prefer to accumulate on concave surfaces such as the inner side of the circular boundary rather than convex surfaces such as the outer side.  This gives rise to a discontinuous drop in density between the inner and outer parts of a circular boundary that is proportional to the curvature $R^{-1}$.}
				\label{fig:circular density in out}
			\end{figure}

			We are interested in how the curvature of a boundary affects the accumulation of weakly active particles.  Let us consider the densities both inside and outside the circular boundary at $r=R$ and compare them with the density at a flat wall.  We focus on $R\gg\lambda$ when the curvature is small.  To start, recall that the density at a flat wall (Eq.\ (\ref{eq:1d density up to 4th})) is $\rho_{\textrm{flat}}\simeq\rho_{\textrm{bulk}}(1+\epsilon^2)$.  For outer part of the circular boundary, we have
			\begin{equation}
				\rho_{\textrm{out}}(R,\theta)-\rho_{\textrm{flat}}\simeq-\epsilon^2\rho_{\textrm{bulk}}\frac{\lambda}{\sqrt{2}R}.
				\label{eq:convex boundary density}
			\end{equation}
			For the inner part of the circular boundary, we have to be a bit more careful since there is not an infinite bulk.  We can mimic a bulk inside the region by maintaining the density at the center $r=0$ to be equal to the bulk density outside or $\rho_{\textrm{in}}(0,\theta)=\rho_{\textrm{bulk}}$.  The resulting density profile inside the circular boundary is
			\begin{equation}
				\rho_{\textrm{in}}(r,\theta)\simeq\rho_{\textrm{bulk}}\left[1+2\epsilon^2\frac{I_0\left(\frac{\sqrt{2}r}{\lambda}\right)-1}{I_0\left(\frac{\sqrt{2}R}{\lambda}\right)+I_2\left(\frac{\sqrt{2}R}{\lambda}\right)}\right].
				\label{eq:circular density inside}
			\end{equation}
			When compared to the flat wall, we have
			\begin{equation}
				\rho_{\textrm{in}}(R,\theta)-\rho_{\textrm{flat}}\simeq\epsilon^2\rho_{\textrm{bulk}}\frac{\lambda}{\sqrt{2}R}.
				\label{eq:concave boundary density}
			\end{equation}
			The $\epsilon^2$ contributions to the densities inside (Eq.\ (\ref{eq:circular density inside})) and outside (Eq.\ (\ref{eq:circular density outside})) the circular boundary are shown in Figure \ref{fig:circular density in out}.  On both sides of the circular boundary, the correction to density is proportional to $R^{-1}$, the curvature of the boundary.  In addition, the sign of the correction tells us that active particles accumulate more on concave surfaces than on convex ones.  This preference to accumulate on concave rather than convex surfaces is shown by a discontinuous drop in density going from inside to outside the circular boundary.  
			
			The key observation here is that our approach, even though applied to weakly active particles, can recover some results beyond the weak limit such as the curvature dependence of the density of active particles near curved boundaries \cite{Duzgan Sellinger,Solon et al laplace,Wittmann et al,Nikola et al,Sandford et al,Fily et al}.
			
		\subsection{\label{subsec:wedge}Active particles inside a wedge-shaped region}
			In Section \ref{subsec:corner}, we considered active particles confined by two walls meeting at a right angle.  We here consider a more general and difficult problem of two walls meeting at an arbitrary angle $2\alpha$ (Figure \ref{fig:various geometries}c).  This problem is inspired by experiments and simulations \cite{Galajda et al, Kaiser et al} that showed that active particles could be trapped or directed by wedge-shaped obstacles.  For simplicity, we will focus on the particular case of weakly active particles trapped within a single wedge whose sides extend indefinitely.  The zero current boundary conditions $J_{\theta}(r,\pm\alpha,\boldsymbol{\eta})=0$ along the walls of the wedge give the following condition on the coefficients
			\begin{equation}
				\frac{1}{\tilde{r}}\frac{\partial C_{\boldsymbol{m}}^{(n)}(\tilde{r},\pm\alpha)}{\partial\theta}=w_{\theta}(\tilde{r},\pm\alpha).
			\end{equation}
			In order to make any progress on finding the coefficients in this geometry, we have to make use of the Kontorovich-Lebedev and Mellin transforms (Appendix \ref{subapp:KL M transforms}).  Details for computing the coefficients up to $\epsilon^2$ can be found in Appendix \ref{subapp:wedge}.  We will focus on two quantities: the average propulsion $\langle\boldsymbol{\eta}\rangle(r,\theta)$ and the density $\rho(r,\theta)$ within the wedge.  The general expressions for arbitrary wedge angle $2\alpha$ are quite cumbersome.  Note that $2\alpha=\pi$ and $2\alpha=\pi/2$ correspond to a single wall and two walls meeting at a right angle, respectively.  We have already considered these in Section \ref{sec:cartesian} and it is quite easy to check that the general solution Eq.\ (\ref{eq:wedge correction general}) reduces to those cases.  From here on, we will consider angles of the form $2\alpha=\pi/2^{l-1}$ with $l=3,4\dots$, for which we can make some analytical progress.
			
			\begin{figure*}
				\centering
				\includegraphics[scale=0.5]{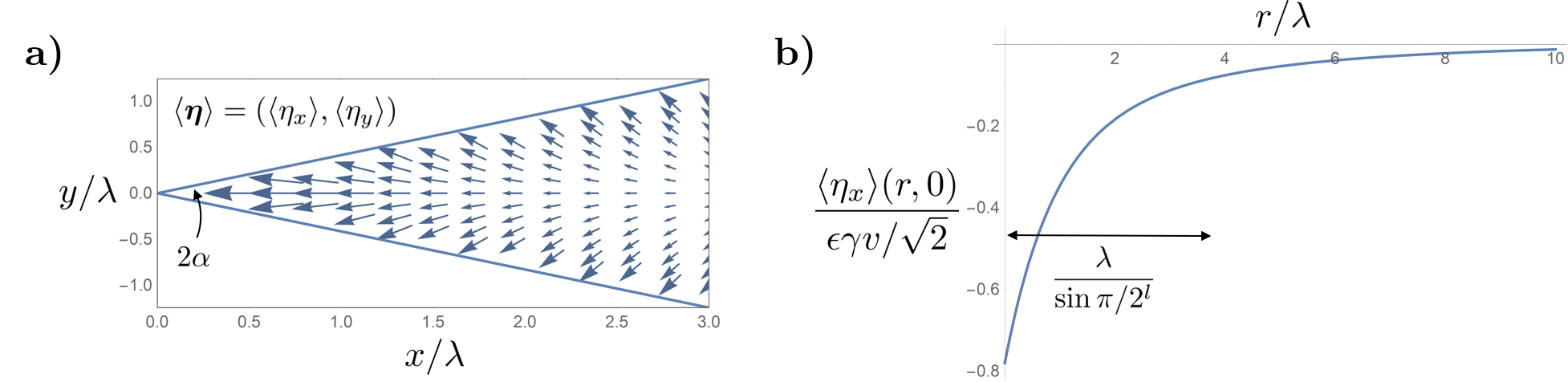}
				\caption{\textbf{a)} Average propulsion $\langle\boldsymbol{\eta}\rangle$ within a wedge-shaped region with angle $2\alpha=\pi/8$.  \textbf{b)} $x$-component of the average propulsion $\langle\eta_x\rangle$ along the center of the wedge $\theta=0$, also for $2\alpha=\pi/8$.  This propulsion towards the tip decays rapidly over a length scale $\frac{\lambda}{\sin\pi/2^l}$.}
				\label{fig:wedge tip orientation}
			\end{figure*}
			
			To start, the components of the average propulsion $\langle\boldsymbol{\eta}\rangle(r,\theta)=(\langle\eta_x\rangle(r,\theta),\langle\eta_y\rangle(r,\theta))$ within the wedge are given by
			\begin{widetext}
				\begin{subequations}
					\begin{align}
						\langle\eta_x\rangle(r,\theta)&\simeq-\frac{\epsilon\gamma v}{\sqrt{2}}\sin\frac{\pi}{2^l}\sum_{k=0}^{2^{l-2}-1}\left\{e^{-\frac{\sqrt{2}r}{\lambda}\sin\left[\frac{\left(2k+1\right)\pi}{2^l}-\theta\right]}+e^{-\frac{\sqrt{2}r}{\lambda}\sin\left[\frac{\left(2k+1\right)\pi}{2^l}+\theta\right]}\right\},\\
						\langle\eta_y\rangle(r,\theta)&\simeq\frac{\epsilon\gamma v}{\sqrt{2}}\cos\frac{\pi}{2^l}\sum_{k=0}^{2^{l-2}-1}(-1)^k\left\{e^{-\frac{\sqrt{2}r}{\lambda}\sin\left[\frac{\left(2k+1\right)\pi}{2^l}-\theta\right]}-e^{-\frac{\sqrt{2}r}{\lambda}\sin\left[\frac{\left(2k+1\right)\pi}{2^l}+\theta\right]}\right\}.
					\end{align}
				\end{subequations}
			\end{widetext}
			The average propulsion $\langle\boldsymbol{\eta}\rangle$ for $l=4$ or $2\alpha=\pi/8$ is shown in Figure \ref{fig:wedge tip orientation}a.  Note the interesting combination of exponentials in the expressions for $\langle\eta_x\rangle$ and $\langle\eta_y\rangle$.  For the simple case of a right-angled corner (Section \ref{subsec:corner}), we found that we could essentially treat each wall as independent up to $\epsilon^2$, that is, each wall contributed a single exponential decay $e^{-\frac{\sqrt{2}x}{\lambda}}$ or $e^{-\frac{\sqrt{2}y}{\lambda}}$ away from itself.  Here, for wedge angles smaller than $\pi/2$, the walls near the tip will influence each other and we find a multitude of exponentials with different length scales.  This has some consequences on the propulsion and accumulation of the active particles.  Let us focus on the propulsion along the center of the wedge $\theta=0$ on which we have $\langle\eta_y\rangle(r,0)=0$ and
			\begin{equation}
				\langle\eta_x\rangle(r,0)=-\frac{\epsilon\gamma v}{\sqrt{2}}\sin\frac{\pi}{2^l}\sum_{k=0}^{2^{l-2}-1}e^{-\frac{\sqrt{2}r}{\lambda}\sin\frac{(2k+1)\pi}{2^l}}.
				\label{eq:average eta_x along center}
			\end{equation}
			This is plotted in Figure \ref{fig:wedge tip orientation}b for $l=4$ or $2\alpha=\pi/8$.  The longest length scale is $\frac{\lambda}{\sin\pi/2^l}$, which grows with decreasing wedge angle.  Beyond this distance, the active particles along the center will be at least a distance $\lambda$ from the sides of the wedge and will effectively not interact with the boundaries.  Within this distance, however, the boundaries will on average orient the active particles towards the tip.  Thus, as the wedge angle decreases, active particles farther and farther from the tip will have some orientation towards it, which in turn should increase the density near the tip.
			
\begin{figure}
				\centering
				\includegraphics[scale=0.5]{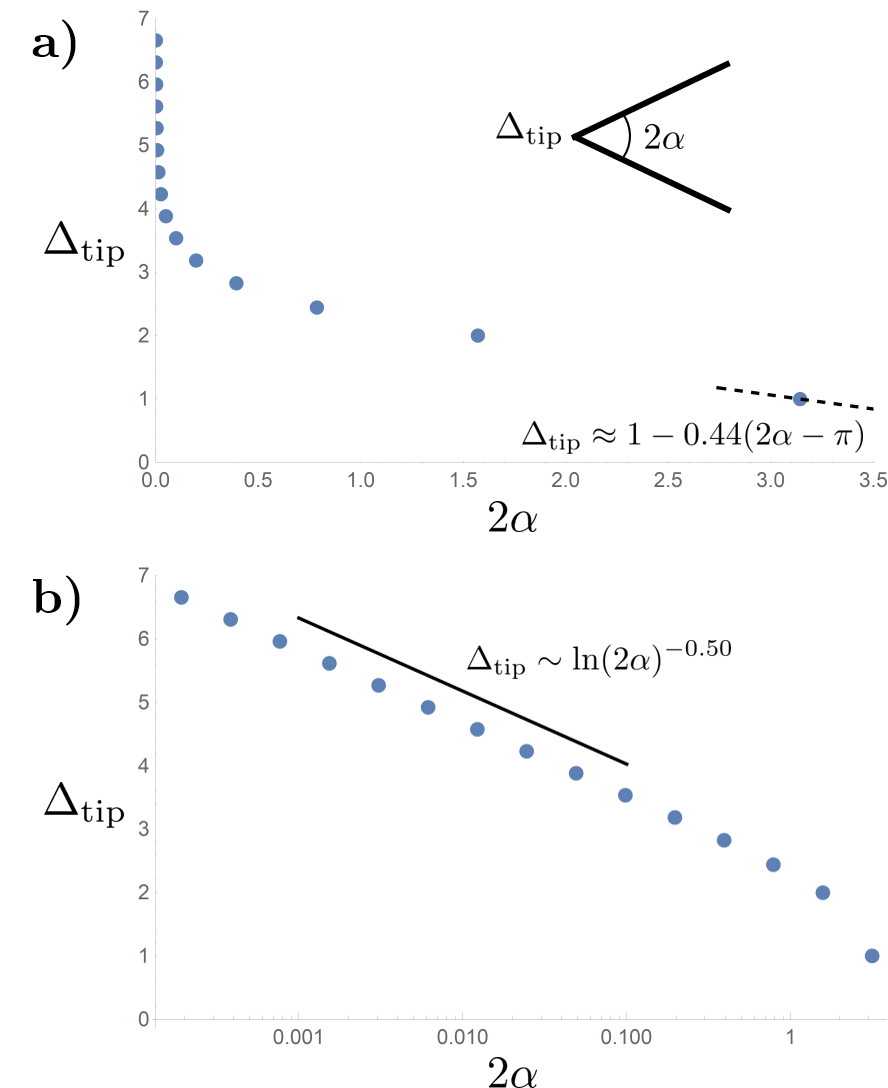}
				\caption{\textbf{a)} Dependence of the correction to wedge tip density $\Delta_{\textrm{tip}}=\lim\limits_{r\rightarrow0}\Delta(r,\theta)$ on the wedge angle $2\alpha=\pi/2^{l-1}$ for $l=1,2,\dots,15$.  The black dotted line shows the linear dependence close to $2\alpha=\pi$, which was obtained using the approach developed in Section \ref{sec:corrugated wall}.  \textbf{b)} Weak scaling of $\Delta_{\textrm{tip}}$ with $2\alpha$ for small wedge angles.}
				\label{fig:wedge tip density}
			\end{figure}

			To study the effect of wedge angle on the density near the tip, we write the density as $\rho(r,\theta)\simeq\rho_{\textrm{bulk}}[1+\epsilon^2\Delta(r,\theta)]$.  The correction to the density $\Delta(r,\theta)$ is given by
			\begin{widetext}
				\begin{align}
					\begin{split}
						\Delta(r,\theta)=&\int_{-\infty}^{\infty}a(s)\cosh s\theta\left(\frac{r}{\lambda}\right)^{-is}ds\\
						&+\sum_{k=0}^{2^{l-2}-1}(-1)^k\cos\left\{\frac{[2k+1-(-1)^k]\pi}{2^l}\right\}\left\{e^{-\frac{\sqrt{2}r}{\lambda}\sin\left[\frac{(2k+1)\pi}{2^l}-\theta\right]}+e^{-\frac{\sqrt{2}r}{\lambda}\sin\left[\frac{(2k+1)\pi}{2^l}+\theta\right]}\right\},
					\end{split}
					\label{eq:wedge correction}
				\end{align}
				where
				\begin{align}
					a(s)=\frac{\sqrt{2}^{-is}\Gamma(is+1)}{2\pi s\sinh\frac{s\pi}{2^l}}\left[1+2\sum_{k=1}^{2^{l-2}-1}\left(\sin^2\frac{\pi}{2^l}+(-1)^k\cos^2\frac{\pi}{2^l}\right)\left(\sin\frac{k\pi}{2^{l-1}}\right)^{-is+1}\right].
					\label{eq:wedge a}
				\end{align}
			\end{widetext}
			For angles $2\alpha=\pi/2^{l-1}$ with $l\ge3$, there is no known closed form solution and we have to numerically evaluate the integral in Eq.\ (\ref{eq:wedge correction}).  Let us focus on how the density at the tip of the wedge $\Delta_{\textrm{tip}}=\lim\limits_{r\rightarrow0}\Delta(r,\theta)$ depends on the angle of the wedge $2\alpha$.  This dependence is shown in Figure \ref{fig:wedge tip density}.
			
			There are a couple of features to note.  The first is the increase in the density at the tip as we decrease the angle of the wedge, as shown in Figure \ref{fig:wedge tip density}a.  This is due to the effect of the sides of the wedge on the average propulsion $\langle\boldsymbol{\eta}\rangle$ discussed earlier (Eq.\ \ref{eq:average eta_x along center}).  As the wedge angle decreases, active particles farther and farther from the tip are on average directed towards it, thus increasing the density.  In addition to the increase in density for small wedge angles, we find that when the wedge is nearly a flat wall ($2\alpha\approx\pi$), the density has a linear dependence given by $\Delta_{\textrm{tip}}\approx1-0.44(2\alpha-\pi)$.  We will show how this is obtained using an approach developed in Section \ref{sec:corrugated wall}.
			
			The second feature is the scaling of the tip density $\Delta_{\textrm{tip}}$ with the wedge angle $2\alpha$.  As shown in Figure \ref{fig:wedge tip density}b, we find an interesting weak dependence of the form $\Delta_{\textrm{tip}}\sim\ln(2\alpha)^{-0.50}$.  This weak dependence is due to passive Brownian diffusion and can actually be obtained through a relatively simple scaling argument.  This argument goes as follows.  As we discussed earlier, there is a length scale $\lambda_l\sim\frac{\lambda}{\sin\pi/2^l}$ over which the active particles on average propel towards the tip.  This increases the density at the tip by $\rho_{\textrm{bulk}}\epsilon^2\Delta_{\textrm{tip}}$.  Thus, the diffusive flux over this length scale away from the tip goes as $J_{\textrm{diff}}\sim D_p\frac{\rho_{\textrm{bulk}}\epsilon^2\Delta_{\textrm{tip}}}{\lambda_l}$.  To estimate the advective flux, we compute a characteristic propulsion towards the tip by averaging $\langle\eta_x\rangle(r,0)$ (Eq.\ \ref{eq:average eta_x along center}) over the region $0<r\lesssim\lambda_l$.  Thus the advective flux towards the tip goes as $J_{\textrm{swim}}\sim\frac{1}{\gamma}\left|\frac{1}{\lambda_l}\int_{0}^{\lambda_l}\langle\eta_x\rangle(r,0)dr\right|\rho_{\textrm{bulk}}$.  To a good approximation, we can take the upper limit of the integral to $\infty$ since $\langle\eta_x\rangle(r,0)$ decays rapidly over $0<r\lesssim\lambda_l$ (Figure \ref{fig:wedge tip orientation}b).  Finally, the diffusive and advective fluxes in steady-state should balance ($J_{\textrm{diff}}\sim J_{\textrm{swim}}$) and so we obtain
			\begin{align}
				\begin{split}
					\Delta_{\textrm{tip}}&\sim\frac{1}{\gamma D_p\epsilon^2}\left|\int_{0}^{\infty}\langle\eta_x\rangle(r,0)dr\right|\\
					&=\sin\frac{\pi}{2^l}\sum_{k=0}^{2^{l-2}-1}\frac{1}{\sin\frac{(2k+1)\pi}{2^l}}\\
					&\approx\frac{1}{2}\int_{\frac{\pi}{2^l}}^{\frac{\pi}{2}}\frac{du}{\sin u}\approx-\frac{1}{2}\ln\frac{\pi}{2^{l+1}},
				\end{split}
			\end{align}
			where we assumed that $2^l\gg1$ for small wedge angles and approximated the sum as an integral.  Thus, up to a constant shift, we see that $\Delta_{\textrm{tip}}\sim\ln\left(\frac{\pi}{2^{l-1}}\right)^{-1/2}=\ln(2\alpha)^{-1/2}$, in close agreement with our numerical estimates in Figure \ref{fig:wedge tip density}b.
		
	\section{\label{sec:corrugated wall}Active particles near a corrugate wall}
		We now turn to one last example involving impenetrable walls.  We study how active particles behave near a corrugated wall (Figure \ref{fig:various geometries}d) and show how we can formulate the boundary condition for such a wall.  This example is inspired by experiments on asymmetric gears in bacterial baths showing that active particles can generate tangential forces on an asymmetric boundary \cite{Sokolov et al,Leonardo et al}, in addition to simulations showing that asymmetric boundaries can transport active particles \cite{Ghosh et al,Ai,Ai et al}.  Thus, we expect that a combination of asymmetry and activity should lead to directional motion of either the boundary or the particles.  However, one of the counterintuitive results we will find here is that there is actually no net transport of noninteracting active particles along an asymmetric corrugated wall if the amplitude of the corrugation is too small.  More precisely, the net tangential current of noninteracting active particles near an asymmetric corrugate wall does not decrease to zero linearly as the wall becomes flatter and flatter.
		
		To set up the problem, suppose we have a boundary with a shape $y=h(x)$ with period $2L$ and characteristic amplitude $\delta$ such that $|h(x)|\lesssim\delta$.  We can decompose the shape into Fourier modes as
		\begin{equation}
			h(x)=\delta\sum_{k=-\infty}^{\infty}h_ke^{\frac{i\pi kx}{L}},
		\end{equation}
		where $h_{k=0}=0$, that is, the shape of the boundary oscillates around $y=0$.  Assuming that the amplitude of the shape is small compared to the length scale of accumulation or $\delta\ll\lambda$, we can write the distribution as
		\begin{equation}
			\tilde{\rho}(\tilde{\boldsymbol{r}},\tilde{\boldsymbol{\eta}})=\sum_{n=0}^{\infty}\epsilon^n\sum_{\boldsymbol{m}}C_{\boldsymbol{m}}^{(n)}(\tilde{\boldsymbol{r}})e^{-\tilde{\boldsymbol{\eta}}^2}H_{m_x}(\tilde{\eta}_x)H_{m_y}(\tilde{\eta}_y),
		\end{equation}
		where we now expand the coefficients as
		\begin{equation}
			C_{\boldsymbol{m}}^{(n)}(\tilde{\boldsymbol{r}})\simeq a_{\boldsymbol{m}}^{(n)}(\tilde{y})+\tilde{\delta}\sum_{k=-\infty}^{\infty}b_{\boldsymbol{m};k}^{(n)}(\tilde{y})e^{\frac{i\pi k\tilde{x}}{\tilde{L}}}.
			\label{eq:corrugated c coefficient}
		\end{equation}
		The functions $a_{\boldsymbol{m}}^{(n)}(\tilde{y})$ are simply the solutions for a flat wall with no corrugation, which we have already computed in Section \ref{sec:cartesian}.  The resulting equation for the unknown coefficients $b_{\boldsymbol{m};k}^{(n)}(\tilde{y})$ is
		\begin{equation}
			\frac{d^2b_{\boldsymbol{m};k}^{(n)}}{d\tilde{y}^2}-\left[2(m_x+m_y)+\frac{\pi^2k^2}{\tilde{L}^2}\right]b_{\boldsymbol{m};k}^{(n)}=\frac{i\pi k}{\tilde{L}}w_x+\frac{dw_y}{d\tilde{y}},
			\label{eq:corrugated coefficient equation}
		\end{equation}
		where
		\begin{subequations}
			\begin{align}
				w_x&=b_{m_x-1,m_y;k}^{(n-1)}+2(m_x+1)b_{m_x+1,m_y;k}^{(n-1)},\\
				w_y&=b_{m_x,m_y-1;k}^{(n-1)}+2(m_y+1)b_{m_x,m_y+1;k}^{(n-1)}.
			\end{align}
		\end{subequations}
		For the boundary condition, we require the normal component of the current to be zero at the boundary or $\boldsymbol{J}(x,h(x))\cdot\hat{\boldsymbol{n}}=0$, where $\hat{\boldsymbol{n}}$ is the normal to the boundary.  Assuming that $\delta\ll\lambda$, we can linearize this boundary condition to get (see Appendix \ref{app:corrugated wall})
		\begin{widetext}
			\begin{align}
				\begin{split}
					&-\left[a_{m_x-1,m_y}^{(n-1)}(0)+2(m_x+1)a_{m_x+1,m_y}^{(n-1)}(0)\right]\frac{i\pi k}{\tilde{L}}h_k+\left[\frac{da_{m_x,m_y-1}^{(n-1)}(0)}{d\tilde{y}}+2(m_y+1)\frac{da_{m_x,m_y+1}^{(n-1)}(0)}{d\tilde{y}}-\frac{d^2a_{\boldsymbol{m}}^{(n)}(0)}{d\tilde{y}^2}\right]h_k\\
					&+b_{m_x,m_y-1;k}^{(n-1)}(0)+2(m_y+1)b_{m_x,m_y+1;k}^{(n-1)}(0)-\frac{db_{\boldsymbol{m};k}^{(n)}(0)}{d\tilde{y}}=0.
				\end{split}
				\label{eq:corrugated boundary condition}
			\end{align}
		\end{widetext}
		Determining the coefficients $b_{\boldsymbol{m};k}^{(n)}(\tilde{y})$ is quite straightforward and the expressions can be found in Appendix \ref{app:corrugated wall}.
		
		Let us start with the simplest case of $h(x)=\delta\cos\frac{\pi x}{L}$ or $h_{k=\pm1}=\frac{1}{2}$.  In particular, consider the density $\rho(x,h(x))$ along the boundary when the amplitude of the boundary is small $\delta\ll\lambda$ and the wavelength is large $L\gg\lambda$.  For this slow varying boundary, the density along the boundary to linear order in the amplitude is
		\begin{equation}
			\rho(x,h(x))\simeq\rho_{\textrm{bulk}}\left[1+\epsilon^2-\epsilon^2\delta\frac{\pi^2\lambda}{\sqrt{2}L^2}\cos\frac{\pi x}{L}\right].
		\end{equation}
		Notice that the last term, which captures the effect of corrugation, is proportional to $\frac{d^2h}{dx^2}$.  Thus, the change in density is related to the local curvature of the boundary with more active particles accumulating on the concave sections than on the convex sections.  The correction due to the corrugation can be written as $\epsilon^2\rho_{\textrm{bulk}}\frac{\lambda}{\sqrt{2}R}$, where $R=\left[1+\left(\frac{dh}{dx}\right)^2\right]^{\frac{3}{2}}\left|\frac{d^2h}{dx^2}\right|^{-1}\simeq\frac{L^2}{\pi^2\delta}$ is the radius of curvature at the maxima and minima of the corrugated boundary.  Note that this is in exact agreement with the result for the densities outside and inside a large circular boundary with radius $R\gg\lambda$ (Eqs.\ (\ref{eq:convex boundary density}) and (\ref{eq:concave boundary density})).  
		
\begin{figure*}
			\centering
			\includegraphics[scale=0.55]{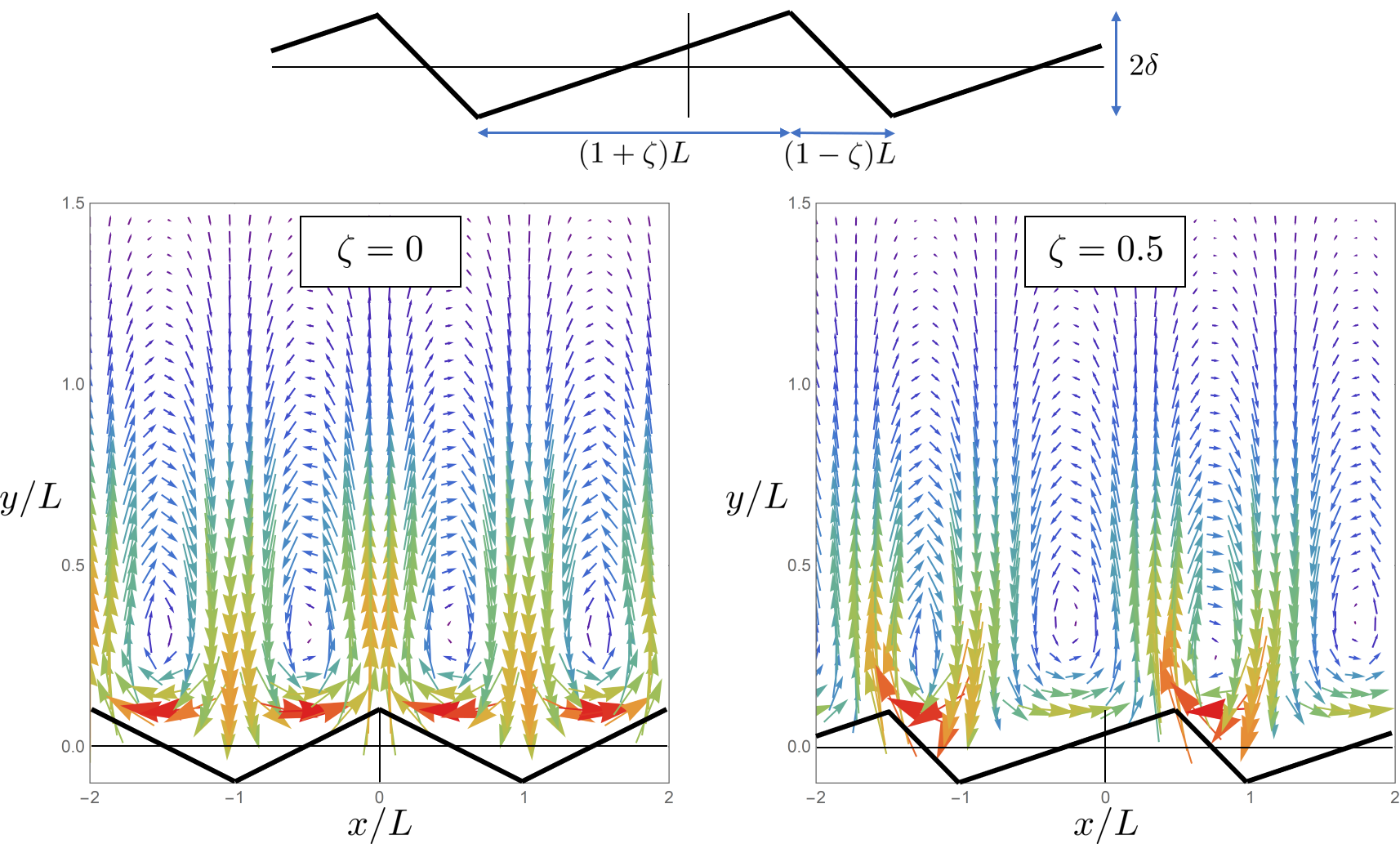}
			\caption{\textbf{Top:} Sawtooth-shaped boundary defined by Eq.\ (\ref{eq:sawtooth shape}).  \textbf{Bottom:} Currents $\boldsymbol{J}(x,y)=(J_x(x,y),J_y(x,y))$ due to sawtooth-shaped boundaries with asymmetry $\zeta=0$ (\textbf{left}) and $\zeta=0.5$ (\textbf{right}) computed using the first 20 modes of $h(x)$.}
			\label{fig:corrugated currents}
		\end{figure*}

		Let us now consider a more complex example.  Suppose we have a sawtooth-shaped boundary (Figure \ref{fig:corrugated currents}, top) given by
		\begin{equation}
			h(x)=\begin{cases}
				-\delta+\frac{2\delta}{(1+\zeta)L}(x+L), & -L<x<\zeta L\\
				\delta-\frac{2\delta}{(1-\zeta)L}(x-\zeta L), & \zeta L<x<L
			\end{cases}.
			\label{eq:sawtooth shape}
		\end{equation}
		The asymmetry is controlled by $\zeta$, where $\zeta=0$ corresponds to a symmetric sawtooth.  Unlike the simple case of a cosine-shaped boundary, the sawtooth is not twice differentiable near the sharp tips, and so curvature is not well-defined.  In Section \ref{subsec:wedge}, we studied how the density at the tip of a wedge depended on the angle of the wedge.  Using a sawtooth-shaped boundary, we can obtain the dependence of the density on angles near $2\alpha=\pi$.  Taking the slow-varying symmetric sawtooth with $\delta\ll\lambda$ and $L\gg\lambda$, the active particles near the tip at $x=0$ effectively see a wedge with angle $2\alpha\simeq\pi+\frac{4\delta}{L}$, where $\delta>0$ and $\delta<0$ correspond to convex and concave, respectively.  Writing the density as $\rho(x,y)=\rho_{\textrm{bulk}}\left[1+\epsilon^2\Delta(x,y)\right]$, just as we did for the wedge, we find that the correction to density at the tip is (see Appendix \ref{app:corrugated wall})
		\begin{equation}
			\Delta_{\textrm{tip}}=\Delta(0,\delta)\approx1+\frac{4\delta}{L}S\simeq1-0.44(2\alpha-\pi).
		\end{equation}  
		
		Before we conclude this section, we briefly discuss the currents of noninteracting weakly active particles in the presence of a corrugated wall.  The explicit forms of the currents can be found in Appendix \ref{app:corrugated wall}.  For a boundary with an asymmetric shape, one expects there to be a net flux of active particles along the boundary.  For example in suspensions of bacteria, it has been seen that swimming bacteria can be directed by funnels and can rotate gears with asymmetric teeth \cite{Galajda et al,Sokolov et al,Leonardo et al}.  However, for our case of an asymmetric sawtooth boundary and noninteracting weakly active particles, we find the surprising result that there is no net drift along the wall to linear order in the amplitude $\delta$ of the corrugation.  Mathematically, this is easily explained by noticing that the coefficient equation (Eq.\ (\ref{eq:corrugated coefficient equation})) and boundary condition (Eq.\ (\ref{eq:corrugated boundary condition})) are all independent for each mode $k$.  Since each mode is a symmetric sine or cosine wave, none of them contribute to a net drift.  In addition, there is also no net tangential force on the boundary by the same reasoning.  
		
		In order to get a net tangential drift or force along the boundary, we need to couple modes with different $k$, which can be done by introducing nonlinearities.  There are two possible ways to do this.  The first way is going beyond the linearized boundary condition (Eq.\ (\ref{eq:corrugated boundary condition})) and considering higher orders in the amplitude $\delta$.  In fact, noting that the transformation $\delta\rightarrow-\delta$ should simply flip the direction of drift, the drift of noninteracting weakly active particles due to an asymmetric corrugated boundary should scale as $\delta^3$ for small amplitudes of corrugation.  This nonlinear scaling with amplitude of corrugation has been seen in simulations of noninteracting active particles \cite{Ghosh et al}.  The calculation for going beyond the linearized boundary condition is rather involved and will be reserved for a future work.  The second possible way is including interactions such as alignment between the active particles.  Interactions may make it easier for an asymmetric boundary to induce net fluxes.  In fact, it has been seen in simulations of aligning active particles in corrugated channels that the net currents along the channels can actually be linear in the amplitude of corrugation \cite{Ai et al}, contrasting our result for noninteracting active particles.
		
		Note that while there is no net tangential drift to linear order $\delta$, there is still a local circulation of active particles (Figure \ref{fig:corrugated currents}, bottom).  These local fluxes of active particles towards the concave parts and away from the convex parts of the boundary are responsible for the increases and decreases of the densities in those parts, respectively.
		
	\section{\label{sec:spherical}A problem in spherical coordinates: absorption of active particles around a sphere}	
		For the final case, we will consider an absorbing boundary as opposed to a impenetrable boundary, which has been our focus thus far.  In particular, let us consider a uniform bath of weakly active particles in which we place an absorbing sphere (Figure \ref{fig:various geometries}e).  We are interested in determining the steady-state rate at which these active particles are absorbed and how that rate differs from that of passive Brownian particles.  For the absorption of passive Brownian particles, this is known as the Smoluchowski problem \cite{Smoluchowski}.  
		
		In spherical coordinates, we write the dimensionless distribution as
		\begin{align}
			\begin{split}
				\tilde{\rho}(\tilde{r},\theta,\phi,\tilde{\boldsymbol{\eta}})=\sum_{n=0}^{\infty}\epsilon^n&\sum_{\boldsymbol{m}}C_{\boldsymbol{m}}^{(n)}(\tilde{r},\theta,\phi)\\
				&\times e^{-\tilde{\boldsymbol{\eta}}^2}H_{m_x}(\tilde{\eta}_x)H_{m_y}(\tilde{\eta}_y)H_{m_z}(\tilde{\eta}_z)
			\end{split}
		\end{align}
		The coefficients satisfy
		\begin{align}
			\begin{split}
				&\frac{1}{\tilde{r}^2}\frac{\partial}{\partial\tilde{r}}\left(\tilde{r}^2\frac{\partial C_{\boldsymbol{m}}^{(n)}}{\partial\tilde{r}}\right)-2(m_x+m_y+m_z)C_{\boldsymbol{m}}^{(n)}\\
				&+\frac{1}{\tilde{r}^2\sin\theta}\frac{\partial}{\partial\theta}\left(\sin\theta\frac{\partial C_{\boldsymbol{m}}^{(n)}}{\partial\theta}\right)+\frac{1}{\tilde{r}^2\sin^2\theta}\frac{\partial^2C_{\boldsymbol{m}}^{(n)}}{\partial\phi^2}\\
				&=\frac{1}{\tilde{r}^2}\frac{\partial}{\partial\tilde{r}}(\tilde{r}^2w_r)+\frac{1}{\tilde{r}\sin\theta}\frac{\partial}{\partial\theta}(\sin\theta w_{\theta})+\frac{1}{\tilde{r}\sin\theta}\frac{\partial w_{\phi}}{\partial\phi},
			\end{split}
		\end{align}
		where the components of $\boldsymbol{w}$ in spherical coordinates are
		\begin{subequations}
			\begin{align}
				w_r&=w_x\sin\theta\cos\phi+w_y\sin\theta\sin\phi+w_z\cos\theta,\\
				w_{\theta}&=w_x\cos\theta\cos\phi+w_y\cos\theta\sin\phi-w_z\sin\theta,\\
				w_{\phi}&=-w_x\sin\phi+w_y\cos\phi,
			\end{align}
		\end{subequations}
		and
		\begin{subequations}
			\begin{align}
				w_x&=C_{m_x-1,m_y,m_z}^{(n-1)}+2(m_x+1)C_{m_x+1,m_y,m_z}^{(n-1)},\\
				w_y&=C_{m_x,m_y-1,m_z}^{(n-1)}+2(m_y+1)C_{m_x,m_y+1,m_z}^{(n-1)},\\
				w_z&=C_{m_x,m_y,m_z-1}^{(n-1)}+2(m_z+1)C_{m_x,m_y,m_z+1}^{(n-1)}.
			\end{align}
		\end{subequations}
		If the sphere has a radius $R$, then the absorbing boundary condition at $r=R$ gives us the condition on the coefficients  $C_{\boldsymbol{m}}^{(n)}(\tilde{R},\theta,\phi,\tilde{\boldsymbol{\eta}})=0$.  Details of the solution can be found in Appendix \ref{app:absorber}.  Taking $\rho(r,\theta,\phi,\boldsymbol{\eta})$ and integrating out $\boldsymbol{\eta}$, we obtain to order $\epsilon^2$ the density
		\begin{align}
			\frac{\rho(r,\theta,\phi)}{\rho_{\textrm{bulk}}}\simeq1-\frac{R}{r}+\epsilon^2\frac{\lambda}{\lambda+\sqrt{2}R}\left[1-e^{-\frac{\sqrt{2}(r-R)}{\lambda}}\right]\frac{R}{r},
		\end{align}
		where $\rho_{\textrm{bulk}}$ is the uniform density far from the sphere.  The first part is the familiar $r^{-1}$ solution for passive Brownian particles while the second part is the correction due to activity, which elevates the density near the absorbing sphere.  The radial current is
		\begin{equation}
			J_r(r,\theta,\phi)\simeq-\frac{\rho_{\textrm{bulk}}D_pR}{r^2}\left(1+\epsilon^2\frac{\sqrt{2}R}{\lambda+\sqrt{2}R}\right),
		\end{equation}
		from which we calculate the capture rate $\kappa=|\int J_r(R,\theta,\phi)R^2\sin\theta\,d\theta \,d\phi\,|$ as 
		\begin{equation}
			\kappa\simeq\kappa_0\left(1+\epsilon^2\frac{\sqrt{2}R}{\lambda+\sqrt{2}R}\right),
		\end{equation}
		where $\kappa_0=4\pi\rho_{\textrm{bulk}}D_pR$ is the well-known capture rate for passive Brownian particles \cite{Smoluchowski}, which depends on the radius and not the surface area of the sphere.  
		
		The correction due to activity is a new result.  There are two limits: a large target $R\gg\lambda$ and a small target $R\ll\lambda$.  For a large target, the time it takes the weakly active particles to passively diffuse over a distance comparable to the radius of the absorber is much longer than the persistence time ($\tau\ll R^2/D_p$).  On this time scale, the active particles appear effectively diffusive and we find the capture rate $\kappa\simeq4\pi\rho_{\textrm{bulk}}R(D_p+v^2\tau/3)$, which can be interpreted as just that of a diffusing particle with effective diffusivity $D_{\textrm{eff}}=D_p+v^2\tau/3$.  For a small target, however, the time it takes to diffuse over the radius of the absorber is much shorter than the persistence time ($\tau\gg R^2/D_p$).  On this time scale, the propulsions of the active particles appear persistent.  We find $\kappa\simeq4\pi\rho_{\textrm{bulk}}R(D_p+v^2\tau\sqrt{2}R/3\lambda)$.  Note that the correction to the rate due to activity scales as $R^2$, which is related to the surface area or cross-sectional area of the absorber.  This is reminiscent of the capture rate $\kappa\sim\rho_{\textrm{bulk}}vR^2$ for a spherical absorber in an ideal gas of ballistic particles \cite{Hubner Titulaer}, whose mean free paths are longer than the radius of the sphere.  One key difference of course is that while the propulsions of our weakly active particles appear persistent, their motions are still dominated by passive Brownian diffusion.  To summarize, in both cases of a small and large target, we find that activity increases the density and enhances capture rate of active particles near an absorbing boundary.
	
	\section{Discussion and Conclusion}
		We studied how noninteracting weakly active particles, for which activity can be treated perturbatively, behave near various types of boundaries in different geometries; for example, active particles moving on a line or in a wedge-shaped region, interacting with a corrugated wall, or absorbing around a sphere.  The key to making progress on this problem is to include passive Brownian diffusion, which allows us not only to cleanly formulate boundary conditions for the different types of boundaries but also to solve the problem systematically.  In other words, by treating activity perturbatively, we can take the solutions for passive Brownian particles, which are often known, and use them to iteratively compute the corrections due to the activity.  We formulated a relatively simple series solution for the distribution of active particles that consists of an expansion in powers of the P\'{e}clet number, which characterizes the strength of activity, and an expansion in terms of Hermite polynomials.  This series solution reduces the Fokker-Planck equation for the distribution of active particles to a simpler partial differential equation and in some cases, to an even simpler ordinary differential equation.  We summarize below some of our main results for the different geometries.  
		
		By considering the simple cases of noninteracting weakly active particles confined by one or two impenetrable walls in 1D (Section \ref{subsec:particles on line}), we found that the active particles on average propel towards nearby walls.  This leads to accumulation and an increase in pressure exerted on the walls.  In particular, we found that the pressure follows the ideal gas law but instead of being proportional to the density in the bulk, it is proportional to the increased density at the walls.  
		
		We also determined how the curvature of a boundary affects the accumulation of weakly active particles (Section \ref{sec:polar}).  For the case of a circular boundary, we found that the accumulation is proportional to the curvature of the boundary.  This last result has been shown to hold in the limit of strong activity \cite{Duzgan Sellinger,Solon et al laplace,Wittmann et al,Nikola et al,Sandford et al,Fily et al}.  Thus, our approach for studying the limit of weakly active particles can potentially gain us insight into the opposite limit.  For the case of a wedge-shaped region, we found that as the wedge angle decreases, active particles farther and farther from the tip gain on average some propulsion towards it.  Interestingly, while this propulsion does increase the density near the tip, we found that the accumulation has a rather weak dependence on the wedge angle.  This is due to passive Brownian diffusion which tends to smooth out variations in density.
		
		Finally, we have also obtained novel results for weakly active particles near a corrugated boundary and around an absorbing sphere.  Near a corrugated boundary (Section \ref{sec:corrugated wall}), particularly one shaped like an asymmetric sawtooth, we found that there is surprisingly no net transport of noninteracting weakly active particles along the boundary to linear order in the amplitude $\delta$ of the corrugation.  This is due to the linearity of the boundary condition and the Fourier modes that make up the shape of the wall.  We argued that in order to observe net currents we have to introduce nonlinearities such as going beyond the linearized boundary condition or including interactions between the active particles.  For our case of noninteracting weakly active particles, we expect to see net currents at order $\delta^3$.  A nonlinear dependence on $\delta$ has been seen in simulations of noninteracting active particles \cite{Ghosh et al}.  For interacting active particles, it is more difficult to determine the dependence.  However, it has been observed in simulations that for active particles with aligning interactions, the net current is linear in $\delta$ \cite{Ai et al}, suggesting that interactions may enhance the transport of active particles.  For an absorbing sphere placed in a bath of active particles (Section \ref{sec:spherical}), we computed the rate at which the weakly active particles are absorbed.  This is the active version of the Smoluchowski problem \cite{Smoluchowski} for passive Brownian particles.  We found that activity elevates the density near the sphere and enhances the absorption rate.  Thus, activity may be useful in enhancing the self-assembly of colloidal structures \cite{Mallory et al}.
		
		Before we end this story, there are some interesting future directions to consider.  The first direction is finding the exact solution for the distribution of noninteracting active particles near the simplest case of a flat wall.  By ``exact'', we mean a closed-form expression for the distribution or, at a minimum, for all the coefficients in our series solution.  With the current approach, one can systematically compute higher and higher orders.  However, the expressions, though straightforward, become increasingly cumbersome to write down.  The goal would be to find a pattern in the coefficients that one can exploit.  Finding a clean way to do this could aid us in finding more exact solutions in other interesting geometries.
		
		The second direction is going beyond the linearized boundary condition for a corrugated boundary.  As we found, there is no net transport of noninteracting active particles or net tangential force along an asymmetrically-shaped boundary to linear order in the amplitude of the corrugation.  In order to observe net tangential currents or forces, one will need to consider higher orders in the amplitude.  It would be interesting to perform this calculation and to analytically compute how fast active particles are transported by an asymmetric wall or how fast an asymmetric wall is pushed like a gear by active particles.  
		
		Finally, it would be interesting to extend the approach developed here to more realistic models of active particles.  This includes studying other models of active particles such as active Brownian particles, which typically model many types of self-propelled colloids \cite{Palacci et al,Paxton et al}, and run-and-tumble particles, which typically model bacteria \cite{Berg et al}.  More generally, it may be interesting to study models where the correlations are not exponential or the persistence times have a broad distribution, as has been seen in some bacterial systems \cite{Figueroa-Morales et al}.  An important question is whether there are critical differences between the many models of active particles, for example, when interacting with boundaries.  In addition to studying different models of active particles, it would also be interesting to include interactions between particles in our approach.  It has been seen that a simple repulsive interaction can have significant effects on the density and pressure of active particles \cite{Buttinoni et al,Ginot et al}.  Similarly, as was discussed, interactions may affect the transport of active particles in corrugated channels \cite{Ghosh et al,Ai et al}.
	
	\section*{Acknowledgments}
		This work was supported primarily by the MRSEC Program of the National Science Foundation under Award Number DMR-1420073.  The author greatly thanks Alexander Grosberg for his invaluable comments and critical reading of the manuscript.
	
	\appendix
	\section{Mathematical preliminaries}
	
		\subsection{\label{subapp:hermite polynomials}Hermite polynomials}
			The dynamics of the active force $\eta$ (Eq.\ \ref{eq:langevin eta}) can be mapped to an overdamped particle in a quadratic potential.  Naturally, Hermite polynomials should come in handy.  Consider the ODE
			\begin{equation}
				\frac{d^2F}{d\tilde{\eta}^2}+2\frac{d}{d\tilde{\eta}}(\tilde{\eta}F)+\kappa^2F=0.
			\end{equation}
			Note that the first two terms are the active parts of the Fokker-Planck equation for the distribution active particles (Eq.\ (\ref{eq:fokker-planck})).  Taking $F(\tilde{\eta})=e^{-\tilde{\eta}^2}H(\tilde{\eta})$, we have
			\begin{equation}
				\frac{d^2H}{d\tilde{\eta}^2}-2\tilde{\eta}\frac{dH}{d\tilde{\eta}}+\kappa^2H=0.
			\end{equation}
			The solutions satisfying the condition that $F(\tilde{\eta})$ decays sufficiently quickly as $|\tilde{\eta}|\rightarrow0$ are Hermite polynomials $H_m(\tilde{\eta})$ with eigenvalues $\kappa^2=2m$.  The first few are
			\begin{subequations}
				\begin{align}
					H_0(\tilde{\eta})&=1\\
					H_1(\tilde{\eta})&=2\tilde{\eta}\\
					H_2(\tilde{\eta})&=4\tilde{\eta}^2-2\\
					H_3(\tilde{\eta})&=8\tilde{\eta}^3-12\tilde{\eta}\\
					H_4(\tilde{\eta})&=16\tilde{\eta}^4-48\tilde{\eta}^2+12
				\end{align}
			\end{subequations}
			These satisfy the orthogonality relation
			\begin{equation}
				\int_{-\infty}^{\infty}d\tilde{\eta}\,e^{-\tilde{\eta}^2}H_n(\tilde{\eta})H_m(\tilde{\eta})=\sqrt{\pi}\,2^nn!\,\delta_{n,m}.
			\end{equation}
			A few useful recursion relations are
			\begin{equation}
				2\tilde{\eta}H_m(\tilde{\eta})=H_{m+1}(\tilde{\eta})+2mH_{m-1}(\tilde{\eta}),
			\end{equation}
			and
			\begin{equation}
				\frac{dH_m}{d\tilde{\eta}}=2mH_{m-1}.
			\end{equation}
			For higher dimensions, we will instead have the PDE
			\begin{equation}
				\nabla_{\tilde{\eta}}^2F+2\boldsymbol{\nabla}_{\tilde{\eta}}\cdot(\tilde{\boldsymbol{\eta}}F)+\kappa^2F=0,
			\end{equation}
			the eigenfunctions of which are simply
			\begin{equation}
				F(\tilde{\boldsymbol{\eta}})=e^{-\tilde{\boldsymbol{\eta}}^2}\prod_{i=1}^{d}H_{m_i}(\tilde{\eta}_i)
			\end{equation}
			with eigenvalues $\kappa^2=2(m_1+\dots+m_d)$.  This easy generalization to higher dimensions is one benefit of our approach.

		\subsection{Modified Bessel functions}
			For problems in polar coordinates, the coefficient equation will often be of the form
			\begin{equation}
				\frac{1}{\tilde{r}}\frac{\partial}{\partial\tilde{r}}\left(\tilde{r}\frac{\partial F}{\partial \tilde{r}}\right)+\frac{1}{\tilde{r}^2}\frac{\partial^2F}{\partial\theta^2}-\kappa^2F=0,
			\end{equation}
			where $\kappa^2=2(m_x+m_y)$.  Writing $F(\tilde{r},\theta)=G(\tilde{r})e^{i\mu\theta}$, we have the ODE
			\begin{equation}
				\frac{1}{\tilde{r}}\frac{d}{d\tilde{r}}\left(\tilde{r}\frac{dG}{d\tilde{r}}\right)-\left(\frac{\mu^2}{\tilde{r}^2}+\kappa^2\right)G=0,	
			\end{equation}
			The solutions to this ODE are the modified Bessel functions of the first and second kinds $I_{\mu}(\kappa\tilde{r}),K_{\mu}(\kappa\tilde{r})$.  A few useful recursion relations are
			\begin{subequations}
				\begin{align}
					\frac{2\mu}{\kappa\tilde{r}}I_{\mu}(\kappa\tilde{r})&=I_{\mu-1}(\kappa\tilde{r})-I_{\mu+1}(\kappa\tilde{r}),\\
					\frac{2}{\kappa}\frac{dI_{\mu}(\kappa\tilde{r})}{d\tilde{r}}&=I_{\mu-1}(\kappa\tilde{r})+I_{\mu+1}(\kappa\tilde{r}),\\
					-\frac{2\mu}{\kappa\tilde{r}}K_{\mu}(\kappa\tilde{r})&=K_{\mu-1}(\kappa\tilde{r})-K_{\mu+1}(\kappa\tilde{r}),\\
					-\frac{2}{\kappa}\frac{dK_{\mu}(\kappa\tilde{r})}{d\tilde{r}}&=K_{\mu-1}(\kappa\tilde{r})+K_{\mu+1}(\kappa\tilde{r}).
				\end{align}
				\label{eq:bessel identities}%
			\end{subequations}
			There are some useful asymptotic forms.  For $\kappa\tilde{r}\gg1$, we have
			\begin{subequations}
				\begin{align}
					I_{\mu}(\kappa\tilde{r})&\simeq\frac{1}{\sqrt{2\pi\kappa\tilde{r}}}e^{\kappa\tilde{r}}\left(1-\frac{4\mu^2-1}{8\kappa\tilde{r}}\right),\\
					K_{\mu}(\kappa\tilde{r})&\simeq\sqrt{\frac{\pi}{2\kappa\tilde{r}}}e^{-\kappa\tilde{r}}\left(1+\frac{4\mu^2-1}{8\kappa\tilde{r}}\right).
				\end{align}
			\end{subequations}
			For $\kappa\tilde{r}\ll1$, we have
			\begin{subequations}
				\begin{align}
					I_{\mu}(\kappa\tilde{r})&\simeq\frac{1}{\Gamma(\mu+1)}\left(\frac{\kappa\tilde{r}}{2}\right)^{\mu},\\
					K_{\mu}(\kappa\tilde{r})&\simeq\begin{cases}
						-\ln\left(\frac{\kappa\tilde{r}}{2}\right)-\gamma, & \mu=0\\
						\frac{\Gamma(\mu)}{2}\left(\frac{2}{\kappa\tilde{r}}\right)^{\mu}, & \mu>0
					\end{cases}.
				\end{align}
			\end{subequations}
			
			For problems in spherical coordinates, modified spherical Bessel functions will instead be used.  The PDE we will be dealing with is of the form
			\begin{align}
				\begin{split}
					&\frac{1}{\tilde{r}^2}\frac{\partial}{\partial\tilde{r}}\left(\tilde{r}^2\frac{\partial F}{\partial\tilde{r}}\right)-\kappa^2F\\
					&+\frac{1}{\tilde{r}^2\sin\theta}\frac{\partial}{\partial\theta}\left(\sin\theta\frac{\partial F}{\partial\theta}\right)+\frac{1}{\tilde{r}^2\sin^2\theta}\frac{\partial^2F}{\partial\phi^2}=0,
				\end{split}
			\end{align}
			where $\kappa^2=2(m_x+m_y+m_z)$.  Defining $F(r,\theta,\phi)=G(r)Y_l^m(\theta,\phi)$, where $Y_l^m$ are spherical harmonics, we obtain the ODE
			\begin{equation}
				\frac{1}{\tilde{r}^2}\frac{\partial}{\partial\tilde{r}}\left(\tilde{r}^2\frac{\partial G}{\partial\tilde{r}}\right)-\left[\frac{l(l+1)}{\tilde{r}^2}+\kappa^2\right]G=0.
			\end{equation}
			The solutions to this ODE are the modified spherical Bessel functions of the first and second kinds $i_l(\kappa\tilde{r}),k_l(\kappa\tilde{r})$.  For our purposes, we only use the latter, the first few of which are
			\begin{subequations}
				\begin{align}
					k_0(\kappa\tilde{r})&=\frac{e^{-\kappa\tilde{r}}}{\kappa\tilde{r}},\\
					k_1(\kappa\tilde{r})&=\frac{e^{-\kappa\tilde{r}}(\kappa\tilde{r}+1)}{\kappa^2\tilde{r}^2},\\
					k_2(\kappa\tilde{r})&=\frac{e^{-\kappa\tilde{r}}(\kappa^2\tilde{r}^2+3\kappa\tilde{r}+3)}{\kappa^3\tilde{r}^3}.
				\end{align}
			\end{subequations}
			Two useful recursion relations are
			\begin{subequations}
				\begin{align}
					-\frac{2l+1}{\kappa\tilde{r}}k_l(\kappa\tilde{r})&=k_{l-1}(\kappa\tilde{r})-k_{l+1}(\kappa\tilde{r}),\\
					-\frac{2l+1}{\kappa}\frac{dk_l(\kappa\tilde{r})}{d\tilde{r}}&=lk_{l-1}(\kappa\tilde{r})+(l+1)k_{l+1}(\kappa\tilde{r}).
				\end{align}
			\end{subequations}

		\subsection{\label{subapp:KL M transforms}Kontorovich-Lebedev and Mellin transforms}
			In wedge-like geometries $0<r<\infty$ and $\theta_1<\theta<\theta_2$, we no longer have periodicity in $\theta$.  In addition, the density must remain finite as $r\rightarrow0$ or $r\rightarrow\infty$.  This requires the use of modified Bessel functions with purely imaginary order $K_{i\nu}$.  This gives rise to the \textbf{Kontorovich-Lebedev (KL) transforms}, which are often used for various problems in wedge-shaped geometries \cite{Kontorovich and Lebedev,Forristall Ingram,Kang et al,Fowkes et al,Smith}.  The pair of transforms is given by
			\begin{subequations}
				\begin{align}
					F(\nu,\theta)&=\int_{0}^{\infty}f(r,\theta)K_{i\nu}(\kappa\tilde{r})\frac{d\tilde{r}}{\tilde{r}},\label{eq:KL forward}\\
					f(\tilde{r},\theta)&=\frac{2}{\pi^2}\int_{-\infty}^{\infty}F(\nu,\theta)K_{i\nu}(\kappa\tilde{r})\nu\sinh(\pi\nu)d\nu.\label{eq:KL backward}
				\end{align}
			\end{subequations}
			A table of such transforms can be found in \cite{Oberhettinger KL}.  It is useful to note that $K_{i\nu}$ satisfies the same recursion relations as $K_{\mu}$ (Eqs.\ (\ref{eq:bessel identities})).  To use the transforms, we start by noting that $K_{i\nu}(\kappa\tilde{r})$ satisfies
			\begin{equation}
				\frac{1}{\tilde{r}}\frac{d}{d\tilde{r}}\left(\tilde{r}\frac{dK_{i\nu}}{d\tilde{r}}\right)-\left(\kappa^2-\frac{\nu^2}{\tilde{r}^2}\right)K_{i\nu}=0.
				\label{eq:K i nu ode}
			\end{equation}
			Thus, given a PDE of the form
			\begin{equation}
				\frac{1}{\tilde{r}}\frac{\partial}{\partial\tilde{r}}\left(\tilde{r}\frac{\partial C}{\partial\tilde{r}}\right)+\frac{1}{\tilde{r}^2}\frac{\partial^2C}{\partial\theta^2}-\kappa^2C=0,
			\end{equation}
			the KL transform reduces this PDE to the simple ODE
			\begin{equation}
				\frac{d^2\widehat{C}}{d\theta^2}=\nu^2\widehat{C},
			\end{equation}
			which has the general solution
			\begin{equation}
				\widehat{C}(\nu,\theta)=a(\nu)e^{\nu\theta}+b(\nu)e^{-\nu\theta}.
			\end{equation}
			The functions $a(\nu),b(\nu)$ can be determined by applying the KL transform to the boundary condition on $C(r,\theta)$.  A useful identity \cite{Kang et al} for doing so is
			\begin{equation}
				\int_{0}^{\infty}K_{i\nu}(\kappa\tilde{r})d\tilde{r}=\frac{\pi}{2\kappa\cosh\nu\pi/2},
				\label{eq:K inu integral identity}
			\end{equation}
			which can be derived from the integral definition
			\begin{equation}
				K_{i\nu}(\kappa\tilde{r})=\int_{0}^{\infty}e^{-\kappa\tilde{r}\cosh t}\cos(\nu t)dt.
			\end{equation}
			
			Another transform is the \textbf{Mellin transform}, which is applicable to the case of $\kappa=0$ in Eq.\ (\ref{eq:K i nu ode}).  The pair of transformations is
			\begin{subequations}
				\begin{align}
					F(z,\theta)&=\int_{0}^{\infty}f(\tilde{r},\theta)\tilde{r}^{z-1}d\tilde{r},\\
					f(\tilde{r},\theta)&=\int_{c-i\infty}^{c+i\infty}F(z,\theta)\tilde{r}^{-z}\frac{dz}{2\pi i},
				\end{align}
			\end{subequations}
			where $c$ is chosen such that there are no poles for $\re(z)>c$.  Since for our situation density must be finite, we can set $c=0$ for physical reasons; otherwise we will have divergences as $\tilde{r}\rightarrow0$.  A table of Mellin transforms can be found in \cite{Oberhettinger M}.
	
	\section{\label{app:expansion arbitrary dimension}Series solution in arbitrary dimensions}
		We show here the series solution in $d$ dimensions.  Just as before, we expand the density in powers of $\epsilon$ as
		\begin{equation}
			\tilde{\rho}(\tilde{\boldsymbol{r}},\tilde{\boldsymbol{\eta}})=\sum_{n=0}^{\infty}\epsilon^n\tilde{\rho}^{(n)}(\tilde{\boldsymbol{r}},\tilde{\boldsymbol{\eta}}).
		\end{equation}
		Substituting this into the dimensionless Fokker-Planck equation (Eq.\ (\ref{eq:dimensionless fokker-planck})), we arrive at
		\begin{equation}
			\nabla_r^2\tilde{\rho}^{(n)}+\nabla_{\eta}^2\tilde{\rho}^{(n)}+2\boldsymbol{\nabla}_{\eta}\cdot\big(\tilde{\boldsymbol{\eta}}\tilde{\rho}^{(n)}\big)=2\tilde{\boldsymbol{\eta}}\cdot\boldsymbol{\nabla}_r\tilde{\rho}^{(n-1)}.
			\label{eq:dimensionless fokker-planck order}
		\end{equation}
		The expansion in Hermite polynomials is the same as in 1D, except now we have a Hermite polynomial for each component of $\tilde{\boldsymbol{\eta}}$.  Thus, writing each order of the density as
		\begin{equation}
			\tilde{\rho}^{(n)}(\tilde{\boldsymbol{r}},\tilde{\boldsymbol{\eta}})=\sum_{\boldsymbol{m}}C_{\boldsymbol{m}}^{(n)}(\tilde{\boldsymbol{r}})e^{-\tilde{\boldsymbol{\eta}}^2}\prod_{i=1}^{d}H_{m_i}(\tilde{\eta}_i),
		\end{equation}
		we reduce the problem to solving for the coefficients $C_{\boldsymbol{m}}^{(n)}(\tilde{\boldsymbol{r}})=C_{m_1,\dots,m_d}^{(n)}(\tilde{\boldsymbol{r}})$, which satisfy a Helmholtz-type equation
		\begin{equation}
			\nabla_r^2C_{\boldsymbol{m}}^{(n)}-2\left(\sum_{i=1}^{d}m_i\right)C_{\boldsymbol{m}}^{(n)}=\boldsymbol{\nabla}\cdot\boldsymbol{w},
			\label{eq:coef equation d-dim}
		\end{equation}
		where the components of $\boldsymbol{w}$ are
		\begin{equation}
			w_{\alpha}=C_{\boldsymbol{m};m_{\alpha}-1}^{(n-1)}+2(m_{\alpha}+1)C_{\boldsymbol{m};m_{\alpha}+1}^{(n-1)}.
		\end{equation}
		Here, $C_{\boldsymbol{m};m_{\alpha}-1}^{(n-1)}$ denotes the coefficient $C_{m_1,\dots,m_{\alpha}-1,\dots,m_d}^{(n-1)}$.  The currents $\tilde{\boldsymbol{J}}_r=2\epsilon\tilde{\boldsymbol{\eta}}\tilde{\rho}-\boldsymbol{\nabla}_r\tilde{\rho}$ and $\tilde{\boldsymbol{J}}_{\eta}=-2\tilde{\boldsymbol{\eta}}\tilde{\rho}-\boldsymbol{\nabla}_{\eta}\tilde{\rho}$ are
		\begin{subequations}
			\begin{align}
				\tilde{\boldsymbol{J}}_r=\sum_{n=0}^{\infty}\epsilon^n&\sum_{\boldsymbol{m}}\left[\boldsymbol{w}-\boldsymbol{\nabla}_rC_{\boldsymbol{m}}^{(n)}\right]e^{-\tilde{\boldsymbol{\eta}}^2}\prod_{i=1}^{d}H_{m_i}(\tilde{\eta}_i),\\
				\begin{split}
					\tilde{J}_{\eta,\alpha}=\sum_{n=0}^{\infty}\epsilon^n&\sum_{\boldsymbol{m}}\left[-2(m_{\alpha}+1)C_{\boldsymbol{m};m_{\alpha}+1}^{(n)}\right]\\
					&\times e^{-\tilde{\boldsymbol{\eta}}^2}\prod_{i=1}^{d}H_{m_i}(\tilde{\eta}_i).
				\end{split}
			\end{align}
		\end{subequations}
		
		It is worth noting that for most of the problems we solve here, we only need to consider a few coefficients.  We briefly summarize the general procedure in 1D.  Passive particles are characterized by $C_0^{(0)}$.  Using this, we can determine the next nonzero coefficient $C_1^{(1)}$.  Continuing, we will have $C_0^{(2)},C_2^{(2)}$ followed by $C_1^{(3)},C_3^{(3)}$, and so on.  In other words, the nonzero coefficients $C_m^{(n)}$ for most of our problems will often alternate between even and odd $m$ as we go to higher and higher orders $n$.
		
		It is also worth noting that if we are interested in, for example, the density $\tilde{\rho}(\tilde{\boldsymbol{r}})$, then integrating out the active force $\tilde{\boldsymbol{\eta}}$ and using the orthogonality of Hermite polynomials will leave us with only the $\boldsymbol{m}=(0,0,\dots,0)$ terms or
		\begin{equation}
			\tilde{\rho}(\tilde{\boldsymbol{r}})=\pi^{d/2}\sum_{n=0}^{\infty}\epsilon^nC_{0,0,\dots,0}^{(n)}(\tilde{\boldsymbol{r}}).
		\end{equation}
	
	\section{\label{app:cartesian}Cartesian coordinates}
		To summarize for 1D, we expand the density as
		\begin{equation}
			\rho(\tilde{x},\tilde{\eta})=\sum_{n=0}^{\infty}\epsilon^n\sum_{m=0}^{\infty}C_m^{(n)}(\tilde{x})e^{-\tilde{\eta}^2}H_m(\tilde{\eta}),
		\end{equation}
		where the coefficients $C_m^{(n)}(\tilde{x})$ satisfy the ODE
		\begin{equation}
			\frac{d^2C_m^{(n)}}{d\tilde{x}^2}-2mC_m^{(n)}=\frac{d}{d\tilde{x}}\left[C_{m-1}^{(n-1)}+2(m+1)C_{m+1}^{(n-1)}\right].
		\end{equation}
		The current $\tilde{J}(\tilde{x},\tilde{\eta})$ along $x$ is
		\begin{align}
			\begin{split}
				\tilde{J}_x=\sum_{n=0}^{\infty}\epsilon^n&\sum_{m=0}^{\infty}\left[C_{m-1}^{(n-1)}+2(m+1)C_{m+1}^{(n-1)}-\frac{dC_m^{(n)}}{d\tilde{x}}\right]\\
				&\times e^{-\tilde{\eta}^2}H_m(\tilde{\eta}).
			\end{split}
		\end{align}
	
		\subsection{\label{subapp:semi-infinite}1D semi-infinite domain: one wall}
			By setting $\tilde{J}(\tilde{x},\tilde{\eta})=0$ and using the orthogonality of Hermite polynomials, we find that the zero current boundary condition for a single wall at $\tilde{x}=0$ gives the following condition on the coefficients
			\begin{equation}
				\frac{dC_m^{(n)}(0)}{d\tilde{x}}=C_{m-1}^{(n-1)}(0)+2(m+1)C_{m+1}^{(n-1)}(0).
			\end{equation}
			To start, we note that the $\underline{\boldsymbol{n=0}}$ order corresponds to a passively diffusing particle whose spacial density will be constant everywhere and the dimensionless active force $\tilde{\eta}$ will be Gaussian distributed in steady state.  Thus, we have
			\begin{equation}
				C_m^{(0)}(\tilde{x})=\mathcal{N}\delta_{m,0},
			\end{equation}
			where $\mathcal{N}=\rho_{\textrm{bulk}}\sqrt{2D_p\tau/\pi}$ is the normalization.  Using the zeroth order solution, we can compute the next order $\underline{\boldsymbol{n=1}}$ to get
			\begin{equation}
				C_m^{(1)}(\tilde{x})=-\frac{\mathcal{N}\sqrt{2}}{2}e^{-\sqrt{2}\tilde{x}}\delta_{m,1}.
			\end{equation}
			For $\underline{\boldsymbol{n=2}}$, we have
			\begin{equation}
				C_m^{(2)}(\tilde{x})=\mathcal{N}e^{-\sqrt{2}\tilde{x}}\delta_{m,0}+\frac{\mathcal{N}}{2}\left[\sqrt{2}e^{-2\tilde{x}}-e^{-\sqrt{2}\tilde{x}}\right]\delta_{m,2}.
			\end{equation}
			For the next two orders, we will only show the solutions necessary for obtaining the $\epsilon^4$ correction to density in Eq.\ (\ref{eq:1d density up to 4th}).  For $\underline{\boldsymbol{n=3}}$, we have
			\begin{equation}
				C_1^{(3)}(\tilde{x})=\mathcal{N}\left[\left(\frac{\sqrt{2}}{4}+1-\frac{\tilde{x}}{2}\right)e^{-\sqrt{2}\tilde{x}}-2\sqrt{2}e^{-2\tilde{x}}\right].
			\end{equation}
			At this order, the other nonzero solution is for $m=3$.  For $\underline{\boldsymbol{n=4}}$, we have
			\begin{equation}
				C_0^{(4)}(\tilde{x})=2\mathcal{N}\sqrt{2}\left[\left(\frac{\tilde{x}}{4}-1\right)e^{-\sqrt{2}\tilde{x}}+e^{-2\tilde{x}}\right].
			\end{equation}
			At this order, the other nonzero solutions are for $m=2,4$.
			
		\subsection{\label{subapp:finite domain}1D finite domain: two walls}
			The approach here is the similar to Section \ref{subapp:semi-infinite}.  The only difference is we now have zero current boundary conditions at the two walls at $\tilde{x}=\pm\tilde{L}$ or
			\begin{equation}
				\frac{dC_m^{(n)}(\pm\tilde{L})}{d\tilde{x}}=C_{m-1}^{(n-1)}(\pm\tilde{L})+2(m+1)C_{m+1}^{(n-1)}(\pm\tilde{L}),
			\end{equation}
			and a finite number of particles $N$ between the walls.  For $\underline{\boldsymbol{n=0}}$ we have
			\begin{equation}
				C_m^{(0)}(\tilde{x})=\mathcal{N}\delta_{m,0},
			\end{equation}
			where $\mathcal{N}=\frac{N}{2L}\sqrt{\frac{2D_p\tau}{\pi}}$.  For $\underline{\boldsymbol{n=1}}$, we get
			\begin{equation}
				C_m^{(1)}(\tilde{x})=\frac{\mathcal{N}\sqrt{2}\sinh\sqrt{2}\tilde{x}}{2\cosh\sqrt{2}\tilde{L}}\delta_{m,1}.
			\end{equation}
			For $\underline{\boldsymbol{n=2}}$ and beyond, we have to enforce that the number of particles between the two walls does not change as we go to higher orders.  This condition is
			\begin{equation}
				\int_{-\tilde{L}}^{\tilde{L}}C_0^{(n)}(\tilde{x})d\tilde{x}=0,
			\end{equation}
			for $n>0$.  Thus, we have for $m=0$
			\begin{equation}
				C_0^{(2)}(\tilde{x})=\mathcal{N}\left(\frac{\cosh\sqrt{2}\tilde{x}}{\cosh\sqrt{2}\tilde{L}}-\frac{\tanh\sqrt{2}\tilde{L}}{\sqrt{2}\tilde{L}}\right).
			\end{equation}
			The other nonzero solution is for $m=2$.
	
		\subsection{\label{subapp:pressure}Ramp potentials and pressure}
			\begin{figure*}
				\centering
				\includegraphics[scale=0.4]{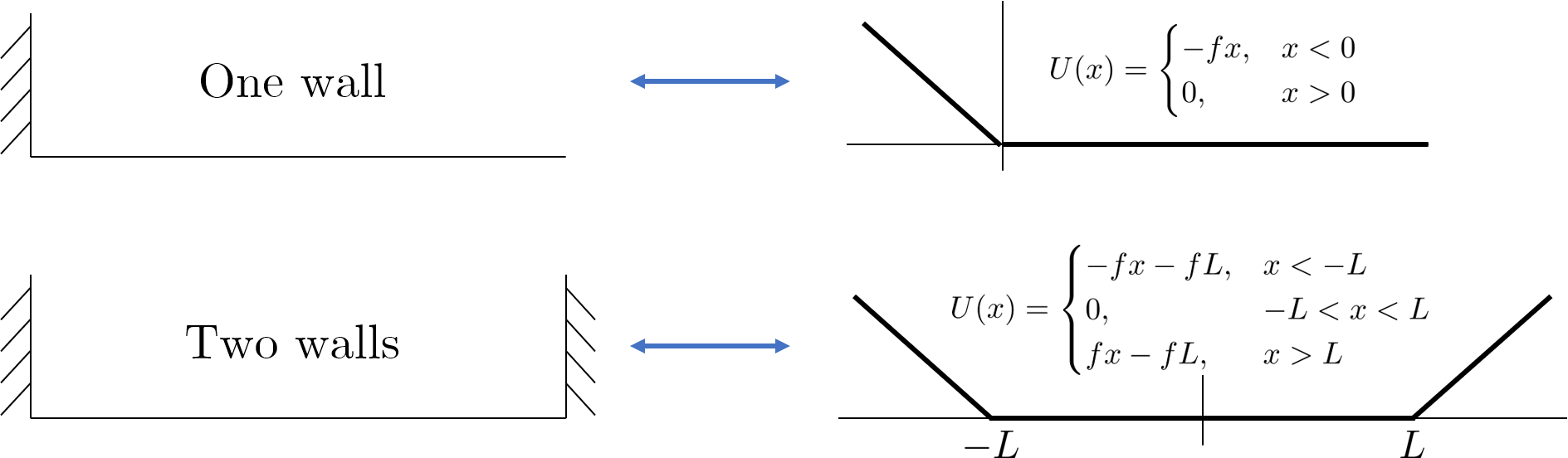}
				\caption{Representation of solid boundaries as ramp potentials}
				\label{fig:ramp potentials}
			\end{figure*}
		
			To compute the pressure on solid walls, we start by representing the boundaries as soft confining potentials (Figure \ref{fig:ramp potentials}) and then taking the limit as those potentials become steep.  We will focus on the case of two walls at $x=\pm L$ since the case of one wall can be obtained from the limit $L\rightarrow\infty$.  Consider the confining potential
			\begin{equation}
				U(x)=\begin{cases}
					-f(x+L), & x<-L\\
					0, & -L<x<L\\
					f(x-L), & x>L
				\end{cases}
			\end{equation}
			We split the density into three pieces $\tilde{\rho}_A$, $\tilde{\rho}_B$, and $\tilde{\rho}_C$ for the regions $\tilde{x}<-\tilde{L}$, $-\tilde{L}<\tilde{x}<\tilde{L}$, and $\tilde{L}<\tilde{x}$, respectively.  We write the densities in the form
			\begin{subequations}
				\begin{align}
					\tilde{\rho}_A(\tilde{x},\tilde{\eta})&=\sum_{n=0}^{\infty}\epsilon^n\sum_{m=0}^{\infty}A_m^{(n)}(\tilde{x})e^{-\tilde{\eta}^2}H_m(\tilde{\eta}),\\
					\tilde{\rho}_B(\tilde{x},\tilde{\eta})&=\sum_{n=0}^{\infty}\epsilon^n\sum_{m=0}^{\infty}B_m^{(n)}(\tilde{x})e^{-\tilde{\eta}^2}H_m(\tilde{\eta}),\\
					\tilde{\rho}_C(\tilde{x},\tilde{\eta})&=\sum_{n=0}^{\infty}\epsilon^n\sum_{m=0}^{\infty}C_m^{(n)}(\tilde{x})e^{-\tilde{\eta}^2}H_m(\tilde{\eta}).
				\end{align}
			\end{subequations}
			The coefficient equations corresponding to each region are
			\begin{subequations}
				\begin{align}
					\frac{d^2A_m^{(n)}}{d\tilde{x}^2}-\tilde{f}\frac{dA_m^{(n)}}{d\tilde{x}}-2mA_m^{(n)}&=\frac{dw_A}{d\tilde{x}},\\
					\frac{d^2B_m^{(n)}}{d\tilde{x}^2}-2mB_m^{(n)}&=\frac{dw_B}{d\tilde{x}},\\
					\frac{d^2C_m^{(n)}}{d\tilde{x}^2}+\tilde{f}\frac{dC_m^{(n)}}{d\tilde{x}}-2mC_m^{(n)}&=\frac{dw_C}{d\tilde{x}},
				\end{align}
			\end{subequations}
			where $\tilde{f}=\frac{2f\tau}{\gamma\sqrt{2D_p\tau}}$ and
			\begin{subequations}
				\begin{align}
					w_A&=A_{m-1}^{(n-1)}+(m+1)A_{m+1}^{(n-1)},\\
					w_B&=B_{m-1}^{(n-1)}+(m+1)B_{m+1}^{(n-1)},\\
					w_C&=C_{m-1}^{(n-1)}+(m+1)C_{m+1}^{(n-1)}.
				\end{align}
			\end{subequations}
			The boundary conditions (continuity in density and current) at $\tilde{x}=\pm\tilde{L}$ are
			\begin{subequations}
				\begin{align}
					A_m^{(n)}(-\tilde{L})&=B_m^{(n)}(-\tilde{L}),\\
					\tilde{f}A_m^{(n)}(-\tilde{L})-\frac{dA_m^{(n)}(-\tilde{L})}{d\tilde{x}}&=-\frac{dB_m^{(n)}(-\tilde{L})}{d\tilde{x}},\\
					B_m^{(n)}(\tilde{L})&=C_m^{(n)}(\tilde{L}),\\
					-\frac{dB_m^{(n)}(\tilde{L})}{d\tilde{x}}&=-\tilde{f}C_m^{(n)}(\tilde{L})-\frac{dC_m^{(n)}(\tilde{L})}{d\tilde{x}}.
				\end{align}
			\end{subequations}
			Just as before, $\underline{\boldsymbol{n=0}}$ corresponds to a passive particle diffusing within the confining potential and the coefficients are
			\begin{subequations}
				\begin{align}
					A_m^{(0)}(\tilde{x})&=\mathcal{N}e^{\tilde{f}(\tilde{x}+\tilde{L})}\delta_{m,0},\\
					B_m^{(0)}(\tilde{x})&=\mathcal{N}\delta_{m,0},\\
					C_m^{(0)}(\tilde{x})&=\mathcal{N}e^{-\tilde{f}(\tilde{x}-\tilde{L})}\delta_{m,0},
				\end{align}
			\end{subequations}
			which is related to the Boltzmann distribution.  For $\underline{\boldsymbol{n=1}}$, we get
			\begin{subequations}
				\begin{align}
					A_m^{(1)}(\tilde{x})&=\left[a_1^{(1)}e^{\kappa_{1,+}(\tilde{x}+\tilde{L})}-\frac{\mathcal{N}\tilde{f}}{2}e^{\tilde{f}(\tilde{x}+\tilde{L})}\right]\delta_{m,1},\\
					B_m^{(1)}(\tilde{x})&=b_1^{(1)}\sinh\sqrt{2}\tilde{x}\delta_{m,1},\\
					C_m^{(1)}(\tilde{x})&=\left[c_1^{(1)}e^{-\kappa_{1,+}(\tilde{x}-\tilde{L})}+\frac{\mathcal{N}\tilde{f}}{2}e^{-\tilde{f}(\tilde{x}-\tilde{L})}\right]\delta_{m,1},
				\end{align}
			\end{subequations}
			where
			\begin{subequations}
				\begin{align}
					a_1^{(1)}&=-c_1^{(1)}=\frac{\mathcal{N}\tilde{f}\sqrt{2}}{2(\sqrt{2}-\kappa_{1,-}\tanh\sqrt{2}\tilde{L})},\\
					b_1^{(1)}&=-\frac{\mathcal{N}\tilde{f}\kappa_{1,-}}{2(\sqrt{2}\cosh\sqrt{2}\tilde{L}-\kappa_{1,-}\sinh\sqrt{2}\tilde{L})},
				\end{align}
			\end{subequations}
			and
			\begin{equation}
				\kappa_{1,\pm}=\frac{1}{2}\left(\tilde{f}\pm\sqrt{\tilde{f}^2+8m}\right).
			\end{equation}
			For $\underline{\boldsymbol{n=2}}$, we will focus on the $m=0$ contribution since we are only interested in the pressure, which only requires knowing the density.  We have the general solutions
			\begin{subequations}
				\begin{align}
					\begin{split}
						A_0^{(2)}(\tilde{x})=&\ a_0^{(2)}e^{\tilde{f}(\tilde{x}+\tilde{L})}-\mathcal{N}\tilde{f}(\tilde{x}+\tilde{L})e^{\tilde{f}(\tilde{x}+\tilde{L})}\\
						&\ +\kappa_{1,+}a_1^{(1)}e^{\kappa_{1,+}(\tilde{x}+\tilde{L})},
					\end{split}
					\\
					B_0^{(2)}(\tilde{x})=&\ b_0^{(2)}+b_1^{(1)}\sqrt{2}\cosh\sqrt{2}\tilde{x},\\
					\begin{split}
						C_0^{(2)}(\tilde{x})=&\ c_0^{(2)}e^{-\tilde{f}(\tilde{x}-\tilde{L})}+\mathcal{N}\tilde{f}(\tilde{x}-\tilde{L})e^{-\tilde{f}(\tilde{x}-\tilde{L})}\\
						&\ -\kappa_{1,+}c_1^{(1)}e^{-\kappa_{1,+}(\tilde{x}-\tilde{L})},
					\end{split}
				\end{align}
			\end{subequations}
			where we wish to determine $a_0^{(2)},b_0^{(2)},c_0^{(2)}$.  In addition to continuity in density and current, we require that all higher orders $n>0$ do not change the number of particles between the walls.  This condition is
			\begin{equation}
				\int_{-\infty}^{-\tilde{L}}A_0^{(2)}(\tilde{x})d\tilde{x}+\int_{-\tilde{L}}^{\tilde{L}}B_0^{(2)}(\tilde{x})d\tilde{x}+\int_{\tilde{L}}^{\infty}C_0^{(2)}(\tilde{x})d\tilde{x}=0.
			\end{equation}
			Thus, we find
			\begin{subequations}
				\begin{align}
					\begin{split}
						a_0^{(2)}&=c_0^{(2)}\\
						&=-\frac{\mathcal{N}}{2(1+\tilde{f}\tilde{L})}\left[2+\tilde{f}^2+\frac{\sqrt{2}\tilde{f}^3\tilde{L}}{\sqrt{2}-\kappa_{1,-}\tanh\sqrt{2}\tilde{L}}\right],
					\end{split}
					\\
					b_0^{(2)}&=-\frac{\mathcal{N}}{1+\tilde{f}\tilde{L}}\left[1-\frac{\tilde{f}^2\kappa_{1,-}\tanh\sqrt{2}\tilde{L}}{2(\sqrt{2}-\kappa_{1,-}\tanh\sqrt{2}\tilde{L})}\right].
				\end{align}
			\end{subequations}
			Knowing the density, we can then compute the pressure, say for the wall at $x=L$, to obtain
			\begin{widetext}
				\begin{equation}
					P=\int_{L}^{\infty}f\rho(x)dx\simeq\frac{Nf}{2(1+\frac{fL}{k_BT})}\left[1+\frac{\epsilon^2}{1+\frac{fL}{k_BT}}\left(\frac{fL}{k_BT}-\frac{\frac{2f^2\tau}{k_BT\gamma}\tanh\frac{\sqrt{2}L}{\lambda}}{\sqrt{\frac{f^2\tau}{k_BT\gamma}}+\sqrt{\frac{f^2\tau}{k_BT\gamma}+4}+2\tanh\frac{\sqrt{2}L}{\lambda}}\right)\right],
				\end{equation}
			\end{widetext}
			where we used the Einstein relation $D_p\gamma=k_BT$.  Note that for a finite sized system, the pressure depends on the confining potential (in this case, on $f$).  This dependence disappears if we take $L\rightarrow\infty$, that is, when we have a bulk where active particles are unaffected by the boundary.  In the limit of a solid boundary ($f\rightarrow\infty$), this pressure becomes
			\begin{equation}
				P\underset{f\rightarrow\infty}{\simeq}\frac{N}{2L}k_BT\left[1+\epsilon^2\left(1-\frac{\tanh\sqrt{2}\tilde{L}}{\sqrt{2}\tilde{L}}\right)\right].
			\end{equation}
	
		\subsection{\label{subapp:exactly solvable}Exactly solvable model of 1D run-and-tumble with passive diffusion}
			In the main text, we found that the pressure of active particles on a solid boundary is approximately given by $P\simeq\rho_{\textrm{wall}}k_BT$.  We show here, using an exactly solvable model of run-and-tumble particles, that this relation for pressure is exact.  The steady-state Fokker-Planck equations for 1D run-and-tumble particles with passive diffusion are
			\begin{subequations}
				\begin{align}
					0&=\frac{d}{dx}\left[\left(-v+\frac{U'}{\gamma}\right)\rho_+\right]+D_p\frac{d^2\rho_+}{dx^2}-\alpha\rho_++\alpha\rho_-,\label{eq:rnt + FP}\\
					0&=\frac{d}{dx}\left[\left(v+\frac{U'}{\gamma}\right)\rho_-\right]+D_p\frac{d^2\rho_-}{dx^2}+\alpha\rho_+-\alpha\rho_-,\label{eq:rnt - FP}
				\end{align}
			\end{subequations}
			where $\rho_+,\rho_-$ are the densities of left- and right-moving particles, $v$ is the swim speed, $\alpha$ is the tumble rate, and $U$ is a confining potential that is zero in a region $-L<x<L$ between the walls.  To determine the pressure, we manipulate Eqs.\ (\ref{eq:rnt + FP}) and (\ref{eq:rnt - FP}) a bit.  We start by adding the two equations to get
			\begin{equation}
				0=\frac{d}{dx}\left[-\bar{v}(x)\rho+\frac{U'}{\gamma}\rho\right]+D_p\frac{d^2\rho}{dx^2},
			\end{equation}
			where $\bar{v}\rho=v(\rho_+-\rho_-)$ and $\rho=\rho_++\rho_-$.  In 1D, the current in $x$ must be zero, and so we have the relation
			\begin{equation}
				\bar{v}\rho=\frac{U'}{\gamma}\rho+D_p\frac{d\rho}{dx}.
				\label{eq:rnt vbar rho}
			\end{equation}	
			Multiplying Eqs.\ (\ref{eq:rnt + FP}) and (\ref{eq:rnt - FP}) by $-v$ and subtracting, we have
			\begin{equation}
				0=\frac{d}{dx}\left(v^2\rho-\frac{U'}{\gamma}\bar{v}\rho\right)-D_p\frac{d^2}{dx^2}\left(\bar{v}\rho\right)+2\alpha\bar{v}\rho.
			\end{equation}
			Substituting in the relation (Eq.\ \ref{eq:rnt vbar rho}) for $\bar{v}\rho$, we have
			\begin{align}
				\begin{split}
					0=&\ \frac{d}{dx}\left[v^2\rho-\left(\frac{U'}{\gamma}\right)^2\rho-\frac{U'}{\gamma}D_p\frac{d\rho}{dx}\right]\\
					&-D_p\frac{d^2}{dx^2}\left(\frac{U'}{\gamma}\rho+D_p\frac{d\rho}{dx}\right)+2\alpha\left(\frac{U'}{\gamma}\rho+D_p\frac{d\rho}{dx}\right).
				\end{split}
			\end{align}
			Finally, integrating from deep inside a wall to a point $-L<x_0<L$ between the walls, we find the mechanical pressure on the left wall
			\begin{align}
				\begin{split}
					P&=-\int_{-\infty}^{x_0}U'\rho dx\\
					&=\left(D_p+\frac{v^2}{2\alpha}\right)\gamma\rho(x_0)-\frac{D_p^2\gamma}{2\alpha}\frac{d^2\rho(x_0)}{dx^2}.
				\end{split}
			\end{align}
			This is valid in the limit of hard walls since it is independent of $U(x)$, and so what remains is determining $\rho(x)$ between two hard walls at $x=\pm L$ with $U=0$.  The zero current boundary conditions are
			\begin{subequations}
				\begin{align}
					v\rho_+(\pm L)-D_p\frac{d\rho_+(\pm L)}{dx}&=0\\
					-v\rho_-(\pm L)-D_p\frac{d\rho_-(\pm L)}{dx}&=0
				\end{align}
			\end{subequations}
			Defining $q_{\pm}=\frac{\partial\rho_{\pm}}{\partial x}$, Eqs.\ (\ref{eq:rnt + FP}) and (\ref{eq:rnt - FP}) can be rewritten as
			\begin{equation}
				\frac{d}{dx}\pmat{\rho_+\\\rho_-\\q_+\\q_-}{1.2}=\bmat{0 & 0 & 1 & 0\\0 & 0 & 0 & 1\\\frac{\alpha}{D_p} & -\frac{\alpha}{D_p} & \frac{v}{D_p} & 0\\-\frac{\alpha}{D_p} & \frac{\alpha}{D_p} & 0 & -\frac{v}{D_p}}{1.2}\pmat{\rho_+\\\rho_-\\q_+\\q_-}{1.2}.
			\end{equation}
			The eigenvalues of the matrix are $0,0,\pm\frac{1}{D_p}\sqrt{v^2+2\alpha D_p}$.  The general solutions obeying the symmetry $\rho_+(x)=\rho_-(-x)$ are therefore
			\begin{subequations}
				\begin{align}
					\rho_+(x)&=c_0+c_1e^{\frac{x}{\lambda}}+c_2e^{-\frac{x}{\lambda}},\\
					\rho_-(x)&=c_0+c_2e^{\frac{x}{\lambda}}+c_1e^{-\frac{x}{\lambda}},
				\end{align}
			\end{subequations}
			where $\lambda=\frac{D_p}{\sqrt{v^2+2\alpha D_p}}$.  Applying the zero current boundary conditions at $x=\pm L$ and fixing the number of particles as $N$, we find
			\begin{align}
				\begin{split}
					\rho_{\pm}(x)=&\ \frac{N}{2L+\frac{v^2\lambda}{\alpha D_p}\tanh\frac{L}{\lambda}}\\
					&\times\left[\frac{1}{2}+\frac{v^2\cosh\frac{x}{\lambda}}{4\alpha D_p\cosh\frac{L}{\lambda}}\pm\frac{v\sinh\frac{x}{\lambda}}{4\alpha\lambda\cosh\frac{L}{\lambda}}\right].
				\end{split}
			\end{align}
			The density is thus
			\begin{align}
				\begin{split}
					\rho(x)&=\rho_+(x)+\rho_-(x)\\
					&=\frac{N}{2L+\frac{v^2\lambda}{\alpha D_p}\tanh\frac{L}{\lambda}}\left(1+\frac{v^2\cosh\frac{x}{\lambda}}{2\alpha D_p\cosh\frac{L}{\lambda}}\right),
				\end{split}
			\end{align}
			which gives us the pressure
			\begin{align}
				\begin{split}
					P&=\frac{ND_p\gamma}{2L+\frac{v^2\lambda}{\alpha D_p}\tanh\frac{L}{\lambda}}\left(1+\frac{v^2}{2\alpha D_p}\right)\\
					&=\rho(\pm L)D_p\gamma=\rho_{\textrm{wall}}D_p\gamma
				\end{split}
			\end{align}
			Using Einstein's relation $D_p\gamma=k_BT$, we have $P=\rho_{\textrm{wall}}k_BT$.
		
		\subsection{\label{subapp:2D corner}2D right-angled corner}
			In this case, we write the density as
			\begin{equation}
				\rho(\tilde{\boldsymbol{r}},\tilde{\boldsymbol{\eta}})=\sum_{n=0}^{\infty}\epsilon^n\sum_{\boldsymbol{m}}C_{\boldsymbol{m}}^{(n)}(\tilde{\boldsymbol{r}})e^{-\tilde{\boldsymbol{\eta}}^2}H_{m_x}(\tilde{\eta}_x)H_{m_y}(\tilde{\eta}_y).
			\end{equation}
			The coefficients satisfy
			\begin{equation}
				\frac{\partial^2C_{\boldsymbol{m}}^{(n)}}{\partial\tilde{x}^2}+\frac{\partial^2C_{\boldsymbol{m}}^{(n)}}{\partial\tilde{y}^2}-2(m_x+m_y)C_{\boldsymbol{m}}^{(n)}=\frac{\partial w_x}{\partial\tilde{x}}+\frac{\partial w_y}{\partial\tilde{y}},
			\end{equation}
			where
			\begin{subequations}
				\begin{align}
					w_x&=C_{m_x-1,m_y}^{(n-1)}+2(m_x+1)C_{m_x+1,m_y}^{(n-1)},\\
					w_y&=C_{m_x,m_y-1}^{(n-1)}+2(m_y+1)C_{m_x,m_y+1}^{(n-1)}.
				\end{align}
			\end{subequations}
			The zero current boundary conditions for each wall are
			\begin{subequations}
				\begin{align}
					\frac{\partial C_{\boldsymbol{m}}^{(n)}(0,\tilde{y})}{\partial\tilde{x}}=w_x(0,\tilde{y}),
					\\
					\frac{\partial C_{\boldsymbol{m}}^{(n)}(\tilde{x},0)}{\partial\tilde{y}}=w_y(\tilde{x},0).
				\end{align}
			\end{subequations}
			Note that part of the solution will be the sum of distributions of each wall if it were by itself since the the coefficient equation and boundary conditions are linear.  This observation will help us get started.  There are however additional terms due to the walls meeting near the origin, which we highlighted in Section \ref{subsec:corner}.  For $\underline{\boldsymbol{n=0}}$, we have the usual constant density
			\begin{equation}
				C_{m_x,m_y}^{(0)}(\tilde{x},\tilde{y})=\mathcal{N}\delta_{m_x,0}\delta_{m_y,0},
			\end{equation}
			where for 2D the normalization is $\mathcal{N}=\rho_{\textrm{bulk}}\frac{2D_p\tau}{\pi}$.  For $\underline{\boldsymbol{n=1}}$,
			\begin{align}
				\begin{split}
					C_{m_x,m_y}^{(1)}(\tilde{x},\tilde{y})=&-\frac{\mathcal{N}\sqrt{2}}{2}e^{-\sqrt{2}\tilde{x}}\delta_{m_x,1}\delta_{m_y,0}\\
					&-\frac{\mathcal{N}\sqrt{2}}{2}e^{-\sqrt{2}\tilde{y}}\delta_{m_x,0}\delta_{m_y,1}.
				\end{split}
			\end{align}
			For $\underline{\boldsymbol{n=2}}$,
			\begin{align}
				\begin{split}
					C_{m_x,m_y}^{(2)}(\tilde{x},\tilde{y})=&\ \mathcal{N}\left(e^{-\sqrt{2}\tilde{x}}+e^{-\sqrt{2}\tilde{y}}\right)\delta_{m_x,0}\delta_{m_y,0}\\
					&+\frac{\mathcal{N}}{2}\left(\sqrt{2}e^{-2\tilde{x}}-e^{-\sqrt{2}\tilde{x}}\right)\delta_{m_x,2}\delta_{m_y,0}\\
					&+\frac{\mathcal{N}}{2}\left(\sqrt{2}e^{-2\tilde{y}}-e^{-\sqrt{2}\tilde{y}}\right)\delta_{m_x,0}\delta_{m_y,2}\\
					&+\frac{\mathcal{N}}{2}e^{-\sqrt{2}\tilde{x}}e^{-\sqrt{2}\tilde{y}}\delta_{m_x,1}\delta_{m_y,1}.
				\end{split}
			\end{align}
			Note that the $(m_x,m_y)=(1,1)$ term does not result from the sum of solutions for the individual walls.  For the next two orders, we only show the terms necessary for obtaining the density Eq.\ (\ref{eq:corner density up to 4th}).  For $\underline{\boldsymbol{n=3}}$, we have
			\begin{align}
				\begin{split}
					C_{1,0}^{(3)}(\tilde{x},\tilde{y})=&\ \mathcal{N}\left[\left(\frac{\sqrt{2}}{4}+2-\frac{\tilde{x}}{2}\right)e^{-\sqrt{2}\tilde{x}}-2\sqrt{2}e^{-2\tilde{x}}\right]\\
					&-\frac{\mathcal{N}\sqrt{2}}{2}e^{-\sqrt{2}\tilde{x}}e^{-\sqrt{2}\tilde{y}},
				\end{split}
				\\
				\begin{split}
					C_{0,1}^{(3)}(\tilde{x},\tilde{y})=&\ \mathcal{N}\left[\left(\frac{\sqrt{2}}{4}+2-\frac{\tilde{y}}{2}\right)e^{-\sqrt{2}\tilde{y}}-2\sqrt{2}e^{-2\tilde{y}}\right]\\
					&-\frac{\mathcal{N}\sqrt{2}}{2}e^{-\sqrt{2}\tilde{x}}e^{-\sqrt{2}\tilde{y}}.
				\end{split}
			\end{align}
			Finally, for $\underline{\boldsymbol{n=4}}$, 
			\begin{align}
				\begin{split}
					C_{0,0}^{(4)}(\tilde{x},\tilde{y})=&\ 2\mathcal{N}\sqrt{2}\left[e^{-2\tilde{x}}+\left(\frac{\tilde{x}}{4}-1\right)e^{-\sqrt{2}\tilde{x}}\right]\\
					&+2\mathcal{N}\sqrt{2}\left[e^{-2\tilde{y}}+\left(\frac{\tilde{y}}{4}-1\right)e^{-\sqrt{2}\tilde{y}}\right]\\
					&+\mathcal{N}e^{-\sqrt{2}\tilde{x}}e^{-\sqrt{2}\tilde{y}}.
				\end{split}
			\end{align}
	
	\section{\label{app:polar}Polar coordinates}
		For problems that require polar coordinates, we write the distribution of active particles as
		\begin{align}
			\tilde{\rho}(\tilde{r},\theta,\tilde{\boldsymbol{\eta}})=\sum_{n=0}^{\infty}\epsilon^n\sum_{\boldsymbol{m}}C_{\boldsymbol{m}}^{(n)}(\tilde{r},\theta)e^{-\tilde{\boldsymbol{\eta}}^2}H_{m_x}(\tilde{\eta}_x)H_{m_y}(\tilde{\eta}_y).
		\end{align}
		The coefficient equation we want to solve in those cases is of the form
		\begin{align}
			\begin{split}
				&\frac{1}{\tilde{r}}\frac{\partial}{\partial\tilde{r}}\left(\tilde{r}\frac{\partial C_{\boldsymbol{m}}^{(n)}}{\partial\tilde{r}}\right)+\frac{1}{\tilde{r}^2}\frac{\partial^2C_{\boldsymbol{m}}^{(n)}}{\partial\theta^2}-2(m_x+m_y)C_{\boldsymbol{m}}^{(n)}\\
				&=\frac{1}{\tilde{r}}\frac{\partial}{\partial\tilde{r}}\left(\tilde{r}w_r\right)+\frac{1}{\tilde{r}}\frac{\partial w_{\theta}}{\partial\theta},
			\end{split}
		\end{align}
		where
		\begin{subequations}
			\begin{align}
				\begin{split}
					w_r=&\ \left[C_{m_x-1,m_y}^{(n-1)}+2(m_x+1)C_{m_x+1,m_y}^{(n-1)}\right]\cos\theta\\
					&+\left[C_{m_x,m_y-1}^{(n-1)}+2(m_y+1)C_{m_x,m_y+1}^{(n-1)}\right]\sin\theta,
				\end{split}
				\\
				\begin{split}
					w_{\theta}=&\, -\left[C_{m_x-1,m_y}^{(n-1)}+2(m_x+1)C_{m_x+1,m_y}^{(n-1)}\right]\sin\theta\\
					&+\left[C_{m_x,m_y-1}^{(n-1)}+2(m_y+1)C_{m_x,m_y+1}^{(n-1)}\right]\cos\theta.
				\end{split}
			\end{align}
		\end{subequations}
		The radial and tangential currents are given by
		\begin{subequations}
			\begin{align}
				\begin{split}
					\tilde{J}_r(\tilde{r},\theta,\tilde{\boldsymbol{\eta}})=\sum_{n=0}^{\infty}\epsilon^n&\sum_{\boldsymbol{m}}\left[w_r-\frac{\partial C_{\boldsymbol{m}}^{(n)}}{\partial\tilde{r}}\right]\\
					&\times e^{-\tilde{\boldsymbol{\eta}}^2}H_{m_x}(\tilde{\eta}_x)H_{m_y}(\tilde{\eta}_y),
				\end{split}
				\\
				\begin{split}
					\tilde{J}_{\theta}(\tilde{r},\theta,\tilde{\boldsymbol{\eta}})=\sum_{n=0}^{\infty}\epsilon^n&\sum_{\boldsymbol{m}}\left[w_{\theta}-\frac{1}{\tilde{r}}\frac{\partial C_{\boldsymbol{m}}^{(n)}}{\partial\theta}\right]\\
					&\times e^{-\tilde{\boldsymbol{\eta}}^2}H_{m_x}(\tilde{\eta}_x)H_{m_y}(\tilde{\eta}_y).
				\end{split}
			\end{align}
		\end{subequations}
	
		\subsection{\label{subapp:circular}Solution for a circular boundary}
			The zero current boundary condition for a solid circular boundary with radius $R$ is
			\begin{equation}
				\frac{\partial C_{m_x,m_y}^{(n)}(\tilde{R},\theta)}{d\tilde{r}}=w_r(\tilde{R},\theta).
			\end{equation}
			For $\underline{\boldsymbol{n=0}}$, we have
			\begin{equation}
				C_{m_x,m_y}^{(0)}(\tilde{r},\theta)=\mathcal{N}\delta_{m_x,0}\delta_{m_y,0},
			\end{equation}
			both inside and outside the circular boundary.  Inside the circular boundary, the normalization is $\mathcal{N}=\frac{N}{\pi R^2}\cdot\frac{2D_p\tau}{\pi}$, where $N$ is the number of particles.  Outside, the normalization is $\mathcal{N}=\rho_{\textrm{bulk}}\frac{2D_p\tau}{\pi}$.  For $\underline{\boldsymbol{n=1}}$, we have for $\tilde{r}<\tilde{R}$
			\begin{align}
				\begin{split}
					C_{m_x,m_y}^{(1)}(\tilde{r},\theta)=&\ \frac{\mathcal{N}\sqrt{2}I_1(\sqrt{2}\tilde{r})}{I_0(\sqrt{2}\tilde{R})+I_2(\sqrt{2}\tilde{R})}\\
					&\times(\cos\theta\delta_{m_x,1}\delta_{m_y,0}+\sin\theta\delta_{m_x,0}\delta_{m_y,1}),
				\end{split}
			\end{align}
			and for $\tilde{r}>\tilde{R}$,
			\begin{align}
				\begin{split}
					C_{m_x,m_y}^{(1)}(\tilde{r},\theta)=&\, -\frac{\mathcal{N}\sqrt{2}K_1(\sqrt{2}\tilde{r})}{K_0(\sqrt{2}\tilde{R}+K_2(\sqrt{2}\tilde{R}))}\\
					&\times(\cos\theta\delta_{m_x,1}\delta_{m_y,0}+\sin\theta\delta_{m_x,0}\delta_{m_y,1}),
				\end{split}
			\end{align}
			where $I_{\mu},K_{\mu}$ are modified Bessel functions of the first and second kinds, respectively.  For $\underline{\boldsymbol{n=2}}$, we focus on $(m_x,m_y)=(0,0)$ since we are only interested in the density.  For $\tilde{r}<\tilde{R}$, we must make sure that the number of particles remains fixed.  This condition is
			\begin{equation}
				\int_{0}^{R}\int_{0}^{2\pi}C_{0,0}^{(2)}(\tilde{r},\theta)\tilde{r}d\tilde{r}d\theta=0.
			\end{equation}
			Thus, we have
			\begin{equation}
				C_{0,0}^{(2)}(\tilde{r},\theta)=\frac{2\mathcal{N}\left[I_2(\sqrt{2}\tilde{R})-I_0(\sqrt{2}\tilde{R})+I_0(\sqrt{2}\tilde{r})\right]}{I_0(\sqrt{2}\tilde{R})+I_2(\sqrt{2}\tilde{R})}.
			\end{equation}
			For $\tilde{r}>\tilde{R}$,
			\begin{equation}
				C_{0,0}^{(2)}(\tilde{r},\theta)=\frac{2\mathcal{N}K_0(\sqrt{2}\tilde{r})}{K_0(\sqrt{2}\tilde{R})+K_2(\sqrt{2}\tilde{R})}.
			\end{equation}
	
		\subsection{\label{subapp:wedge}Solution for a wedge-shaped region}
			The zero current boundary condition for each wall of the wedge is $\tilde{J}_{\theta}(\tilde{r},\pm\alpha,\tilde{\boldsymbol{\eta}})=0$.  In terms of the coefficients, we have the condition
			\begin{equation}
				\frac{1}{\tilde{r}}\frac{\partial C_{m_x,m_y}^{(n)}(\tilde{r},\pm\alpha)}{\partial\theta}=w_{\theta}(\tilde{r},\pm\alpha).
			\end{equation}
			For $\underline{\boldsymbol{n=0}}$, we should have a uniform density of passive particles or
			\begin{equation}
				C_{m_x,m_y}^{(0)}(\tilde{r},\theta)=\mathcal{N}\delta_{m_x,0}\delta_{m_y,0}.
			\end{equation}
			For higher orders, we have to make use of the Kontorovich-Lebedev (KL) transform.  The KL transform of our coefficients is defined as (for $m_x+m_y>0$)
			\begin{equation}
				\widehat{C}_{\boldsymbol{m}}^{(n)}(\nu,\theta)=\int_{0}^{\infty}C_{\boldsymbol{m}}^{(n)}(\tilde{r},\theta)K_{i\nu}\left(\sqrt{2(m_x+m_y)}\,\tilde{r}\right)\frac{d\tilde{r}}{\tilde{r}}.
			\end{equation}
			For $\underline{\boldsymbol{n=1}}$, the coefficient equation is
			\begin{equation}
				\frac{1}{\tilde{r}}\frac{\partial}{\partial\tilde{r}}\left(\tilde{r}\frac{\partial C_{\boldsymbol{m}}^{(1)}}{\partial\tilde{r}}\right)+\frac{1}{\tilde{r}^2}\frac{\partial^2C_{\boldsymbol{m}}^{(1)}}{\partial\theta^2}-2(m_x+m_y)C_{\boldsymbol{m}}^{(1)}=0.
			\end{equation}
			Applying the KL transform, this coefficient equation becomes
			\begin{equation}
				\frac{\partial^2\widehat{C}_{\boldsymbol{m}}^{(1)}}{\partial\theta^2}=\nu^2\widehat{C}_{\boldsymbol{m}}^{(1)}.
			\end{equation}
			The zero current boundary condition at this order is
			\begin{equation}
				\frac{1}{\tilde{r}}\frac{\partial C_{\boldsymbol{m}}^{(1)}(\tilde{r},\pm\alpha)}{\partial\theta}=\mathcal{N}(\cos\alpha\delta_{m_x,0}\delta_{m_y,1}\mp\sin\alpha\delta_{m_x,1}\delta_{m_y,0}).
			\end{equation}
			Applying the KL transform and using Eq.\ (\ref{eq:K inu integral identity}), the boundary condition becomes
			\begin{align}
				\begin{split}
					\frac{\partial\widehat{C}_{\boldsymbol{m}}^{(1)}(\nu,\pm\alpha)}{\partial\theta}=&\ \frac{\mathcal{N}\pi}{2\sqrt{2}\cosh\nu\pi/2}\\
					&\times(\cos\alpha\delta_{m_x,0}\delta_{m_y,1}\mp\sin\alpha\delta_{m_x,1}\delta_{m_y,0}).
				\end{split}
			\end{align}
			The solution is
			\begin{widetext}
				\begin{equation}
					\widehat{C}_{\boldsymbol{m}}^{(1)}(\nu,\theta)=\ \frac{\mathcal{N}\pi\sqrt{2}\cos\alpha\sinh\nu\theta}{4\nu\cosh\frac{\nu\pi}{2}\cosh\nu\alpha}\delta_{m_x,0}\delta_{m_y,1}-\frac{\mathcal{N}\pi\sqrt{2}\sin\alpha\cosh\nu\theta}{4\nu\cosh\frac{\nu\pi}{2}\sinh\nu\alpha}\delta_{m_x,1}\delta_{m_y,0}.
				\end{equation}
				Taking the inverse KL transform, we have
				\begin{align}
					\begin{split}
						C_{\boldsymbol{m}}^{(1)}(\tilde{r},\theta)=&\ \frac{\mathcal{N}\sqrt{2}\cos\alpha\delta_{m_x,0}\delta_{m_y,1}}{\pi}\int_{0}^{\infty}\frac{\sinh\frac{\nu\pi}{2}\sinh\nu\theta}{\cosh\nu\alpha}K_{i\nu}(\sqrt{2}\tilde{r})d\nu\\
						&-\frac{\mathcal{N}\sqrt{2}\sin\alpha\delta_{m_x,1}\delta_{m_y,0}}{\pi}\int_{0}^{\infty}\frac{\sinh\frac{\nu\pi}{2}\cosh\nu\theta}{\sinh\nu\alpha}K_{i\nu}(\sqrt{2}\tilde{r})d\nu.
					\end{split}
				\end{align}
				For $\underline{\boldsymbol{n=2}}$, we will focus on the $(m_x,m_y)=(0,0)$ solution.  Using the $n=1$ solution and the recursion relations Eqs.\ (\ref{eq:bessel identities}), we have for the coefficient equation
				\begin{align}
					\begin{split}
						\frac{1}{\tilde{r}}\frac{\partial}{\partial\tilde{r}}\left(\tilde{r}\frac{\partial C_{0,0}^{(2)}}{\partial\tilde{r}}\right)+\frac{1}{\tilde{r}^2}\frac{\partial^2C_{0,0}^{(2)}}{\partial\theta^2}=&\ \frac{2\mathcal{N}}{\pi}\int_{0}^{\infty}\sinh\frac{\nu\pi}{2}\left(\frac{\sin\alpha}{\sinh\nu\alpha}+i\frac{\cos\alpha}{\cosh\nu\alpha}\right)\cos[(i\nu-1)\theta]K_{i\nu-1}(\sqrt{2}\tilde{r})d\nu\\
						&+\frac{2\mathcal{N}}{\pi}\int_{0}^{\infty}\sinh\frac{\nu\pi}{2}\left(\frac{\sin\alpha}{\sinh\nu\alpha}-i\frac{\cos\alpha}{\cosh\nu\alpha}\right)\cos[(i\nu+1)\theta]K_{i\nu+1}(\sqrt{2}\tilde{r})d\nu,
					\end{split}
				\end{align}
				and for the boundary condition
				\begin{equation}
					\frac{1}{\tilde{r}}\frac{\partial C_{0,0}^{(2)}(\tilde{r},\pm\alpha)}{\partial\theta}=\pm\frac{2\sqrt{2}\mathcal{N}}{\pi}\int_{0}^{\infty}\sinh\frac{\nu\pi}{2}\left(\frac{\sin^2\alpha\cosh\nu\alpha}{\sinh\nu\alpha}+\frac{\cos^2\alpha\sinh\nu\alpha}{\cosh\nu\alpha}\right)K_{i\nu}(\sqrt{2}\tilde{r})d\nu.
				\end{equation}
				The solution is
				\begin{align}
					\begin{split}
						C_{0,0}^{(2)}(\tilde{r},\theta)=\int_{-\infty}^{\infty}a(s)\cosh s\theta\, \tilde{r}^{-is}ds&+\frac{\mathcal{N}}{\pi}\int_{0}^{\infty}\sinh\frac{\nu\pi}{2}\left(\frac{\sin\alpha}{\sinh\nu\alpha}+i\frac{\cos\alpha}{\cosh\nu\alpha}\right)\cos[(i\nu-1)\theta]K_{i\nu-1}(\sqrt{2}\tilde{r})d\nu\\
						&+\frac{\mathcal{N}}{\pi}\int_{0}^{\infty}\sinh\frac{\nu\pi}{2}\left(\frac{\sin\alpha}{\sinh\nu\alpha}-i\frac{\cos\alpha}{\cosh\nu\alpha}\right)\cos[(i\nu+1)\theta]K_{i\nu+1}(\sqrt{2}\tilde{r})d\nu,
					\end{split}
					\label{eq:wedge correction general}
				\end{align}
				where $a(s)$ satisfies
				\begin{equation}
					\int_{-\infty}^{\infty}a(s)s\sinh s\alpha\, \tilde{r}^{-is-1}ds=\frac{\mathcal{N}\sqrt{2}}{\pi}\int_{0}^{\infty}\sinh\frac{\nu\pi}{2}\left(\frac{\sin^2\alpha\cosh\nu\alpha}{\sinh\nu\alpha}+\frac{\cos^2\alpha\sinh\nu\alpha}{\cosh\nu\alpha}\right)\frac{d^2}{d\tilde{r}^2}K_{i\nu}(\sqrt{2}\tilde{r})d\nu.
				\end{equation}
			\end{widetext}
			We can in principle determine $a(s)$ using the Mellin transform.  This can be done with a change of variables by defining $z=is+1$ and $A(z)=2\pi a\left(\frac{z-1}{i}\right)\frac{z-1}{i}\sinh\left(\frac{z-1}{i}\alpha\right)$.  The left-hand side integral then becomes
			\begin{equation}
				\int_{1-i\infty}^{1+i\infty}A(z)\tilde{r}^{-z}\frac{dz}{2\pi i},
			\end{equation}
			which can be inverted using the Mellin transform to obtain $A(z)$ and hence $a(s)$.
			
			In the main text (Section \ref{subsec:wedge}), we are interested in the average propulsion $\langle\boldsymbol{\eta}\rangle(r,\theta)$ within the wedge and the correction to density $\Delta(r,\theta)$ near the tip of the wedge.  These can be determined from $C_{1,0}^{(1)}$, $C_{0,1}^{(1)}$, and $C_{0,0}^{(2)}$ after restoring dimensions.  We can obtain slightly more simplified forms of these quantities if we consider angles of the form $2\alpha=\pi/2^{l-1}$.  To do that, we make use of the following identities for $l\ge2$
			\begin{widetext}
				\begin{subequations}
					\begin{align}
						\frac{\sinh\nu\pi/2}{\sinh\nu\pi/2^l}\cosh\nu\theta&=\sum_{k=0}^{2^{l-2}-1}\left\{\cosh\nu\left[\frac{(2^{l-1}-1-2k)\pi}{2^l}+\theta\right]+\cosh\nu\left[\frac{(2^{l-1}-1-2k)\pi}{2^l}-\theta\right]\right\},\\
						\frac{\sinh\nu\pi/2}{\sinh\nu\pi/2^l}\sinh\nu\theta&=\sum_{k=0}^{2^{l-2}-1}\left\{\sinh\nu\left[\frac{(2^{l-1}-1-2k)\pi}{2^l}+\theta\right]-\sinh\nu\left[\frac{(2^{l-1}-1-2k)\pi}{2^l}-\theta\right]\right\},\\
						\frac{\sinh\nu\pi/2}{\cosh\nu\pi/2^l}\sinh\nu\theta&=\sum_{k=0}^{2^{l-2}-1}(-1)^k\left\{\cosh\nu\left[\frac{(2^{l-1}-1-2k)\pi}{2^l}+\theta\right]-\cosh\nu\left[\frac{(2^{l-1}-1-2k)\pi}{2^l}-\theta\right]\right\},\\
						\frac{\sinh\nu\pi/2}{\cosh\nu\pi/2^l}\cosh\nu\theta&=\sum_{k=0}^{2^{l-2}-1}(-1)^k\left\{\sinh\nu\left[\frac{(2^{l-1}-1-2k)\pi}{2^l}+\theta\right]+\sinh\nu\left[\frac{(2^{l-1}-1-2k)\pi}{2^l}-\theta\right]\right\},	
					\end{align}
				\end{subequations}
			\end{widetext}
			in addition to \cite{Oberhettinger KL}
			\begin{subequations}
				\begin{align}
					&\int_{0}^{\infty}\cosh a\nu\,K_{i\nu}(\sqrt{2}\tilde{r})d\nu=\frac{\pi}{2}e^{-\sqrt{2}\tilde{r}\cos a},\\
					&\int_{0}^{\infty}\nu\sinh a\nu\,K_{i\nu}(\sqrt{2}\tilde{r})d\nu=\frac{\pi}{2}\sqrt{2}\tilde{r}\sin a\,e^{-\sqrt{2}\tilde{r}\cos a}.
				\end{align}
			\end{subequations}
	
	\section{\label{app:corrugated wall}Solution for a corrugated wall}
		We start with a boundary deformed around $y=0$.  Suppose its shape is given by $h(x)$, which has period $2L$ and characteristic amplitude $\delta$.  This shape can be decomposed into Fourier modes as
		\begin{equation}
			h(x)=\delta\sum_{k=-\infty}^{\infty}h_ke^{\frac{i\pi k}{L}x},
		\end{equation}
		where we assume that $h_0=0$ and
		\begin{equation}
			h_k=\frac{1}{2L\delta}\int_{-L}^{L}h(x)e^{-\frac{i\pi k}{L}x}dx.
		\end{equation}
		We can solve this particular case in Cartesian coordinates.  Just as before, we write the distribution as
		\begin{equation}
			\tilde{\rho}(\tilde{\boldsymbol{r}},\tilde{\boldsymbol{\eta}})=\sum_{n=0}^{\infty}\epsilon^n\sum_{\boldsymbol{m}}C_{\boldsymbol{m}}^{(n)}(\tilde{\boldsymbol{r}})e^{-\tilde{\boldsymbol{\eta}}^2}H_{m_x}(\tilde{\eta}_x)H_{m_y}(\tilde{\eta}_y),
		\end{equation}
		except now we expand the coefficients $C_{\boldsymbol{m}}^{(n)}$ as
		\begin{equation}
			C_{\boldsymbol{m}}^{(n)}(\tilde{\boldsymbol{r}})\simeq a_{\boldsymbol{m}}^{(n)}(\tilde{y})+\tilde{\delta}\sum_{k=-\infty}^{\infty}b_{\boldsymbol{m};k}^{(n)}(\tilde{y})e^{\frac{i\pi k}{\tilde{L}}\tilde{x}},
			\label{eq:coef C delta expansion}
		\end{equation}
		where $a_{\boldsymbol{m}}^{(n)}$ is the solution for a flat boundary, the first two orders of which are
		\begin{subequations}
			\begin{align}
				a_{0,0}^{(0)}(\tilde{y})&=\mathcal{N},\\
				a_{0,1}^{(1)}(\tilde{y})&=-\frac{\mathcal{N}\sqrt{2}}{2}e^{-\sqrt{2}\tilde{y}},\\
				a_{0,0}^{(2)}(\tilde{y})&=\mathcal{N}e^{-\sqrt{2}\tilde{y}}.
			\end{align}
		\end{subequations}
		The coefficients $b_{\boldsymbol{m};k}^{(n)}(\tilde{y})$ satisfy the ODE
		\begin{equation}
			\frac{d^2b_{\boldsymbol{m};k}^{(n)}}{d\tilde{y}^2}-\left[2(m_x+m_y)+\frac{\pi^2k^2}{\tilde{L}^2}\right]b_{\boldsymbol{m};k}^{(n)}=\frac{i\pi k}{\tilde{L}}w_x+\frac{dw_y}{d\tilde{y}},
		\end{equation}
		where
		\begin{subequations}
			\begin{align}
				w_x&=b_{m_x-1,m_y;k}^{(n-1)}+2(m_x+1)b_{m_x+1,m_y;k}^{(n-1)},\\
				w_y&=b_{m_x,m_y-1;k}^{(n-1)}+2(m_y+1)b_{m_x,m_y+1;k}^{(n-1)}.
			\end{align}
		\end{subequations}
		Finally, for the boundary condition, we require that the normal component of the current at the boundary to be zero.  Assuming that the function describing the shape of the boundary is single-valued, the normal to the boundary is
		\begin{equation}
			\hat{\boldsymbol{n}}=\frac{(-\tilde{h}'(\tilde{x}),1)}{\sqrt{1+\tilde{h}'(\tilde{x})^2}}.
		\end{equation}
		Therefore, the zero current boundary condition along the wall, $\tilde{\boldsymbol{J}}(\tilde{x},\tilde{h}(\tilde{x}),\tilde{\boldsymbol{\eta}})\cdot\hat{\boldsymbol{n}}=0$, is
		\begin{equation}
			-\tilde{J}_x(\tilde{x},\tilde{h}(\tilde{x}),\tilde{\boldsymbol{\eta}})\frac{d\tilde{h}}{d\tilde{x}}+\tilde{J}_y(\tilde{x},\tilde{h}(\tilde{x}),\tilde{\boldsymbol{\eta}})=0.
			\label{eq:J_perp=0 bc corrugated wall}
		\end{equation}
		To make progress, we assume that the amplitude of the corrugation is small compared to the accumulation of active particles or $\tilde{\delta}=\delta/\lambda\ll1$ so that we can linearize the boundary condition Eq.\ (\ref{eq:J_perp=0 bc corrugated wall}).  The currents $\tilde{J}_x(\tilde{x},\tilde{h},\tilde{\boldsymbol{\eta}})$ and $\tilde{J}_y(\tilde{x},\tilde{h},\tilde{\boldsymbol{\eta}})$ are given by
		\begin{widetext}
			\begin{subequations}
				\begin{align}
					\tilde{J}_x(\tilde{x},\tilde{h},\tilde{\boldsymbol{\eta}})&=\sum_{n=0}^{\infty}\epsilon^n\sum_{\boldsymbol{m}}\left[C_{m_x-1,m_y}^{(n-1)}(\tilde{x},\tilde{h})+2(m_x+1)C_{m_x+1,m_y}^{(n-1)}(\tilde{x},\tilde{h})-\frac{dC_{\boldsymbol{m}}^{(n)}(\tilde{x},\tilde{h})}{d\tilde{x}}\right]e^{-\tilde{\boldsymbol{\eta}}^2}H_{m_x}(\tilde{\eta}_x)H_{m_y}(\tilde{\eta}_y),\\
					\tilde{J}_y(\tilde{x},\tilde{h},\tilde{\boldsymbol{\eta}})&=\sum_{n=0}^{\infty}\epsilon^n\sum_{\boldsymbol{m}}\left[C_{m_x,m_y-1}^{(n-1)}(\tilde{x},\tilde{h})+2(m_y+1)C_{m_x,m_y+1}^{(n-1)}(\tilde{x},\tilde{h})-\frac{dC_{\boldsymbol{m}}^{(n)}(\tilde{x},\tilde{h})}{d\tilde{y}}\right]e^{-\tilde{\boldsymbol{\eta}}^2}H_{m_x}(\tilde{\eta}_x)H_{m_y}(\tilde{\eta}_y).
				\end{align}
			\end{subequations}
			Substituting these currents into the boundary condition Eq.\ (\ref{eq:J_perp=0 bc corrugated wall}) and using the orthogonality of Hermite polynomials, we obtain for each order $n$
			\begin{align}
				\begin{split}
					&-\left[C_{m_x-1,m_y}^{(n-1)}(\tilde{x},\tilde{h})+2(m_x+1)C_{m_x+1,m_y}^{(n-1)}(\tilde{x},\tilde{h})-\frac{dC_{\boldsymbol{m}}^{(n)}(\tilde{x},\tilde{h})}{d\tilde{x}}\right]\frac{d\tilde{h}}{d\tilde{x}}\\
					&+\left[C_{m_x,m_y-1}^{(n-1)}(\tilde{x},\tilde{h})+2(m_y+1)C_{m_x,m_y+1}^{(n-1)}(\tilde{x},\tilde{h})-\frac{dC_{\boldsymbol{m}}^{(n)}(\tilde{x},\tilde{h})}{d\tilde{y}}\right]=0
				\end{split}
			\end{align}
		\end{widetext}
		Finally, inserting the expansion of $C_{\boldsymbol{m}}^{(n)}(\tilde{x},\tilde{h})$ in terms of $a_{\boldsymbol{m}}^{(n)}(\tilde{h})$ and $b_{\boldsymbol{m};k}^{(n)}(\tilde{h})$ (Eq.\ \ref{eq:coef C delta expansion}) and Taylor expanding everything about $\tilde{\delta}=0$, we can collect all terms of order $\tilde{\delta}$ and use the orthogonality of the Fourier modes to obtain the boundary condition for the coefficients $b_{\boldsymbol{m};k}^{(n)}(\tilde{y})$ shown in the main text (Eq.\ \ref{eq:corrugated boundary condition}).  Note that the zeroth order boundary condition, which corresponds to that of a flat wall, is already satisfied.
		
		Solving for $b_{\boldsymbol{m};k}^{(n)}(\tilde{y})$ is straightforward.  We find
		\begin{subequations}
			\begin{align}
				b_{0,0;k}^{(0)}(\tilde{y})=\,&0\\
				b_{1,0;k}^{(1)}(\tilde{y})=\,&\frac{i\mathcal{N}\pi kh_k}{\tilde{L}\sqrt{2+\frac{\pi^2k^2}{\tilde{L}^2}}}e^{-\sqrt{2+\frac{\pi^2k^2}{\tilde{L}^2}}\tilde{y}}\\
				b_{0,1;k}^{(1)}(\tilde{y})=\,&-\frac{\mathcal{N}h_k\sqrt{2}}{\sqrt{2+\frac{\pi^2k^2}{\tilde{L}^2}}}e^{-\sqrt{2+\frac{\pi^2k^2}{\tilde{L}^2}}\tilde{y}}\\
				\begin{split}
					b_{0,0;k}^{(2)}(\tilde{y})=\,&\frac{\mathcal{N}\pi|k|h_k}{\tilde{L}}\left(1-\frac{\sqrt{2}}{\sqrt{2+\frac{\pi^2k^2}{\tilde{L}^2}}}\right)e^{-\frac{\pi|k|}{\tilde{L}}\tilde{y}}\\
					&+\mathcal{N}h_k\left(\sqrt{2}-\frac{\pi^2k^2}{\tilde{L^2}\sqrt{2+\frac{\pi^2k^2}{\tilde{L}^2}}}\right)e^{-\sqrt{2+\frac{\pi^2k^2}{\tilde{L}^2}}\tilde{y}}
				\end{split}
				\label{eq:corrugated b coefficient}
			\end{align}
		\end{subequations}
		For the asymmetric sawtooth in the main text, the Fourier amplitudes are $h_0=0$ and
		\begin{equation}
			h_k=\frac{2\left[e^{-i\pi k\zeta}-(-1)^k\right]}{\pi^2k^2(1-\zeta^2)}.
		\end{equation}
		The spacial currents described in the main text (Section \ref{sec:corrugated wall} and Figure \ref{fig:corrugated currents}) can be written as
		\begin{subequations}
			\begin{align}
				\begin{split}
					\tilde{J}_x(\tilde{x},\tilde{y})&=\pi\epsilon^2\tilde{\delta}\sum_{k=-\infty}^{\infty}\left[2b_{1,0;k}^{(1)}(\tilde{y})-\frac{i\pi k}{\tilde{L}}b_{0,0;k}^{(2)}(\tilde{y})\right]e^{\frac{i\pi k}{\tilde{L}}\tilde{x}},
				\end{split}\\
				\begin{split}
					\tilde{J}_y(\tilde{x},\tilde{y})&=\pi\epsilon^2\tilde{\delta}\sum_{k=-\infty}^{\infty}\left[2b_{0,1;k}^{(1)}(\tilde{y})-\frac{db_{0,0;k}^{(2)}(\tilde{y})}{d\tilde{y}}\right]e^{\frac{i\pi k}{\tilde{L}}\tilde{x}}.
				\end{split}
			\end{align}
		\end{subequations}
		Since $h_0=0$ and there is no $k=0$ contribution to the currents, it is easy to see that averaging over a period $2\tilde{L}$ gives zero net flux along the wall.
		
		As discussed in the main text, we can obtain the density at the tip of a wedge with angle close to $2\alpha=\pi$.  Consider a symmetric sawtooth-shaped boundary ($\zeta=0$) with a small amplitude ($\delta\ll\lambda$) and a long wavelength ($L\gg\lambda$).  Writing the density as $\rho(x,y)=\rho_{\textrm{bulk}}[1+\epsilon^2\Delta(x,y)]$, we can compute the density at the tip $\Delta(0,\delta)$ up to order $\delta$.  We find
		\begin{align}
			\begin{split}
				\Delta_{\textrm{tip}}\simeq1+\frac{4\delta}{\pi L}&\sum_{k=1}^{\infty}\frac{[1-(-1)^k]}{k}\\
				&\times\left(1-\frac{\sqrt{2}}{\sqrt{2+\frac{\pi^2k^2\lambda^2}{L^2}}}-\frac{\pi k\lambda}{L\sqrt{2+\frac{\pi^2k^2\lambda^2}{L^2}}}\right)
			\end{split}
		\end{align}
		For $L\gg\lambda$, the summation weakly depends on $L$ and can be well approximated by an integral.  We thus have
		\begin{align}
			\begin{split}
				\Delta_{\textrm{tip}}&\approx1+\frac{4\delta}{\pi L}\int_{0}^{\infty}\frac{dt}{t}\left[1-\frac{\sqrt{2}}{\sqrt{2+t^2}}-\frac{t}{\sqrt{2+t{^2}}}\right]\\
				&\approx1-0.44(2\alpha-\pi),
			\end{split}
		\end{align}
		where we used $2\alpha\simeq\pi+\frac{4\delta}{L}$.
	
	\section{\label{app:absorber}Solution for a spherical absorber}
		For problems requiring spherical coordinates, we write the density as
		\begin{align}
			\begin{split}
				\tilde{\rho}(\tilde{r},\theta,\phi,\tilde{\boldsymbol{\eta}})=\sum_{n=0}^{\infty}\epsilon^n&\sum_{\boldsymbol{m}}C_{\boldsymbol{m}}^{(n)}(\tilde{r},\theta,\phi)\\
				&\times e^{-\tilde{\boldsymbol{\eta}}^2}H_{m_x}(\tilde{\eta}_x)H_{m_y}(\tilde{\eta}_y)H_{m_z}(\tilde{\eta}_z).
			\end{split}
		\end{align}
		The coefficients $C_{\boldsymbol{m}}^{(n)}$ satisfy
		\begin{align}
			\begin{split}
				&\frac{1}{\tilde{r}^2}\frac{\partial}{\partial\tilde{r}}\left(\tilde{r}^2\frac{\partial C_{\boldsymbol{m}}^{(n)}}{\partial\tilde{r}}\right)-2(m_x+m_y+m_z)C_{\boldsymbol{m}}^{(n)}\\
				&+\frac{1}{\tilde{r}^2\sin\theta}\frac{\partial}{\partial\theta}\left(\sin\theta\frac{\partial C_{\boldsymbol{m}}^{(n)}}{\partial\theta}\right)+\frac{1}{\tilde{r}^2\sin^2\theta}\frac{\partial^2C_{\boldsymbol{m}}^{(n)}}{\partial\phi^2}\\
				&=\frac{1}{\tilde{r}^2}\frac{\partial}{\partial\tilde{r}}(\tilde{r}^2w_r)+\frac{1}{\tilde{r}\sin\theta}\frac{\partial}{\partial\theta}(\sin\theta w_{\theta})+\frac{1}{\tilde{r}\sin\theta}\frac{\partial w_{\phi}}{\partial\phi},
			\end{split}
		\end{align}
		where the components of $\boldsymbol{w}$ in spherical coordinates are
		\begin{subequations}
			\begin{align}
				w_r&=w_x\sin\theta\cos\phi+w_y\sin\theta\sin\phi+w_z\cos\theta,\\
				w_{\theta}&=w_x\cos\theta\cos\phi+w_y\cos\theta\sin\phi-w_z\sin\theta,\\
				w_{\phi}&=-w_x\sin\phi+w_y\cos\phi,
			\end{align}
		\end{subequations}
		and
		\begin{subequations}
			\begin{align}
				w_x&=C_{m_x-1,m_y,m_z}^{(n-1)}+2(m_x+1)C_{m_x+1,m_y,m_z}^{(n-1)},\\
				w_y&=C_{m_x,m_y-1,m_z}^{(n-1)}+2(m_y+1)C_{m_x,m_y+1,m_z}^{(n-1)},\\
				w_z&=C_{m_x,m_y,m_z-1}^{(n-1)}+2(m_z+1)C_{m_x,m_y,m_z+1}^{(n-1)}.
			\end{align}
		\end{subequations}
		For the absorbing boundary condition, we have $C_{\boldsymbol{m}}^{(n)}(\tilde{R},\theta,\phi)=0$.  For $\underline{\boldsymbol{n=0}}$, we have the usual density profile for passive particles around an absorbing sphere given by
		\begin{equation}
			C_{\boldsymbol{m}}^{(0)}(\tilde{r},\theta,\phi)=\mathcal{N}\left(1-\frac{\tilde{R}}{\tilde{r}}\right)\delta_{m_x,0}\delta_{m_y,0}\delta_{m_z,0},
		\end{equation}
		where $\mathcal{N}=\rho_{\textrm{bulk}}(2D_p\tau/\pi)^{3/2}$.  For $\underline{\boldsymbol{n=1}}$, we have
		\begin{subequations}
			\begin{align}
				C_{1,0,0}^{(n)}(\tilde{r},\theta,\phi)&=\frac{\mathcal{N}}{2\tilde{R}}\left[\frac{k_1(\sqrt{2}\tilde{r})}{k_1(\sqrt{2}\tilde{R})}-\frac{\tilde{R}^2}{\tilde{r}^2}\right]\sin\theta\cos\phi,\\
				C_{0,1,0}^{(1)}(\tilde{r},\theta,\phi)&=\frac{\mathcal{N}}{2\tilde{R}}\left[\frac{k_1(\sqrt{2}\tilde{r})}{k_1(\sqrt{2}\tilde{R})}-\frac{\tilde{R}^2}{\tilde{r}^2}\right]\sin\theta\cos\phi,\\
				C_{0,0,1}^{(1)}(\tilde{r},\theta,\phi)&=\frac{\mathcal{N}}{2\tilde{R}}\left[\frac{k_1(\sqrt{2}\tilde{r})}{k_1(\sqrt{2}\tilde{R})}-\frac{\tilde{R}^2}{\tilde{r}^2}\right]\cos\theta,
			\end{align}
		\end{subequations}
		where $k_{\mu}$ is the modified spherical Bessel function of the second kind.  For $\underline{\boldsymbol{n=2}}$, we will only write the $(m_x,m_y,m_z)=(0,0,0)$ term since we are only interested in the density and current.  We have
		\begin{equation}
			C_{0,0,0}^{(2)}(\tilde{r},\theta,\phi)=\frac{\mathcal{N}\sqrt{2}k_0(\sqrt{2}\tilde{R})}{2\tilde{R}k_1(\sqrt{2}\tilde{R})}\left[\frac{\tilde{R}}{\tilde{r}}-\frac{k_0(\sqrt{2}\tilde{r})}{k_0(\sqrt{2}\tilde{R})}\right].
		\end{equation}
		Integrating out $\tilde{\boldsymbol{\eta}}$, we have for the density
		\begin{align}
			\begin{split}
				\tilde{\rho}(\tilde{r},\theta,\phi)\simeq&\ \pi^{3/2}C_{0,0,0}^{(0)}(\tilde{r},\theta,\phi)+\epsilon^2\pi^{3/2}C_{0,0,0}^{(2)}(\tilde{r},\theta,\phi)\\
				=&\ \mathcal{N}\pi^{3/2}\left(1-\frac{\tilde{R}}{\tilde{r}}\right)\\
				&+\epsilon^2\frac{\mathcal{N}\pi^{3/2}}{1+\sqrt{2}\tilde{R}}\left[1-e^{-\sqrt{2}(\tilde{r}-\tilde{R})}\right]\frac{\tilde{R}}{\tilde{r}}.
			\end{split}
		\end{align}
		The radial current is given by
		\begin{align}
			\tilde{J}_r(\tilde{r},\theta,\phi)\simeq-\frac{\partial C_{0,0,0}^{(0)}}{\partial\tilde{r}}\pi^{3/2}+\epsilon^2\pi^{3/2}\left(w_r-\frac{\partial C_{0,0,0}^{(2)}}{\partial\tilde{r}}\right),
		\end{align}
		where
		\begin{equation}
			w_r=2C_{1,0,0}^{(1)}\sin\theta\cos\phi+2C_{0,1,0}^{(1)}\sin\theta\sin\phi+2C_{0,0,1}^{(1)}\cos\theta.
		\end{equation}
		Substituting everything in, we arrive at
		\begin{equation}
			\tilde{J}_r(\tilde{r},\theta,\phi)=-\frac{\mathcal{N}\pi^{3/2}\tilde{R}}{\tilde{r}^2}\left(1+\epsilon^2\frac{\sqrt{2}\tilde{R}}{1+\sqrt{2}\tilde{R}}\right).
		\end{equation}


\begin{thebibliography}{10}
		\bibitem{Ramaswamy}
			S. Ramaswamy, The Mechanics and Statics of Active Matter, Annu. Rev. Condens. Matter Phys. \textbf{1}, 323 (2010)
		\bibitem{Bechinger et al}
			C. Bechinger, R. Di Leonardo, H. L\"{o}wen, C. Reinhardt, G. Volpe, G. Volpe, Active particles in complex and crowded environments, Rev. Mod. Phys. \textbf{88}, 045006 (2016)
		\bibitem{Silverberg et al}
			J. L. Silverberg, M. Bierbaum, J. P. Sethna, I. Cohen, Collective Motion of Humans in Mosh and Circle Pits at Heavy Metal Concerts, Phys. Rev. Lett. \textbf{110}, 228701 (2013)
		\bibitem{Berg et al}
			H. C. Berg, D. A. Brown, Chemotaxis in \textit{Escherichia coli} analysed by three-dimensional tracking, Nature \textbf{239}, 500 (1972)
		\bibitem{Polin et al}
			M. Polin, I. Tuval, K. Drescher, J. P. Gollub, R. E. Goldstein, \textit{Chlamydomonas} Swims with Two ``Gears'' in a Eukaryotic Version of Run-and-Tumble Locomotion, Science \textbf{325}, 487 (2009)
		\bibitem{Palacci et al}
			J. Palacci, S. Sacanna, S. H. Kim, G. R. Yi, D. J. Pine, P. M. Chaikin, Light-activated self-propelled colloids, Philos. Trans. R. Soc. A \textbf{372}, 20130372 (2014)
		\bibitem{Paxton et al}
			W. F. Paxton, K. C. Kistler, C. C. Olmeda, A. Sen, S. K St Angelo, T. Cao, T. E. Mallouk, P. E. Lammert, V. H. Crespi, Catalytic nanomotors: autonomous movement of striped nanorods, J. Am. Chem. Soc. \textbf{126}, 13424 (2004)
		\bibitem{Walsh et al}
			L. Walsh, C. G. Wagner, S. Schlossberg, C. Olsen, A. Baskaran, N. Menon, Noise and diffusion of a vibrated self-propelled granular particle, Soft Matter, \textbf{13}, 8964 (2017)
		\bibitem{Buttinoni et al}
			I. Buttinoni, J. Bialk\'{e}, F. K\"{u}mmel, H. L\"{o}wen, C. Bechinger, T. Speck, Dynamical Clustering and Phase Separation in Suspensions of Self-Propelled Colloidal Particles, Phys. Rev. Lett. \textbf{110}, 238301 (2013)
		\bibitem{Galajda et al}
			P. Galajda, J. Keymer, P. Chaikin, R. Austin, A Wall of Funnels Concentrates Swimming Bacteria, J. Bacterio. \textbf{189}, 8704 (2007)
		\bibitem{Sokolov et al}
			A. Sokolov, M. M. Apodaca, B. A. Grzybowski, I. S. Aranson, Swimming bacteria power microscopic gears, Proc. Natl. Acad. Sci. \textbf{107}, 969 (2010)
		\bibitem{Leonardo et al}
			R. Di Leonardo, L. Angelani, D. Dell'Arciprete, G. Ruocco, V. Lebba, S. Schippa, M. P. Conte, F. Mecarini, F. De Angelis, E. Di Fabrizio, Bacterial ratchet motors, Proc. Natl. Acad. Sci. \textbf{107}, 9541 (2010)
		\bibitem{Volpe et al}
			G. Volpe, I. Buttinoni, D. Vogt, H.-J. K\"{u}mmerer, C. Bechinger, Microswimmers in patterned environments, Soft Matter \textbf{7}, 8810 (2011)
		\bibitem{Li Tang}
			G. Li, J. X. Tang, Accumulation of Microswimmers near a Surface Mediated by Collision and Rotational Brownian Motion, Phys. Rev. Lett. \textbf{103}, 078101 (2009)
		\bibitem{Lee}
			C. F. Lee, Active particles under confinement: aggregation at the wall and gradient formation inside a channel, New J. Phys. \textbf{15}, 055007 (2013)
		\bibitem{Wagner et al}
			C. G. Wagner, M. F. Hagan, A. Baskaran, Steady-state distributions of active Brownian particles under confinement and forcing, J. Stat. Mech., 043203 (2017)
		\bibitem{Ezhilan et al}
			B. Ezhilan, R. Alonso-Matilla, D. Saintillan, On the distribution and swim pressure of run-and-tumble particles in confinement, J. Fluid. Mech. \textbf{781}, R4 (2015)
		\bibitem{Fisch Kruskal}
			N. J. Fisch, M. D. Kruskal, Separating variables in two-way diffusion equations, J. Math. Phys. \textbf{21}, 740 (1980)
		\bibitem{Beals}
			R. Beals, Partial-range completeness and existence of solutions to two-way diffusion equation, J. Math. Phys. \textbf{22}, 954 (1981)
		\bibitem{Wagner Beals}
			C. G. Wagner, R. Beals, Constructing solutions to two-way diffusion problems, J. Phys. A: Math Theor. \textbf{52}, 115204 (2019)
		\bibitem{Solon et al pressure}
			A. P. Solon, Y. Fily, M. E. Cates, Y. Kafri, M. Kardar, J. Tailleur, Pressure is not a state function for generic active fluids, Nat. Phys. \textbf{11}, 673 (2015)
		\bibitem{Caprini Marconi}
			L. Caprini, U. M. B. Marconi, Active particles under confinement and effective force generation among surfaces, Soft Matter \textbf{14}, 9044 (2018)
		\bibitem{Marconi et al}
			U. M. B. Marconi, A. Sarracino, C. Maggi, A. Puglisi, Self-propulsion against a moving membrane: Enhanced accumulation and drag force, Phys. Rev. E \textbf{96}, 032601 (2017)
		\bibitem{Battle et al}
			C. Battle, C. P. Broedersz, N. Fakhri, V. F. Geyer, J. Howard, C. F. Schmidt, F. C. MacKintosh, Broken detailed balance at mesoscopic scales in active biological systems, Science \textbf{352}, 604 (2016)
		\bibitem{Zakine et al}
			R. Zakine, Y. Zhao, M. Knezevic, A. Daerr, Y. Kafri, J. Tailleur, F. van Wijland, Surface Tensions between Active Fluids and Solid Interfaces: Bare and Dressed, Phys. Rev. Lett. \textbf{124}, 248003 (2020)
		\bibitem{Grosberg and Joanny}
			A. Y. Grosberg, J.-F. Joanny, Nonequilibrium statistical mechanics of mixtures of particles in contact with different thermostats, Phys. Rev. E \textbf{92}, 032118 (2015)
		\bibitem{Duzgan Sellinger}
			A. Duzgan, J. V. Sellinger, Active Brownian particles near straight and curved walls: Pressure and boundary layers, Phys. Rev. E \textbf{97}, 032606 (2018)
		\bibitem{Solon et al laplace}
			A. P. Solon, J. Stenhammar, M. E. Cates, Y. Kafri, J. Tailleur, Generalized thermodynamics of motility-induced phase separation: phase equilibria, Laplace pressure, and change of ensembles, New. J. Phys. \textbf{20}, 075001 (2018)
		\bibitem{Wittmann et al}
			R. Wittmann, F. Smallenburg, J. M. Brader, Pressure, surface tension, and curvature in active systems: a touch of equilibrium, J. Chem. Phys. \textbf{150}, 174908 (2019)
		\bibitem{Nikola et al}
			N. Nikola, A. P. Solon, Y. Kafri, M. Kardar, J. Tailleur, R. Voituriez, Active Particles with Soft and Curved Walls: Equation of State, Ratchets, and Instabilities, Phys. Rev. Lett. \textbf{117}, 098001 (2016)
		\bibitem{Sandford et al}
			C. Sandford, A. Y. Grosberg, J.-F. Joanny, Pressure and flow of exponentially self-correlated active particles, Phys. Rev. E \textbf{96}, 052605 (2017)
		\bibitem{Fily et al}
			Y. Fily, A. Baskaran, M. F. Hagan, Dynamics of self-propelled particles under strong confinement, Soft Matter \textbf{10}, 5609 (2014)
		\bibitem{Kaiser et al}
			A. Kaiser, H. H. Wensink, H. L\"{o}wen, How to capture active particles, Phys. Rev. Lett. \textbf{108}, 268307 (2012)
		\bibitem{Smoluchowski}
			M. von Smoluchowski, Phys. Z. \textbf{17}, 557 (1916); Z. Phys. Chem. \textbf{92}, 129 (1917)
		\bibitem{Hubner Titulaer}
			G. F. Hubner, U. M. Titulaer, The Kinetic Boundary Layer for the Linearized Boltzmann Equation around an Absorbing Sphere, J. Stat. Phys. \textbf{59}, 441 (1990)
		\bibitem{Ghosh et al}
			P. K. Ghosh, V. R. Misko, F. Marchesoni, F. Nori, Self-Propelled Janus Particles in a Ratchet: Numerical Simulations, Phys. Rev. Lett. \textbf{110}, 268301 (2013)
		\bibitem{Ai}
			B. Ai, Ratchet transport powered by chiral active particles, Sci. Rep. \textbf{6}, 18740 (2016)
		\bibitem{Ai et al}
			B. Ai, Y. He, W. Zhong, Entropic Ratchet transport of interacting active Brownian particles, J. Chem. Phys. \textbf{141}, 194111 (2014)
		\bibitem{Mallory et al}
			S. A. Mallory, C. Valeriani, A. Cacciuto, An Active Approach to Colloidal Self-Assembly, Annu. Rev. Phys. Chem. \textbf{69}, 59 (2018)
		\bibitem{Figueroa-Morales et al}
			N. Figueroa-Morales, R. Soto, G. Junot, T. Darnige, C. Douarche, V. A. Martinez, A. Lindner, \'{E}. Cl\'{e}ment, 3D Spacial Exploration by \textit{E.\ coli} Echoes Motor Temporal Variability, Phys. Rev. X \textbf{10}, 021004 (2020)
		\bibitem{Ginot et al}
			F. Ginot, I. Theurkauff, D. Levis, C. Ybert, L. Bocquet, L. Berthier, C. Cottin-Bizonne, Nonequilibrium Equation of State in Suspensions of Active Colloids, Phys. Rev. X \textbf{5}, 011004 (2015)
		\bibitem{Kontorovich and Lebedev}
			M. J. Kontorovich, N. N. Lebedev, A method for the solution of problems in diffraction theory and related topics, Zh. Eksper. Teor. Fiz. \textbf{8}, 1192 (1938)
		\bibitem{Forristall Ingram}
			G. Z. Forristall, J. D. Ingram, Elastodynamics of a wedge, Bull. Seism. Soc. Amer. \textbf{61}, 275 (1971)
		\bibitem{Kang et al}
			K. H. Kang, I. S. Kang, C. M. Lee, Geometry Dependence of Wetting Tension on Charge-Modified Surfaces, Langmuir \textbf{19}, 6881 (2003)
		\bibitem{Fowkes et al}
			N. D. Fowkes, M. J. Hood, Surface Tension Effects in a Wedge, Q. J. Mech. Appl. Math. \textbf{51}, 553 (1998)
		\bibitem{Smith}
			J. H. Smith, Solution of the Inhomogeneous Helmholtz Equation with Lebedev Transforms, J. Math. Phys. \textbf{47}, 442 (1968)
		\bibitem{Oberhettinger KL}
			F. Oberhettinger, Tables of Bessel Transforms, Springer-Verlag (1972)
		\bibitem{Oberhettinger M}
			F. Oberhettinger, Tables of Mellin Transforms, Springer-Verlag (1974)
	\end{thebibliography}
\end{document}